%% file: main.tex
\documentclass[a4paper,titlepage]{article}
\input{Packages.tex}

\makeatother
\begin{document}
\center
\justifying
\input{cover}

\newpage

\tableofcontents
\newpage


\input{Selection_of_the_Generation_Algorithms}
\newpage

\input{Review_of_the_state_of_art}

\newpage

\input{Meshgen2_0}
\newpage

\input{Trimpack}

\printbibliography[
heading=bibnumbered,
title={References}
]

\end{document}

%% file: Packages.tex
\usepackage[utf8]{inputenc}
\usepackage{graphicx}
\usepackage[hidelinks]{hyperref}

\usepackage{gensymb}
\usepackage{float}
\usepackage{tabularx}

\usepackage{xcolor}

\usepackage{geometry}
\usepackage{fancyhdr}
\usepackage{ragged2e}

\usepackage{matlab-prettifier}

\usepackage[
backend=biber,
sorting=ynt
]{biblatex}

\usepackage{nomencl}

\usepackage{enumitem}
\usepackage{enumitem}

\usepackage{multirow}
\usepackage{booktabs}
\usepackage{longtable}
\usepackage{hhline}

\usepackage{amsmath}
\usepackage{bm} 
\usepackage{amsfonts}

\usepackage{caption}
\usepackage{subcaption}

\usepackage{chngcntr}

\usepackage{appendix}

\usepackage{graphicx}

\newenvironment{changemargin}[2]{%
\begin{list}{}{%
\setlength{\topsep}{0pt}%
\setlength{\leftmargin}{#1}%
\setlength{\rightmargin}{#2}%
\setlength{\listparindent}{\parindent}%
\setlength{\itemindent}{\parindent}%
\setlength{\parsep}{\parskip}%
}%
\item[]}{\end{list}}

%% file: cover.tex
 \begin{center}
       \vspace*{1.5cm}
\begin{changemargin}{1cm}{1cm}
\centering
 
       \textbf{\Huge Trimpack}

       \vspace{0.5cm}
       
       \textbf{\Huge Unstructured Triangular Mesh Generation Library}

       \vspace{1.5cm}
       \textsc{
        \large
       Juan M. Tizón \footnote{Corresponding author. E-mail: \href{mailto:jm.tizon@upm.es}{jm.tizon@upm.es} - ORCID-ID: \url{https://orcid.org/0000-0002-8687-6657}},
       Nicolás Becerra\footnote{E-mail:\href{mailto:becerrazun.n@gmail.com}{becerrazun.n@gmail.com} - ORCID-ID: \url{https://orcid.org/0000-0002-4929-2206}},
       Daniel Bercebal,
      Claus-Peter Grabowsky \footnote{E-mail:\href{mailto:clausg@gmx.com}{clausg@gmx.com}}
       \normalsize
       }
        \vspace{1cm}

\textsc{      Departamento de Mecánica de Fluidos y Propulsión Aeroespacial,
    Escuela Técnica Superior de Ingeniería Aeronáutica y del Espacio (ETSIAE), Universidad Politécnica de Madrid (UPM), Pza. del Cardenal Cisneros 3, 28040 Madrid, Spain}
    
    \end{changemargin}

    \vspace{2cm}

    \section*{Abstract}

    \justifying
    \begin{changemargin}{2cm}{2cm}
        
\textit{   
    Trimpack is a library of routines written in Fortran that allow to create unstructured triangular meshes in any domain and with a user-defined size distribution. The user must write a program that uses the elements of the library as if it were a mathematical tool. First, the domain must be defined, using point-defined boundaries, which the user provides. The library internally uses splines to mesh the boundaries with the node distribution function provided by the user. Several meshing methods are available, from simple Dalaunay mesh creation from a point cloud, an incremental Steiner-type algorithm that also generates Dalaunay meshes to an efficient advancing-front type algorithm. This report carries out a bibliographic review of the state of the art in mesh generation corresponding to the period in which Trimpack was written for the first time, which is a very fruitful period in the development of this type of algorithms. Next, MeshGen is described in detail, which is a program written in C ++ that exploits the possibilities of the Trimpack library for the generation of unstructured triangular meshes and that has a powerful graphical interface. Finally, it also explains in detail the content of the Trimpack library that is available under GNU Public license for anyone who wants to use or improve it.
}
    \end{changemargin}
\end{center}

%% file: Selection_of_the_Generation_Algorithms.tex
\section{Introduction}

In Computational Fluid Dynamics the mesh is the support on which the conservation equations are discretized and has an essential influence on the results obtained. 
Because of that, the selection of a correct generation method is a keypoint in any problem solving. In order to choose correctly, first of all we will ask for our necessities, considering that our election must be guide, also by an instructive and formative character of this work:
\begin{itemize}
    \item \textbf{Rapidity:} The intrinsic searching and identification procedures of the generation algorithm mus be as little onerous as possible.
    \item \textbf{Flexibility:} The algorithm must own a good adaptation characteristic to complex geometries.
    \item \textbf{Be able to adapt:} The capacity of adaptation to the solution must be given.
    \item \textbf{Simplicity:} The algorithm must be easy to handle for any user.
    \item \textbf{Development:} A potential development of the selected algorithm must be possible.
    \item \textbf{Integration:} A posterior integration of the algorithm in another data structure must be considered.
\end{itemize}

Considering all these facts, knowing beforehand there is any algorithm which collects all these characteristics in an optimal way, we must move towards an algorithm which offers suitable results in the majority of the exposed points.

This, in fact, is the case of the unstructured grid generation algorithm based on the basic Delaunay triangulation. The principal advantages are the relative programming simplicity, due to the fact that the node selection is based on the Delaunay criteria (see properties of the Delaunay triangulation \ref{chap:delaunay}), and that the algorithm generates a flexible meshing for any sort of flow problem. Another advantage is that the basic Delaunay triangulation is the point of departure for developing many other algorithms, as could be the Divide-and-Conquer algorithm, the Space Marching algorithm or the incremental insertion Delaunay algorithms. Also it is necessary for the initial Delaunay triangulation of the boundaries in the Steiner algorithm, as we have appreciated before. May be the unique disadvantage that it is not very fast, due to the supported onerous searching and control routines, but in summary the whole algorithm offers very suitable results.

Another algorithm that offers many advantages triangulating unstructured meshes is the Advancing Front Method (AFM), originally developed by Peraire. Therefore we will select this algorithm as second alternative generation program in this work. It is faster than the Delaunay one and simpler for programming, due to the fact that it can use the advancing front program structure of the basic Delaunay algorithm. The possibilities that offers the Advancing Front Method as the inherent ability to adapt, which means that the algorithm facilitates and easy adaptation the the solution, and the direct control by the user over the mesh through the background parameters, makes the algorithm very competitive. The adjusting the single grid parameters to obtain an optimal triangulation, due to the sensitivity of the node insertion criteria.

%% file: Review_of_the_state_of_art.tex
\section{Mesh Generation Procedures}
\subsection{MESH TYPES}
 Essentially, in many fields of interest, two types of meshes are used: \emph{structured meshes}  and \emph{unstructured meshes}. In both cases, external contours aligned with the boundary walls of the domain are used, although there is the alternative of using non-conforming borders, that is, in which the mesh and domain contours do not coincide. The latter solution uses Cartesian meshes in which the mesh is adjusted to the domain by successive refinements of uniform meshes, with the undoubted advantage of greatly simplifying the process of generating the mesh in complex domains. However, this type of mesh is not common, because the precise description of the flow in the boundary layers of the domain, has a very pronounced impact on the solution obtained.

The mesh generation process can have a high impact on the total simulation time. In the case of aerospace engineering, numerical simulation of the external aerodynamics of an aircraft may require weeks of work in the mesh generation process, depending on the detail to describe the geometry. The geometry of the fuselage and wings might relatively straightforward, but the calculation of the combustion chamber of the engines have non-trivial complications. Other problems.

The meshes used in CFDs can be classified as follows:
\begin{itemize}
    \item Structured meshes: These are regular discretizations, with mesh lines normally aligned (and perpendicular) to the domain boundaries. This category should include multiblock meshes, in which each mesh block is structured but the connectivity of the blocks to each other is not regular. There is an extra work of data management that, in codes prepared for unstructured mesh, can be generalized as if it were an unstructured mesh. Structured meshes are easily generated in simple domains or within each block. They can be efficiently aligned with the domain contours and current surfaces of the solution, increasing accuracy and confidence in the results. Finally, they use less storage memory and less calculation time. The fundamental disadvantage lies in the difficulty of discretizing complex domains, which may lead to the quality of the mesh in the transition regions not being good or, on the contrary, the work to improve this quality complicates the structure of blocks so much that no advantage is obtained in its use.
	
	\item Unstructured mesh: An unstructured mesh definition can be sketched: a mesh is unstructured when it is not possible to establish a simple application that assigns to each topological component of the mesh (node, edge, face, etc.) the connected elements. For this reason, unstructured meshes are handled by connectivity matrix that contain this information. The two most important qualities of unstructured meshes are adaptability and flexibility. On the one hand, this meshes allow control of mesh size in specific regions of the domain (adaptability), without this having an impact on other regions where refinement is not necessary. In contrast, in structured meshes, local refinements spread through the domain unnecessarily. And the other feature (flexibility) allows the meshing of extraordinarily complex regions almost automatically; that is, it allows the simulation of problems that can hardly be discretized in a structured way or in which the time spent generating the mesh is reduced by one or two orders of magnitude. The disadvantages focus on the lower accuracy because the faces of the volumes are not aligned, in general, with the current lines and that this type of meshes requires the use of more memory and calculation time.
	Hybrid meshes: Meshes that combine structured and unstructured meshes try to get the advantages of each option. This type of strategy is particularly useful when the structured part corresponds to boundary layer discretization in which the slenderness and alignment of the mesh are essential to achieve good results.
	
	\item Unconventional meshes: In this section can be mentioned cartesian meshes with non-conforming contour or gridless meshes in which the geometric substrate is a distributed point cloud without apparent interconnection.
	\end{itemize}
The regularity and smoothness of the mesh has a direct impact on the quality of numerical calculation results in fluid problems and, in general, in nonlinear problems. For this reason, in the last decades of the twentieth century, when numerical algorithms and the power of computers began to allow the resolution of practical problems, methods based on solving partial differential equations to generate meshes became very popular. Thus, the use of equipotential surfaces, for example, guaranteed mesh lines locally orthogonal to the contours, so that streamlines, and perpendicular gradients were adequately described. Nowadays, mesh generation algorithms are applied to very complex and three-dimensional domains. Mesh generation methods based on differential equations have lost popularity and have been replaced by algebraic methods on regions parameterized with algorithms that have their origin in graphic design and CAD systems.
If the spatial discretization used is not sufficiently regular and smooth, the expected truncation error of the numerical algorithms deteriorates and the results obtained lack sufficient precision. The complexity and three-dimensionality of the calculation domains prevent the degree of regularity and smoothness of the mesh from being sufficiently adequate if careful precautions are not taken. An inspection of the quality of the mesh must always be carried out, and for this it is necessary to define a series of indices that characterize that quality and allow the comparison of different situations.
In general, each mesh generation program has its set of quality indices that are based on relationships between various geometric parameters, but all reflect the regularity of the mesh in the following categories:
\begin{itemize}
    \item 	Cell slenderness: To measure anisotropy, various methods are used to calculate relationships between lengths of edges, diagonals, or thickness. Only elongated cells are considered suitable in boundary layer meshes where it is ensured that the current lines are significantly parallel to the direction of .
	
	\item Cell skewness: The marked differences between angles that form the faces and edges, as well as angles very different from $\pi/2$ (above or below), are responsible for notable calculation inaccuracies.
	
	\item Inequality of contiguous cells: For reasons of domain complexity or complex fluid field morphology, the appropriate mesh size can be very different from one region to another. Nerveless transitions must be smooth. Abrupt changes in mesh size should be avoided or placed in regions where fluid variables have small gradients. Otherwise, the results degrade rapidly because the order of approximation drops.
	
Finally, it should be noted that from an operational point of view, in these circumstances of low mesh quality, numerical convergence to the solution is difficult, calculation times are lengthened when more iterations are needed and, eventually, adequate convergence levels are not reached, or the solution simply does not converge.
The quality of the calculation mesh has a direct impact on the accuracy of the solution obtained. In general, structured meshes are preferred, where the distortion is as low as possible, with their main directions aligned with the flow.
\end{itemize}
Consistent numerical analysis should provide results of increasing accuracy as finer discretizations are used. In this sense, the mistake made is related to the size of the mesh and the question that immediately arises is what is the level of refinement that must be used to achieve a certain level of confidence characterized by an estimate of the error committed. This task could be simplified in the extreme if a priori error estimation methods were available, but in CFDs sufficiently reliable indicators have not been defined and the only current possibility, from a practical point of view, is to carry out a convergence study of the solution by successive refinements of the calculation mesh. But, it is not easy to obtain conclusive results in terms of the quality of the solutions obtained, due to the complexity of the domains used, the details of the mesh, the complexity and the nonlinearity of the mathematical models used. All these aspects make up a complex picture, in which it is essential to carry out rigorous convergence studies, since the number of sources of uncertainty is high and their nature very different.
The standard procedure for conducting a mesh size convergence study is to use a different resolution mesh set for one or more case studies. It is possible to define point or integral values of the solution to carry out the comparison between the results obtained with each mesh. At this point, it is expected to be using a verified and validated code with which, if the mesh is fine, the solution obtained is true. Of course, we need to reflect on the true or real solution, in fact the term real numerical solution should be used in an asymptotic sense.

\subsubsection{STRUCTURED GRIDS}
In the most widely used approach the domain is divided into a structured assembly of quadrilateral cells. The structure in the grid is apparent from the fact that each interior nodal point is surrounded by exactly the same number of grid cells (or elements). Note that, in this case, we can immediately identify two directions within the grid by associating a curvilinear coordinate system ($\xi, \eta$) with the grid lines. if we number the nodes consecutively along lines of constant $\eta$, so that the numbers increase as $\xi$ increases, we can easily identify the nearest neighbours of any node J on the grid. Generally, such grids are constructed by mapping the domain of interest into a square and then constructing a rectangular grid over the square. If the equation itself is also mapped, this grid can be used to obtain a solution, otherwise the inverse mapping is applied to obtain the required grid over the original domain.

Various approaches may be regarded as candidates for accomplishing the mapping, such as conformal techniques, the use of differential equations or algebraic methods. All the major discretization procedures for the equation of fluid flow can normally be implemented on grids of this type. A major advantage to the computational fluid dynamics arising from the use of a structured grid is that it can choose an appropriate solution method from among the large number of algorithms which are available. These algorithms have the advantage that they can normally be implemented in a computationally efficient manner. A disadvantage is the fact that it is not possible to guarantee an acceptable grid by applying the mapping method, as described above, to regions of general shape. This difficulty can be alleviated by appropriately sub-dividing the computational domain into blocks and then producing a grid by applying the mapping method to each block separately. This results in an extremely powerful method \cite{4}, but problems can still be caused by the generation of elements of poor quality and by the elapsed time necessary to produce a grid for domains of extremely complex shape.\\

\begin{figure}[ht]
    \centering
    \includegraphics[width=7.5cm, height=4.7cm]{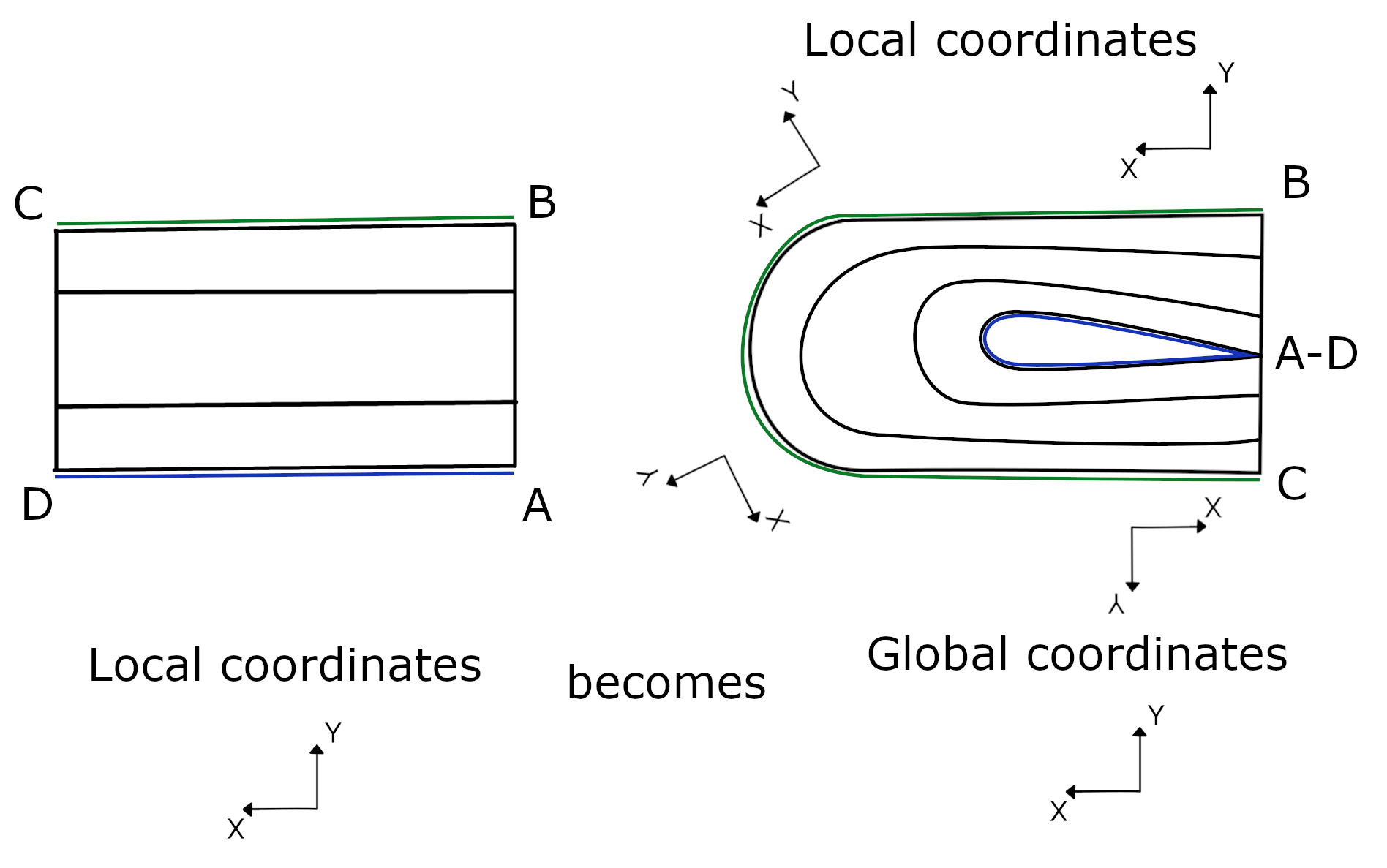}
    \caption{Transformation of a mesh with regular structure}
    \label{fig:structuredgrid}
\end{figure}

\subsubsection{UNSTRUCTURED}
The alternative approach is to divide the computational domain into an unstructured assembly of computational cells. The notable feature of an unstructured grid is that the number of cells surrounding a typical interior node of the grid is not necessarily constant. It will be apparent that quadrilateral cells could again be used in this context, but we will be concentrating our attention in this section upon the use of triangular grids. The nodes and the elements are now numbered and to get the necessary information on the neighbours, we could store the numbers of the nodes which belong to each element. It is apparent that there is no concept of directionality within a grid of this type and that, therefore, solution techniques based upon this concept(e.g. ADI methods) will not be directly applicable. The methods which are normally adopted to generate unstructured triangular grids are based upon either the Delaunay \cite{5} or the advancing front \cite{6} approaches. Discretization methods for the equations of fluid flow which are based upon integral procedures, such as the finite volume or the finite element method, are natural candidates for use with unstructured grids. The principal advantage of the unstructured approach is that it provides a very powerful tool for discretising domains of complex shape (\cite{7}, \cite{8}), especially if triangles are used in two dimensions and tetrahedron are used in three dimensions. In addition, unstructured grid methods naturally offer the possibility of incorporating adaptivity \cite{9}. Disadvantages which follow from adopting the unstructured grid approach are that the number of alternative solution algorithms is currently rather limited and that their computational implementation places large demands on both computer memory and CPU \cite{10}. Further, these algorithms are rather sensitive to the quality of the grid which is being employed and so great care has to be taken in the generation process \cite{11}.

\begin{figure}[H]
    \centering
    \includegraphics[width=6.5cm]{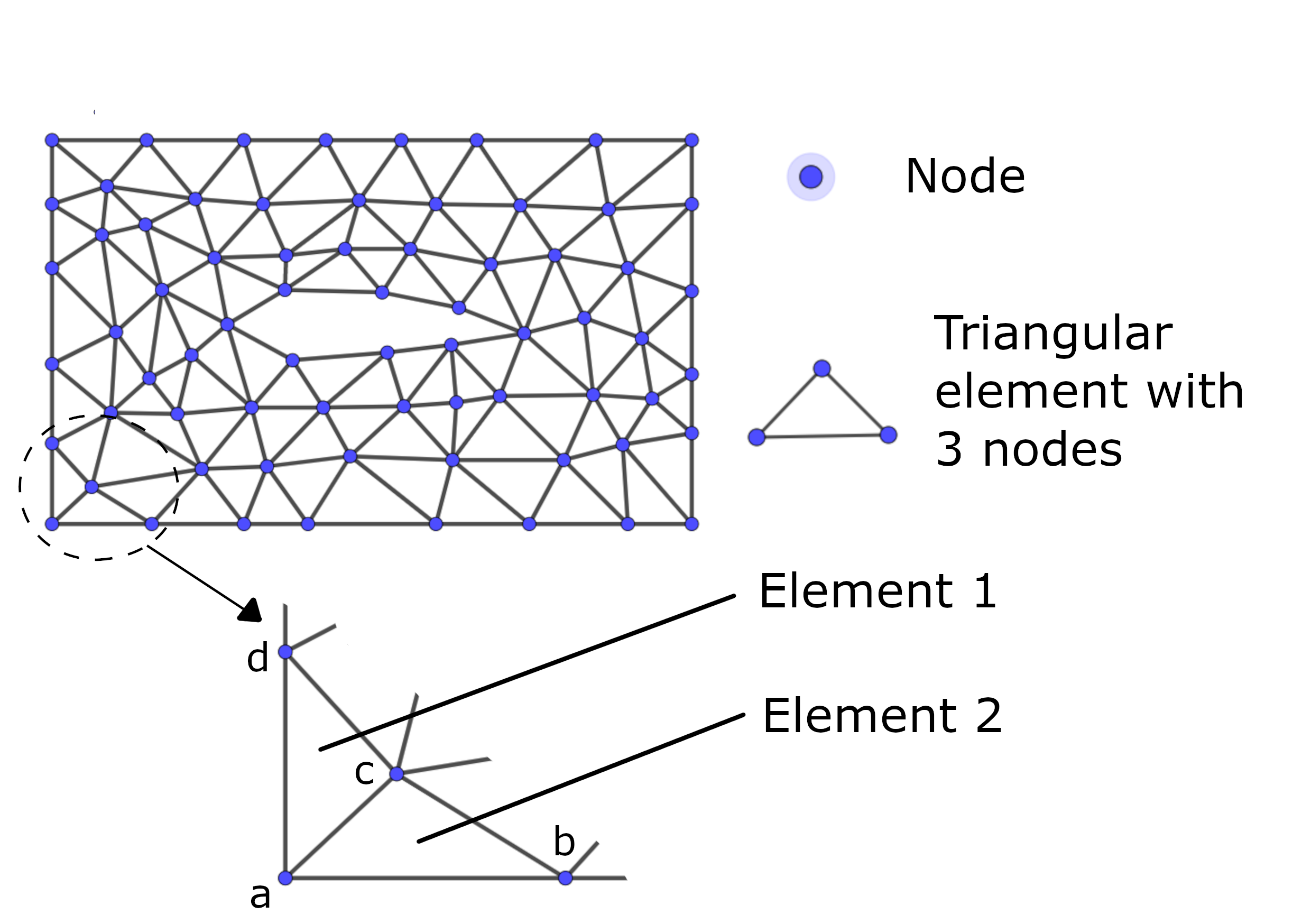}
    \caption{A mesh with an irregular structure}
    \label{fig:unstructuredgrid}
\end{figure}

\newpage

\subsection{GENERATION OF STRUCTURED GRIDS}
Throughout this review, and in the references cited, a numerically generated structured grid is understood to be the organized set of points formed by the intersections of the lines of a boundary-conforming curvilinear coordinate system. The cardinal gesture of such a system is that some coordinate line(surface in three dimensions) is coincident with each segment of the boundary of the physical region. This allows boundary conditions to be represented entirely along coordinate lines without need of interpolation. The use of coordinate line intersections to define the grid points provides an organizational structure which allows all computation to be done on a fixed square grid when the partial differential equations of interest have been transformed so that the curvilinear coordinates replace the Cartesian coordinates as the independent variables \cite{12}.

The basic ideas of the generation of such grids, the necessary transformation relations, and the procedures for the  application in the numerical solution of partial differential equations are assembled in an introductory fashion in Ref. \cite{13}. The various types of boundary-conforming coordinate systems, and the different methods of numerical generation thereof, are discussed in some detail in ref. \cite{14}. Numerous examples of applications to field problems are also cited in this reference. Different types of grids are more appropriate to different physical problems and configurations, and also to different modes of usage. To choose between the various types of coordinates, we must first consider which constraints are needed for a given problem. The fundamental constraint for a general region is its boundary geometry. When the coordinates match the boundary, the need for boundary interpolation disappears and the grid is also aligned with the desired solution near the boundary. Without further requirement in the two-dimensional case, conformal systems are usually the best. In addition to boundary geometry, however, the point wise distribution along the boundary is often required as a further constraint. This distribution is a boundary coordinate system or systems, which together with the geometry forms a complete boundary representation. When the representation is arbitrarily prescribed, conformal transformations are not applicable because of analytic continuation. As the next simple case, orthogonal coordinates are preferable. In two dimension they are generally applicable on both planes and curved surfaces. In three-dimensional regions orthogonal systems are severally restricted and are not generally applicable. The best that can be done in general context is to bound such regions with orthogonal systems so that full orthogonality can be specified at the boundaries. Further boundary constraints can also be imposed with specified derivatives so that rates of entry or exit from a region can be given. In any dimension, the capability to create a smoothly assembled composite mesh for topologically complex configurations would be achieved. In addition to the various boundary constraints, significant advantages can be obtained under the region. The purpose is usually to more fully resolve the numerical solution of a given problem with a fixed number of mesh points. Addition advantages can also be achieved with the constraint that a certain desirable mesh structure be smoothly embedded within the region \cite{15}.

Structured grid generation systems are procedures for generating the curvilinear coordinate system which defines the grid. The systems fall into two basis classes: algebraic systems, in which the coordinates are determined by interpolation, and partial differential equation systems, in which the coordinates are the solution of these equations.
\begin{figure}[H]
    \centering
    \includegraphics[width=10cm]{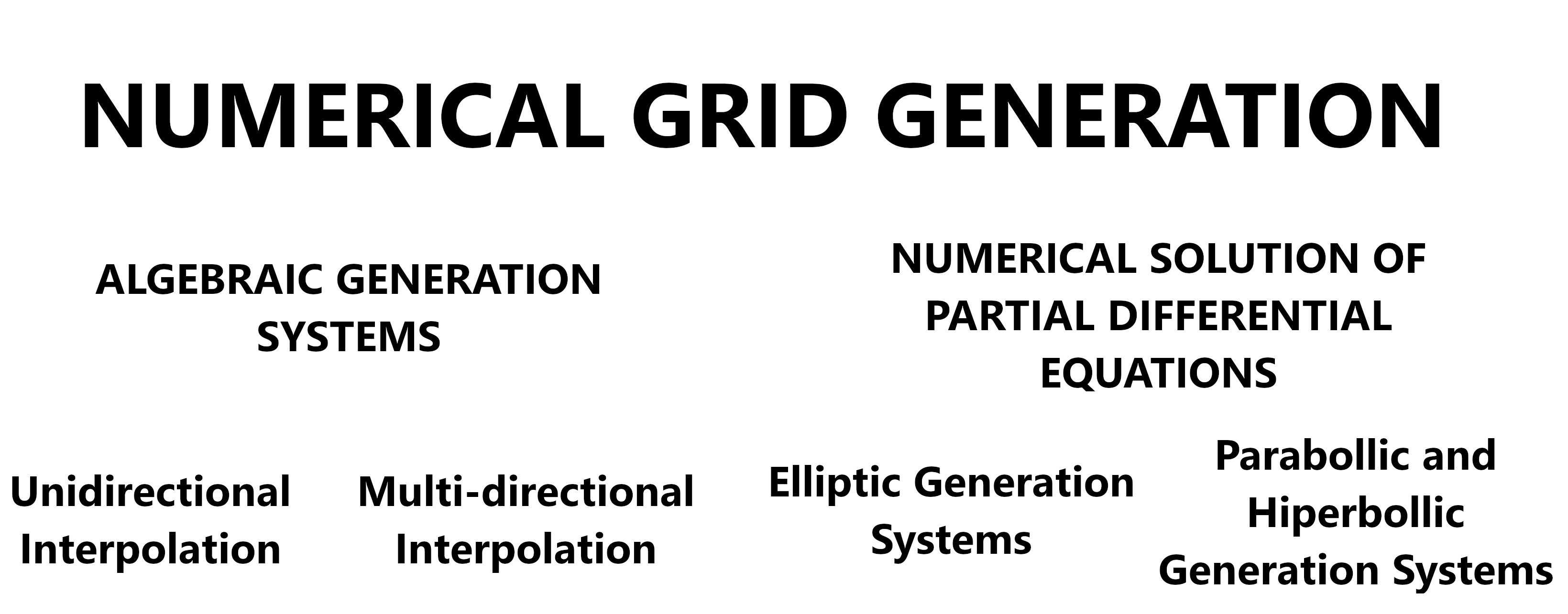}
    \caption{Classes of structured grid generation}
    \label{fig:classes of structured grid}
\end{figure}
\subsubsection{ALGEBRAIC GENERATION SYSTEMS}
The problem of generating a curvilinear coordinate system can be formulated as a problem of generating values of the Cartesian coordinates in the interior of the transformed region from specified values on the boundaries. This, of curse can be done directly by interpolation from the boundaries, and such coordinate generation procedures are referred to as algebraic generation systems \cite{16}. Algebraic generation is the fastest procedure in many cases, and its use is surveyed in refs. (\cite{14},\cite{17},\cite{18}). Algebraic procedures also allow explicit control of the grid point distribution. Some algebraic grid generation systems propagate boundary slope discontinuities into the field, and there is no inherent smoothing mechanism. Some problems in this regard were experienced in ref [19]. However, the use of local interpolation in the multi surface method can prevent this propagation of discontinuities into the field \cite{20}. The algebraic approach is particularly attractive for use with interactive graphics since grids can be produced quickly \cite{12}.

\paragraph*{Unidirectional Interpolation}
Algebraic grid generation is basically an interpolation among boundaries and/or intermediate surfaces in the field. Simple on-dimensional stretching involves only the use of transformation functions and is often applied to a coordinate system generated by other means \cite{14}. Examples of simple stretching along straight lines are given in Refs. (\cite{21},\cite{22},\cite{23},\cite{24},\cite{25},\cite{26}). Another simple application is the normalization of the separation between two boundaries as often has been used (\cite{27},\cite{28},\cite{29}) for recent applications.

Intermediate surface within the region may be necessary with several distorted regions for which interpolation only between boundaries would result in unsatisfactory grids. ref. \cite{30} shows an example where the use of just an intermediate point in the corrected a coordinate system that had overlapped the boundary. The "two-boundary" technique (\cite{31},\cite{32}) and the "outer surface " method (\cite{17}) are examples of one-dimensional Hermite interpolation using only  the opposing boundary surfaces.

The multi-surface method (\cite{14},\cite{17}) is a related unidirectional interpolation procedure. this procedure is constructed from an interpolation of a specified vector field followed by vector normalization at each interpolation point in order to cause a desired telescopic collapse so that the boundaries are matched. In Ref. \cite{33} the multi surface method is applied to generate embedded grids with continuity of coordinate lines slop, using piece wise-linear interpolants. this approach is extended in Ref. \cite{20} to use higher order local interpolants to allow for curvature continuity as well. A collection of subroutines which automatically perform the necessary parts of grid construction using multi-surface procedure has been written and is described in \cite{34}. The multi-surface method has been applied in Refs. (\cite{35}, \cite{36}, \cite{37}) in transonic flow solutions.

\paragraph*{Multi-directional Interpolation}

The various interpolation methods differ primarily in regard to how may and what curves or surfaces are used, and what derivatives, if any, are specified on these surfaces, and secondarily in regard to what type of blending functions are used. Several approaches are discussed in Refs. (\cite{14},\cite{17},\cite{30}). The order of accuracy of the interpolation may be increased either by adding more curves of surfaces, of by adding more information, e.g., specification of higher derivatives on the curves or surfaces.

Transfinite interpolation \cite{30} is among curves of surfaces, transfinite interpolation, originally developed for computer-aided design of sculptured surfaces and solids, involves interpolation among functions defined along curves and surfaces, rather than among point values, and thus matches the function at a non enumerable number of points. (The non enumerable aspect of transfinite interpolation comes from the possible infinity of points defining general boundaries as compared to a tensor product structure, i.e. product of projectors, that uses only corner information defined by discrete sets of value, i.e. piece wise linear function. In higher dimensions this interpolation can be stated as a sequence of uni variate interpolation,k i.e. projection, which are put together as Boolean sum projections. (The Boolean sum is the primary mechanism for defining transfinite interpolation. The multi-directional results are Boolean sums of unidirectional interpolations.) The function specify the values (and perhaps some derivatives of the variables on the curves or surfaces. Values in the interior between these curves or surfaces are determined by interpolation, using specified interpolation functions usually called blending functions. The blending functions are often polynomials, but other functions can be used also. Transfinite interpolation is used to generate grids joined to an analytically generated grid near a corner in Ref. \cite{38}. An other application to two-dimensional airfoil appears in Ref. \cite{39}. Transfinite interpolation is used in Ref. (\cite{40}, \cite{41}, \cite{42}) for three-dimensional grid generation.

\subsubsection{NUMERICAL SOLUTION OF PARTIAL DIFFERENTIAL EQUATIONS}
\begin{figure}[H]
    \centering
    \includegraphics[width=7cm]{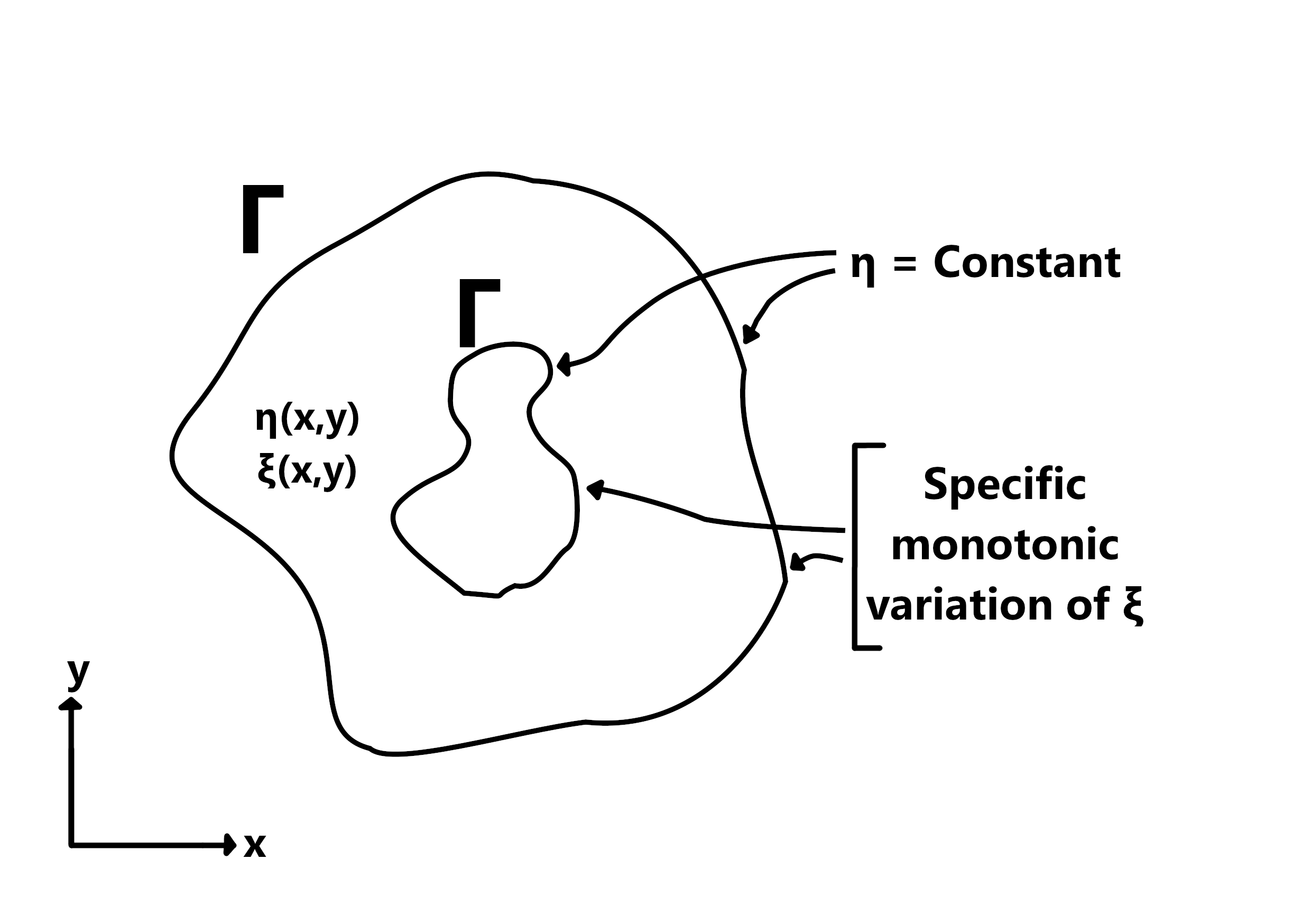}
    \caption{Transformation scheme example}
    \label{fig:transformation1}
\end{figure}

The generation of a boundary-conforming coordinate system is accomplished by the determination of the values of the curvilinear coordinates in the interior of a physical region from specified values( and/or slopes of the coordinate lines intersecting the boundary) on the boundary of the region (see figure \ref{fig:transformation1}).

One coordinate will be constant on each segment of the physical boundary curve (surface in three dimensions=, while the other varies monotonically along the segment \cite{16}.

\begin{figure}[H]
    \centering
    \includegraphics[width=8cm]{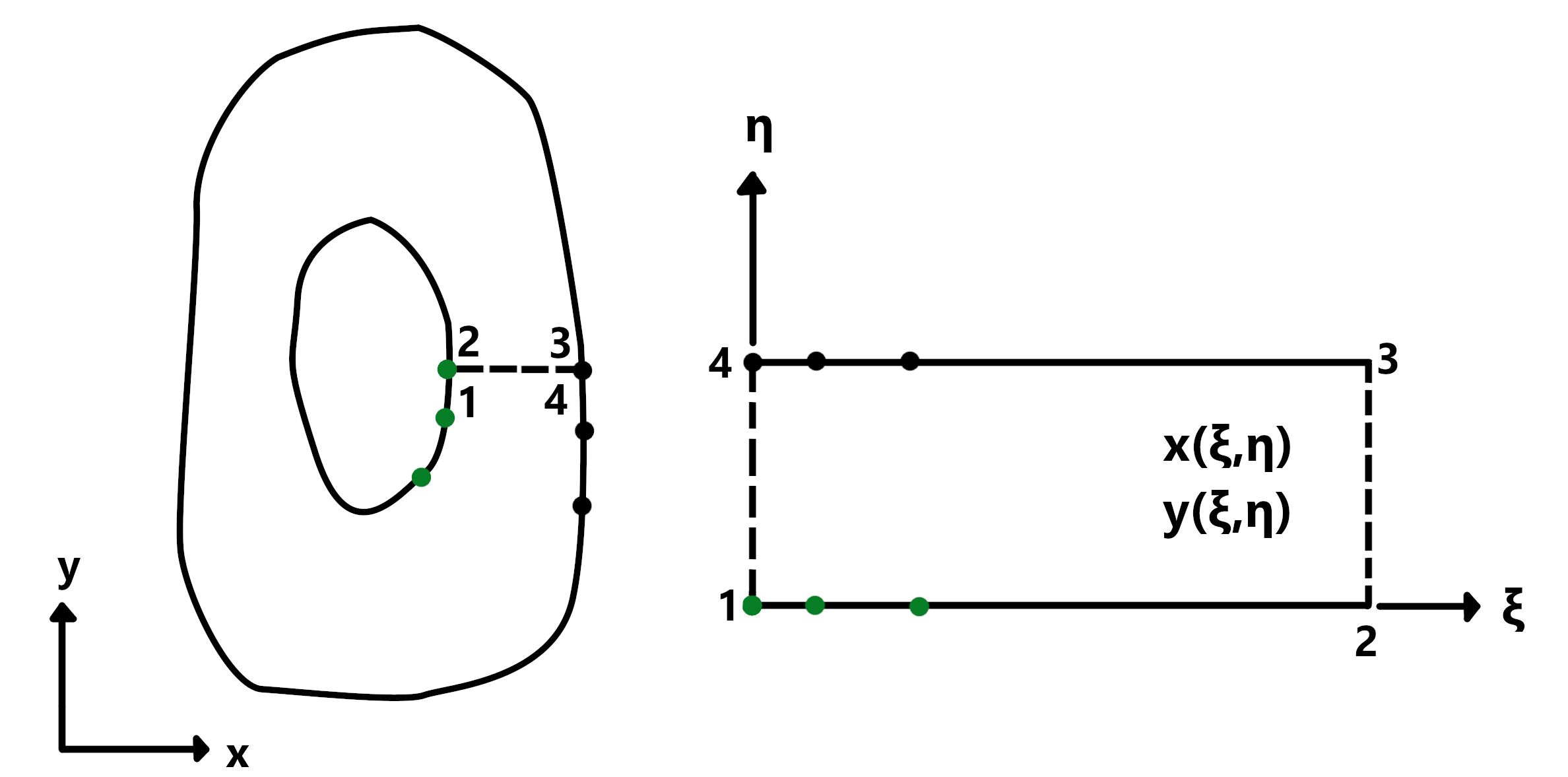}
    \caption{Transformation scheme example}
    \label{fig:transformation2}
\end{figure}

The equivalent problem in the transformed region is the determination of values of the physical (Cartesian or other) coordinates in the interior of the transformed region from specified values and/or slopes on the boundary of this region (see figure \ref{fig:transformation2}).

This is a more amenable problem for computation, since the boundary of the transformed region is comprised of horizontal and vertical segments, so that this region is composed of rectangular blocks which are contiguous, at least in the sense of being joined by re-entrant boundaries (branch cuts) \cite{16}.

The generation of field values of function from boundary values can be done in various ways, e.g., by interpolation between boundaries, etc. the solution of such a boundary-value problem, however, is a classic problem of partial differential equations, so that it is logical to take the coordinates to be solutions of a system of partial differential equations. If the coordinate points (and/or slopes) are specified partial differential equations. If the coordinate points (and/or slopes) are specified on the entire closed boundary of the physical region, the equation must be elliptic, while if the specification is on only a portion of the boundary the equations would be parabolic or hyperbolic. This latter case would occur, for instance, when an inner boundary of physical region is specified, but a surrounding outer boundary is arbitrary.

\paragraph*{ELLIPTIC GENERATION SYSTEMS}
The extremum principles, i.e., that extreme of solutions cannot occur within the field, that are exhibited by some elliptic systems can serve to guarantee a one-to-one mapping between the physical and the transformed regions (\cite{43}, \cite{44}). This, since the variation of the curvilinear coordinate along a physical boundary segment must be monotonic, and is over the same range along facing boundary segments \cite{16}, it clearly follows that extrema of the curvilinear coordinates cannot be allowed in the interior of the physical region, else overlapping of the coordinate system will occur. Note that it is the extremum principles of the partial differential system in the physical space, i.e., with the curvilinear coordinates as the dependent variables, that is relevant since it is the curvilinear coordinates, not the Cartesian coordinates, that must be constant an monotonic on the boundaries. Thus it is the form of the partial differential equations in the physical space, i.e., containing derivatives with respect to the Cartesian coordinates, that is important.

Another important property is regard to coordinate system generation is the inherent smoothness that prevails in the solutions of elliptic systems. Furthermore, boundary slope discontinuities, are not propagated into the field. Finally, the smoothing tendencies of elliptic operators, and the extremum principles, allow grids to be generated for any configurations without overlap of grid lines.

There are thus a number of advantages to using a system of elliptic partial differential equations as a means of coordinate system generation. A disadvantage, of course, is that a system of partial differential equations must e solved to generate the coordinate systems.

The historical progress of the form of elliptic systems used for grid generation has been traced in ref. \cite{14}. Numerous examples of the generation and application of coordinate systems generated from elliptic partial differential equations are covered in the above reference, as well as in Ref. \cite{16}. The best general choice seems to be those based on some consideration of differential geometry \cite{46}. Linearization by replacement of certain metric coefficients in these equations with specified functions can be considered \cite{14}, but this approach is likely to lead to distorted grids with more general shapes and configurations. The most widely used elliptic generation system is that based on the system of Poisson-like equations. This system has been sued in many works as note d in Ref. \cite{14} and discussed also in Ref. \cite{47}. Other uses are given in Refs. (\cite{48}, \cite{49}, \cite{50}, \cite{51}, \cite{52}, \cite{53}).

\begin{figure}[H]
    \centering
    \includegraphics[width=10cm]{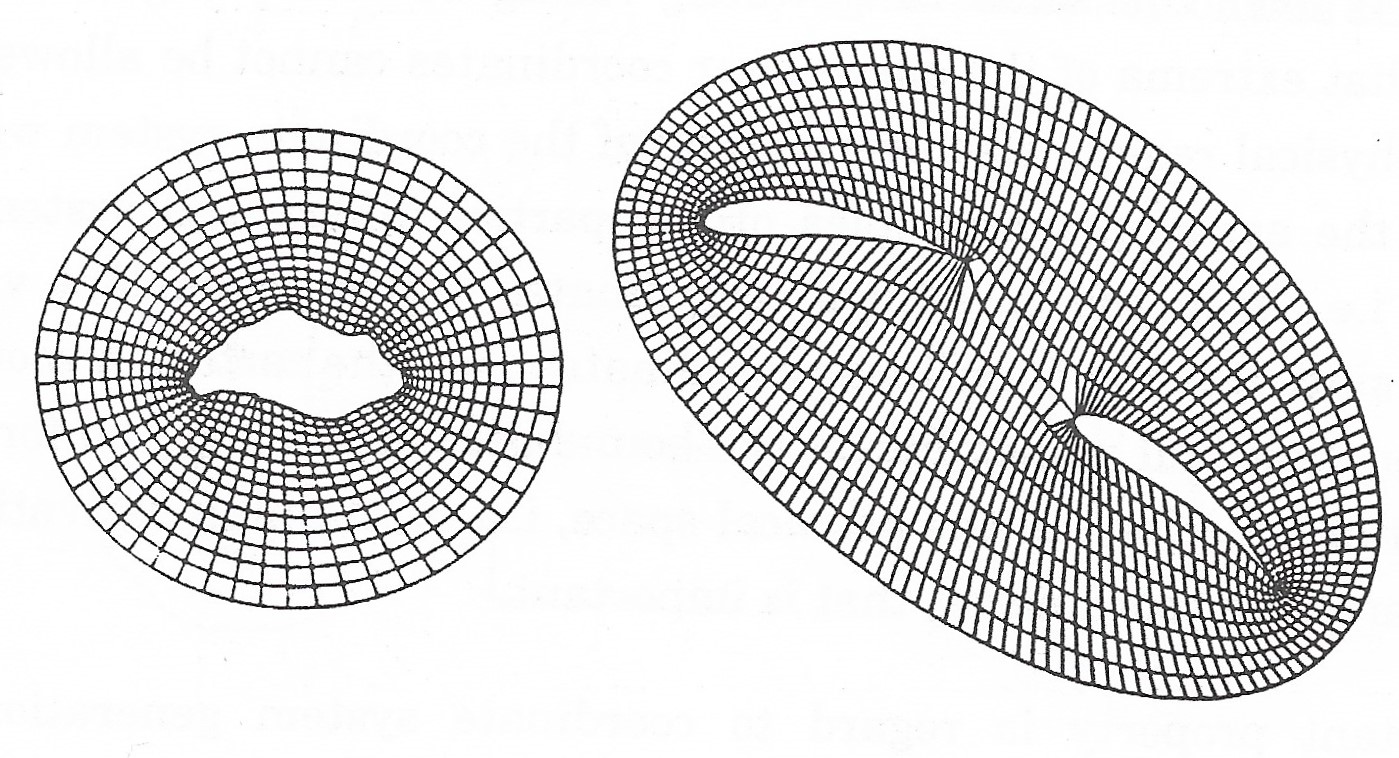}
    \caption{Elliptic grids}
    \label{fig:ellipticgrids}
\end{figure}

\paragraph*{PARABOLIC AND HYPERBOLIC GENERATION SYSTEM}

It is also possible to base a grid generation system on hyperbolic or parabolic partial differential equations, rather that elliptic equations. In each of these cases the grid is generated by numerically solving the partial differential equations, marching in the direction of one curvilinear coordinate between two boundary surfaces in three dimensions. In neither case can the entire boundaries of a general region be specified - only the elliptic equations allow that \cite{16}.\\
The parabolic system can be applied to generate the grid between the two boundaries of a doubly-connected region with each of these boundaries specified. The hyperbolic case, however, allows only one boundary to be specified, and is therefore of interest only for use in calculation on physical unbounded regions where the precise location of a computational outer boundary is not important. Both parabolic and hyperbolic grid generation systems have the advantage of being generally faster that elliptic generation systems, but, as just noted, are applicable only to certain configurations. Hyperbolic generation systems can be used to generate orthogonal grids.
\newpage
\subparagraph{Hyperbolic Grid Generation}
an example of grids generated by this procedure follows (figure \ref{fig:hyperbolic grid}):

\begin{figure}[H]
    \centering
    \includegraphics[width=7cm]{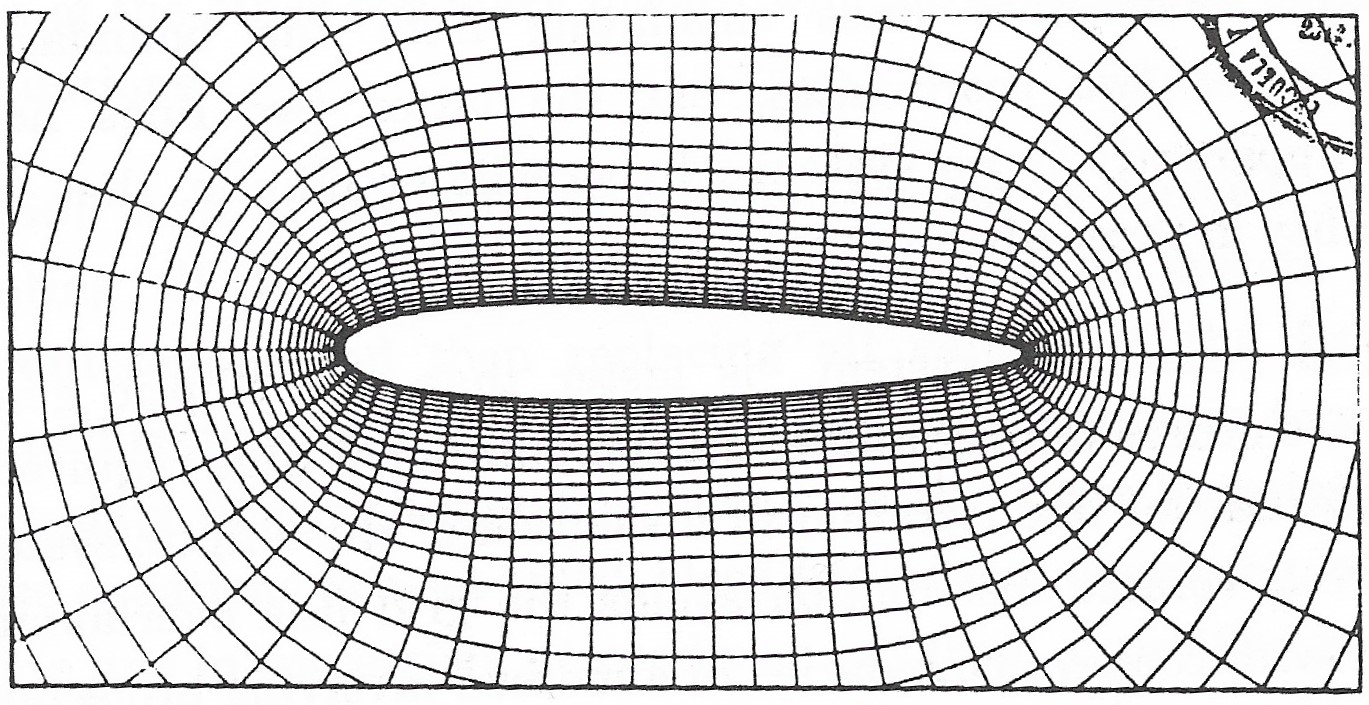}
    \caption{Hyperbolic Grid}
    \label{fig:hyperbolic grid}
\end{figure}

The specification of the cell volume prevents the coordinate system from overlapping even above a concave boundary. In this case line spacing will expand rapidly away from the boundary in order to keep the cell volume from vanishing, as in the following figure \ref{fig:hyperbolic grid 2}.

\begin{figure}[H]
    \centering
    \includegraphics[width=7cm]{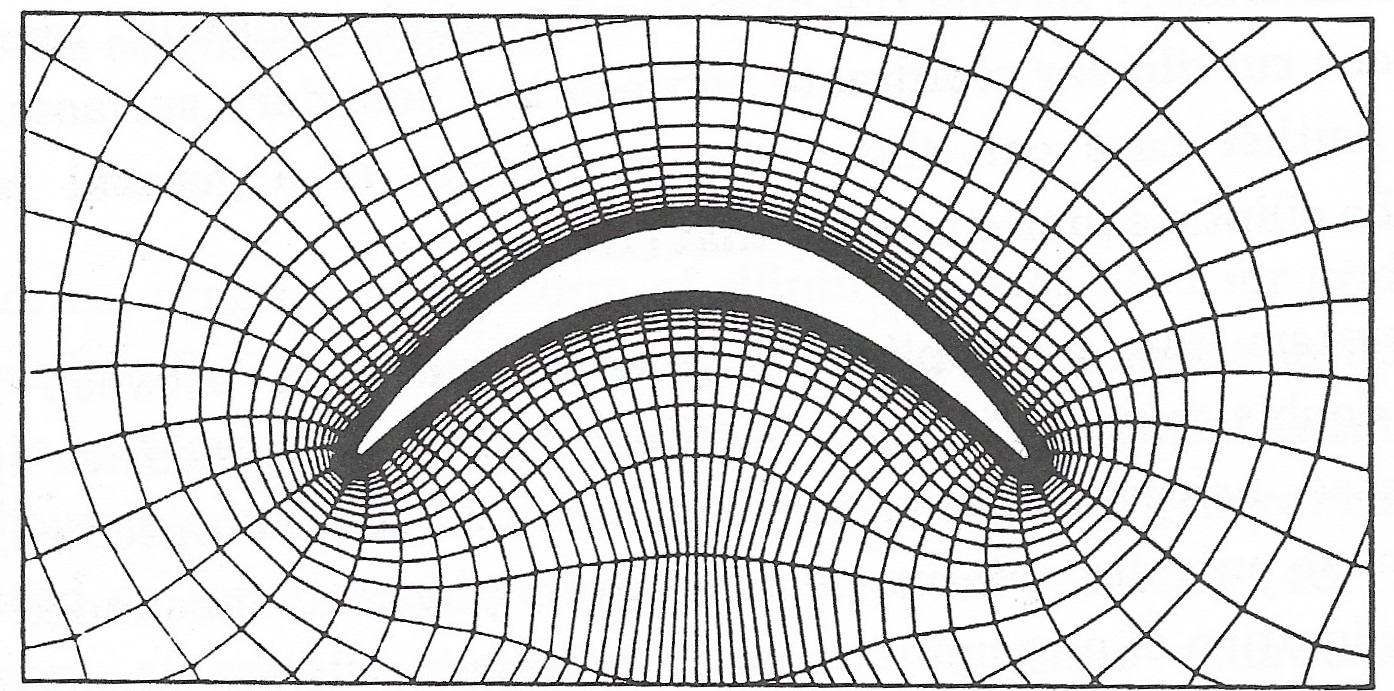}
    \caption{Hyperbolic grid}
    \label{fig:hyperbolic grid 2}
\end{figure}

Although this prevent overlap, the rapid expansion that occurs can lead to problems with truncation error in some cases. This approach is extendable to three dimensions with the coordinate lines emanating from the boundary being orthogonal to the other two coordinates, but the latter two lines not being orthogonal to the other two coordinates, but the latter two lines not being orthogonal. There apparently is no system, hyperbolic or elliptic, that will give complete orthogonality in three dimensions.\\
This hyperbolic grid generation system is not as general as the elliptic generation systems by one or two orders of magnitude, the computational time required being equivalent to about that for one iteration in a solution of the elliptic system. The specification of the cell volume distribution avoids the grid line overlapping that otherwise can occur with concave boundaries in a method involving projection away form a boundary. The grid may, however, be somewhat distorted when concave from a boundary. The grid may, however, be somewhat distorted when concave from a boundary. The grid may, however, be somewhat distorted when concave boundaries are involved. The cell volume specification also allows control of the grid line spacing, but again concave boundaries may cause the intended spacing to occur in the wrong coordinate direction, as in the lower part of this figure, since it is only the volume, and not the spacing in the two separate coordinate directions, that is controlled. As has been noted, the grid constructed to be orthogonal.\\
The hyperbolic generation system is not as general as the elliptic systems, however, since the entire boundary of the region cannot be specified. As noted above, boundary slope discontinuities are propagated into the field, so that the metric elements will be discontinuous along coordinate lines emanating from boundary slope discontinuities. Finally, since hyperbolic partial differential equations can have shock-like solutions in some circumstances, it is possible for very unsuitable grids to result with some specifications of boundary point and cell volume distributions. This result with some specifications of boundary point and cell volume distributions. This result with some specifications of boundary point and cell volume distributions. This in contrast with the elliptic generation systems which tend to emphasize the smoothness because o f the nature of elliptic partial differential equations. In Ref. (\cite{54}, \cite{55}) a hyperbolic generation system is used in which the Jacobian, i.e., the cell area, can be specified. The resulting coordinate system is orthogonal at the boundary and nearly so elsewhere. This generation system has been discussed also in Reef. \cite{14}.

\subparagraph{Parabolic Gird Generation}

 Parabolic grid generation systems may be constructed by modifying elliptic generation systems so that the second derivatives in one coordinate direction do not appear. The solution the can be marched away from a boundary in much the same manner as described above for the hyperbolic systems. Here, however some influence of the other boundary toward which the marching progresses is retained in the equations.

 In Ref. \cite{56} such a parabolic generation system is formed essentially by first representing all derivatives in an elliptic generation system with second-order central differences and then replacing all values on the forward line in one coordinate direction, with values specified in some manner in terms of the values on the preceding lines and specified values on the outer boundary. This reduces the difference equations to a set of 2x2 block tridiagonal equations to be solved on each coordinate line in succession, proceeding away from a specified boundary. Control of the coordinate line spacing can be achieved by certain control functions that are drawn from some analogy with the elliptic system. It is possible to use the functional specification of the forward values to cause the grid to be nearly-orthogonal.

 The parabolic generation system is also faster that the elliptic generation systems to the same degree as is the hyperbolic system, since again only succession of tridiagonal solutions is required. The functional specification of the forward values, with an influence of an outer boundary, introduces a smoothing effect from this second boundary not present in the hyperbolic system. Orthogonality is not achieved as directly as with the hyperbolic system, however the forms of the forward values specification, and of the control functions, have not yet been well-developed.
 
 Ref \cite{57} uses a parabolic system formed by adding a time derivative to the usual elliptic system in order to obtain grid evaluations equations to couple with the flow solution equations for a tie-dependent free surface.
 
 Finally, we will say that all these different generation systems of structured grids have been worked out for the las twenty years and have reached, as we could see, to a high degree of mellowness. This is way the structured grid generation is the usually used in practical flow problems.
\subsection{GENERATION OF UNSTRUCTURED MESH}
First of all, will be resumed the most significant advantages and disadvantages which has the unstructured grid generation over the structured, to make an idea about the convenience of this generation system.

The advantages of unstructured meshes are:
\begin{itemize}
    \item \textbf{Adaptivity}. Unstructured grid generation offers the inherent possibility of adapting the mesh to the solution of the flow problem. With mesh adaptation, will be achieved major accuracy of these solutions, moving points to the regions of interest and removing them from unimportant areas. This means that with the same number of points will be obtained a more accurate solution with an unstructured generated mesh than with a structured one
    \item \textbf{Flesibility}. One of the greatest qualities of the unstructured grid generation is the good coupling of these grids in difficult and complex geometries. In this sense the unstructured grids surpass totally the structured ones. This virtue is one of the fundamenteal reasons of its creation.
\end{itemize}

While the main disadvantages are:
\begin{itemize}
    \item \textbf{Complexity of the generation:}. Complex and laborious searching and identifying algorithms, special data structures and large time-demands on the computer do not turn the unstructured grid generation into a simple task
    \item \textbf{Complexity of the solution}. The solution algorithms are limited, as we could see, to the integral procedures(finite element of finite volume method). The absence of linked matrix structures make the procedures much more complex. Also the computer time and memory demand is extremely high.
\end{itemize}

Now will be presented the different unstructured grid generation algorithms used in this application. There are three basic methods of generating unstructured meshes: \textbf{Delaunay},\textbf{Advanced Frotier Method (AFM)} and \textbf{Steiner}.

\subsubsection{DELAUNAY FUNDAMENTS} \label{chap:delaunay}
\begin{figure}[H]
\centering
\includegraphics[width=10cm]{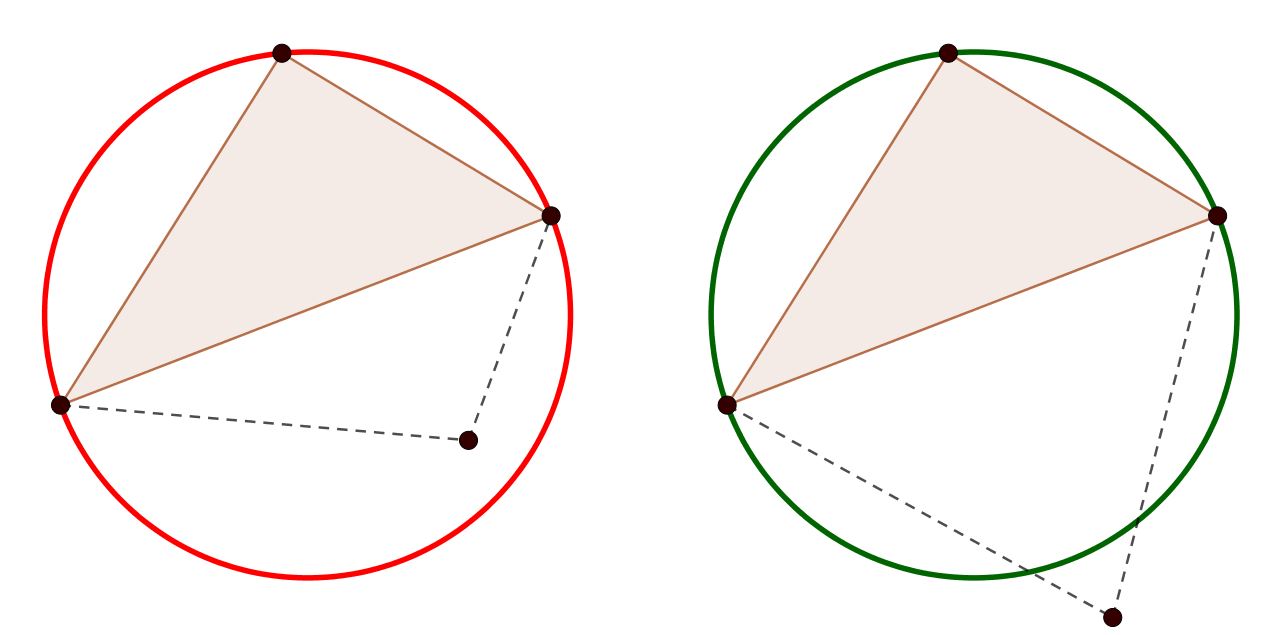}
\caption{Delaunay Condition. Green circumcircle fulfills the condition the red one do not.}
\label{delaunay1}
\end{figure}

Delaunay triangulation, also known as "Delone" triangulation, is a grid of connected and convex triangles that satisfies the Delaunay condition. This condition affirms that no point of the mesh must be contained in the circumcircle of each triangle (\ref{delaunay1}).

\begin{figure}[H]
\begin{minipage}[trim=10 0 0 10mm]{0.45\linewidth}
\centering
\includegraphics[width=6.7cm]{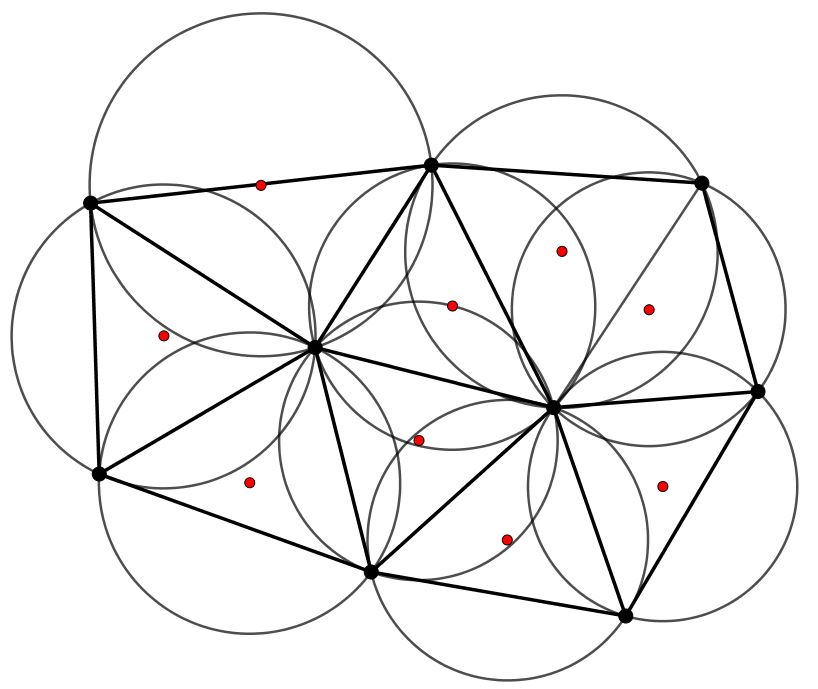}
\caption{Delaunay Diagram Circumcircles}
\label{delaunayfig}
\end{minipage}
\hfill
\begin{minipage}[trim=10 0 0 10mm]{0.45\linewidth}
\includegraphics[width=\linewidth]{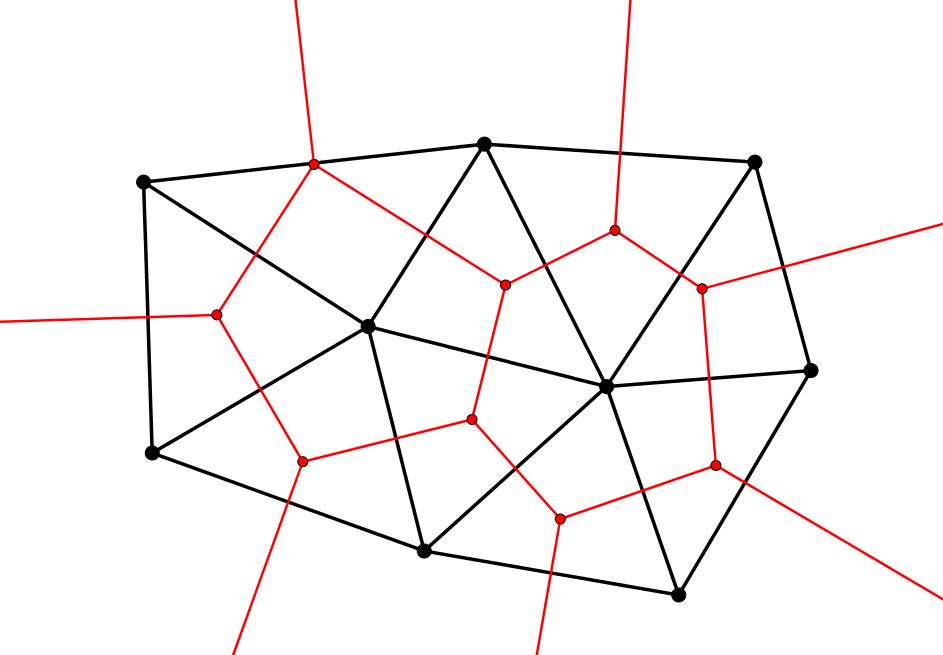}
\caption{Voronoi Diagram}
\label{Voronoifig}
\end{minipage}%
\end{figure}

It is impossible to mention Delaunay without mentioning Voronoi Diagram (or Dirichlet teselation). That is because Voronoi Diagram is the dual graph of Delaunay triangulation for a given number of points. The circumcenters of Delaunay triangles correspond to the vertex of the Voronoi Diagram (\ref{delaunayfig}, \ref{Voronoifig}).
\paragraph*{PROPERTIES OF DELAUNAY TRIANGULATION}

\begin{enumerate}
    \item \textbf{Uniqueness}. The Delaunay triangulation is unique. This assumes that no four nodes are co-circular. The uniqueness follows from the uniqueness of the Voronoi Diagram.
    \item \textbf{Fulfills the Delaunay Criteria}, showed in \ref{delaunay2}. Algebraically can be written like: if point D lies interior to the circumcircle of $\Delta ABC$ which is equal to:
    \begin{figure}[H]
\centering
\includegraphics[width=4cm]{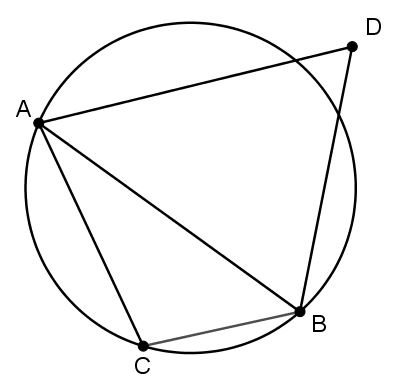}
\caption{Incircle test for $\Delta ABC$ and D (Delaunay Condition true)}
\label{delaunay2}
\end{figure}
    \begin{equation}
        \langle ABC + \langle CDA =  
        \begin{cases}
      < 180\degree & \text{Delaunay condition true} \\
      = 180\degree & \text{A,B,C,D cocircular} \\
      >180\degree & \text{Delaunay condition false}
    \end{cases}
    \end{equation}
    
    \item \textbf{Edge circle property.} A triangulation of nodes is Delaunay if and only if there exists some circle passing through the end position of each and every edge which is point-free. This characterization is very useful because it also provides a mechanism for defining a constrained Delaunay triangulation where certain edges are prescribed a priori. A triangulation of nodes is a constrained Delaunay triangulation if for each and every edge of the mesh there exists some circle passing through its endpoints containing no other node in the triangulation which is visible to the edge.
    In figure \ref{Edge property}, node d is not visible to the segment a-c because of the constrained edge a-b.
    \begin{figure}[H]
\centering
\includegraphics[width=5cm]{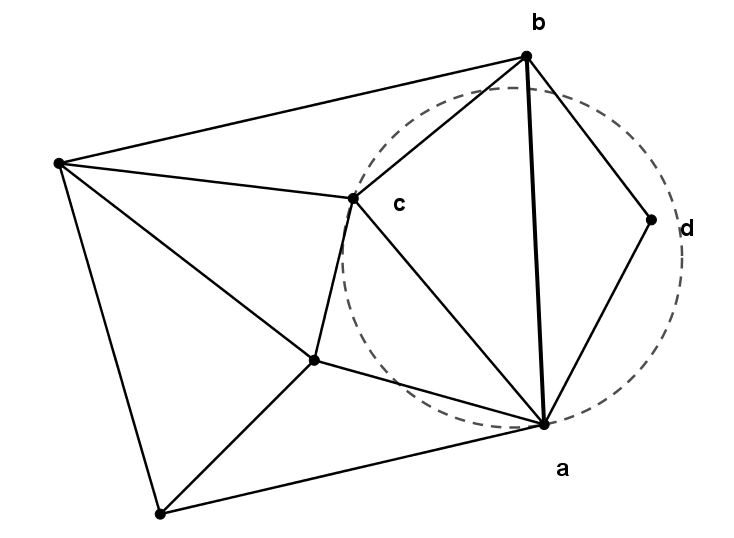}
\caption{Constrained Delaunay triangulation. Node d is not visible to a-c, due constrained segment a-b}
\label{Edge property}
\end{figure}
\item \textbf{Equiangularity property}. Delaunay triangulation maximizes the minimum angle of the triangulation. For this reason Delaunay triangulation of the is called the MaxMin triangulation. This property is also locally true for all adjacent triangle pairs which form a convex quadrilateral. This is basic for the local edge swapping algorithm of Lawson.

\item \textbf{Nearest neighbour property.} 
An edge formed by joining a vertex to its nearest neighbour is an edge of the Delaunay triangulation. This property makes Delaunay triangulation a powerful tool in solving the closest proximity problem. Note that the nearest neighbour do not describe all edges of the Delaunay triangulation. 

\item \textbf{Minimal roughness.}
The Delaunay triangulation is a minimal roughness triangulation for arbitrary sets of scattered data, \cite{96}, \cite{97}. Given arbitrary data $f_i$ all vertices of the mesh and a triangulation of these points, a unique piece wise linear interpolating surface cam be constructed. the Delaunay triangulation has the property that of all triangulations it minimizes the roughness of this surface as measured by the following Sobolev seminorm:
\begin{equation}
\int_{T}[(\frac{\partial f}{\partial x})^2+(\frac{\partial f}{\partial y})^2] \,dx\,dt 
\end{equation}

This is an interesting result as it does not depend on the actual form of the data. This also indicates that Delaunay triangulation approximates well those functions which minimize this Sobolev norm. One example would be the harmonic functions satisfying Laplace's equation with suitable boundary condition which minimize exactly this norm. It is proved that a Delaunay triangulation guarantees a maximum principle for the discrete Laplacian approximation (with linear elements).
\end{enumerate}

\paragraph*{GENERATION OF A BASIC DELAUNAY TRIANGULATION}
It is important her to point out that the most Delaunay triangulation algorithms, start from the principle that the triangulating set of nodes (node cloud) is previously previously specified.

\begin{figure}[H]
    \centering
    \includegraphics[width=12cm]{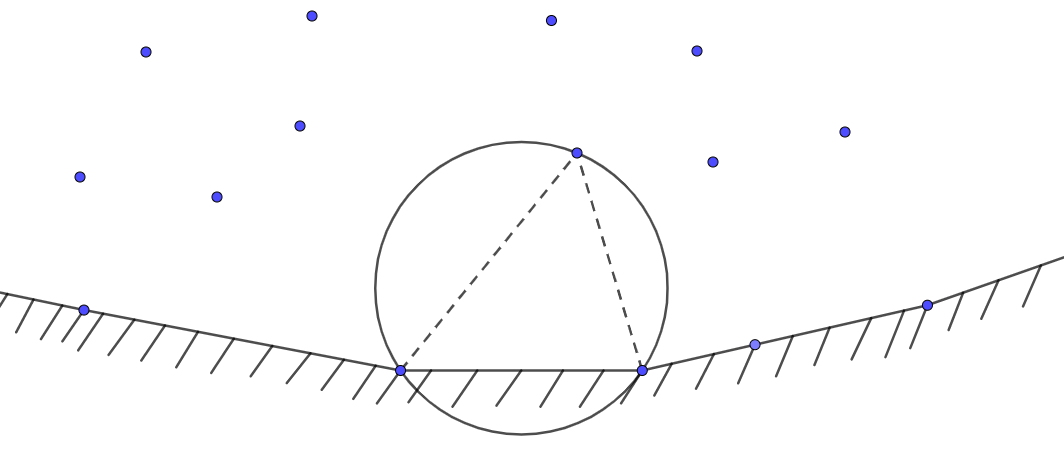}
    \caption{A straightforward iteration procedure selects teh node which generates the smallest circumcircle for a given edges. The absce of nodes inside the circumcircle establishes convergence.}
    \label{fig:circumcircle4}
\end{figure}

The idea is to start a known boundary edge and form a new triangle by joining both endpoints to one of the interior nodes. This may generate up to two additional edges, providing they are not already part of another triangle. After all the boundary edges have been incorporated into triangles, the new edges will appear to be a boundary. This moving boundary is often called an advancing front (no to confuse with the advancing from method o Peraire, which does not generate Delaunay triangulation. This can be done by taking advantages of the in-circle property, the circumcircle of a Delaunay triangle contains no other points. This allows the appropriate point to be selected iteratively as shown in figure \ref{fig:circumcircle4}.

The iteration begins by selecting any node which is on the desired side of the given edge. If there are no such nodes, the given edge is part of a convex hull. Next, the circumcircle is constructed which passes through the edge endpoints and the selected node. Finally, check for nodes inside this circle. If there are any, replace the selected node with the node closest to the circumcenter and repeat the process. When the circumcircle is empty of nodes, connect the edge end points to the selected node.

This seemingly simple algorithm is at heart a very onerous procedure, due to large searching sequences and control routines, extended to all of the nodes. this triangulation algorithm was developed by Tanemura, Ogawa and Ogita (1983)\cite{64} and later rediscovered by Merriam (1991)\cite{65}. See also Ref. \cite{59}

\paragraph*{REMESHING TECHNIQUES} \label{chap:remesh}
To facilitate the understanding of some of the later explained algorithms we will present a few of the most common remeshing techniques used. Many generation algorithms include these techniques during or after the triangulation process to improve the mesh quality. Here we will only mention the basic operation principle and later on, its application to the different algorithms.

\begin{figure}[H]
    \centering
    \includegraphics[width=8cm]{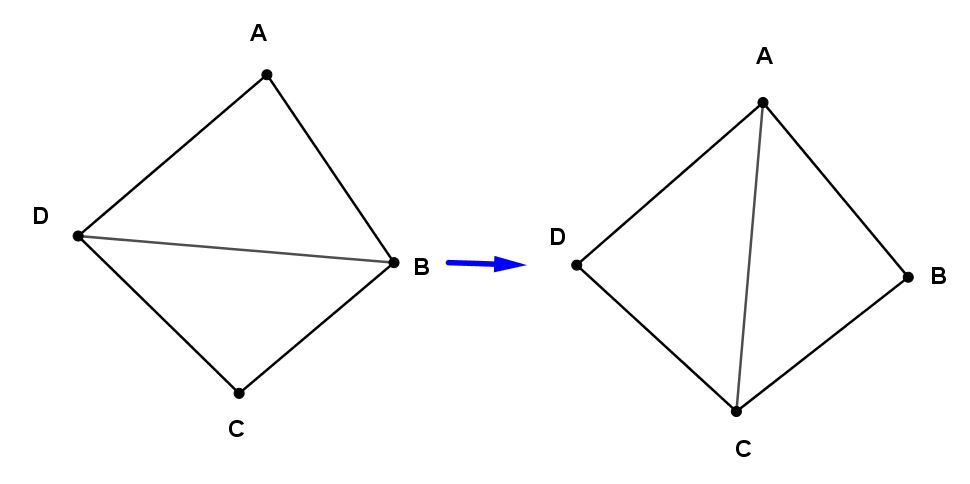}
    \caption{Edge Swapping}
    \label{fig:edgeswapping}
\end{figure}

\subparagraph{EDGE SWAPPING}
The diagonal swapping changes the connectivities among nodes in the grid without altering their position. This process requires a loop over all the element sides excluding those sides on the boundary. For each side AB (fig \ref{fig:edgeswapping}) common to the triangles ABC and ADB we consider the possibility of swapping AB by CI, this replacing he two triangles ABC and ADB by the triangles ADC and BCD. The swapping is performed inf a prescribed regularity criterion is satisfied better by the new configuration than by the existing one. In our implementation, the swapping operation is performed if the minimum angle occurring in the new configuration is larger that in the original one. The edge swapping technique is only permitted with convex quadrilaterals \cite{11}.

\subparagraph{NODE INSERTION}

Node insertion is, as we will see, one of themost common techniques used by the generation algorithms for mesh enhancement, We could officiously classify the different generation algorithms according to its node insertion process.

Previous insertion: The nodes are previously prespecified. That is, the nodes could be specified by means of a node cloud (maybe randomly distributed nodes= or the nodes fixed by another mesh (e.g. Basic Delaunay algorithms, edge swapping algorithms, etc.)

Insertion during the generation: The initial triangulating domain is node.free and the mesh is created node by node following determined criteria. (e.g. advancing from methods, etc.)

Posterior insertion: Additional nodes are added to the already generated grid to improve the mesh quality or to increase the accuracy of the solution (e.g. different Delaunay algorithms, Steiner algorithms, etc.)

Note that much of the mentioned algorithms combine these insertion procedures.

\begin{figure}[H]
    \centering
    \includegraphics[width=10cm]{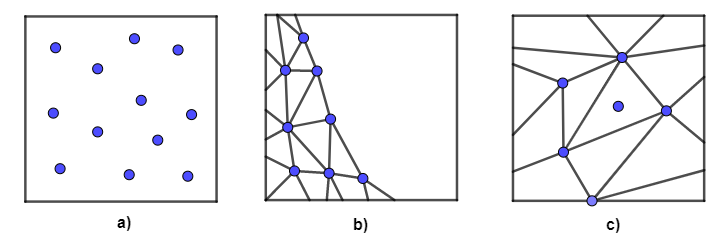}
    \caption{Node insertion (a) previous, (b) during and (c) posterior}
    \label{fig:nodeinsertion}
\end{figure}

\subparagraph{MESH SMOOTHING}

Grid smoothing alters the position of the interior nodes without changing the topology of the grid. The element sides are considered as springs of stiffness proportional to the length of the side. The nodes are moved until the spring system is in equilibrium. The equilibrium positions are found by iteration. Each iteration amounts to performing a loop over the interior points and moving their coordinates to coincide with those of the centroid of the neighbouring pints. Usually three to five iteration are performed. 

\begin{figure}[H]
    \centering
    \includegraphics[width=8cm]{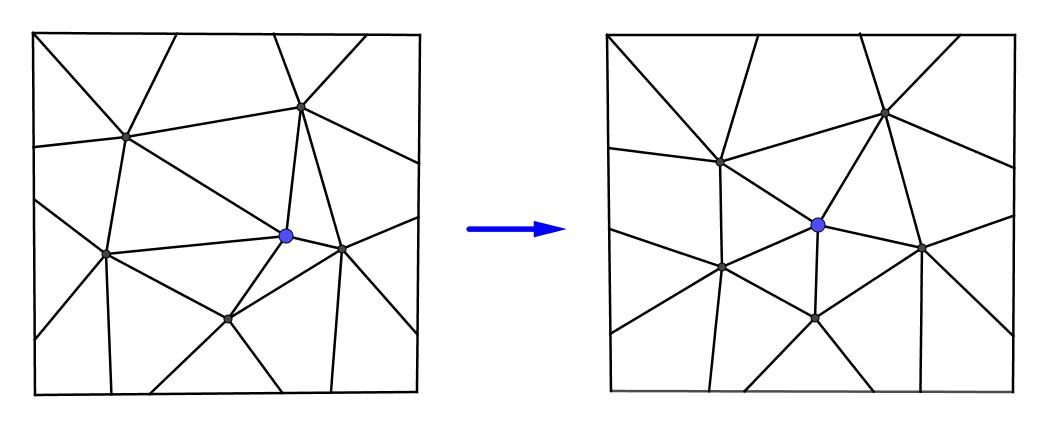}
    \caption{Smoothing}
    \label{fig:smoothing}
\end{figure}

the combined application of these three post-processing algorithms is found to be very effective in improving the smoothness and rgularity of the generated grids.

\subsubsection{DELAUNAY GENERATION ALGORITHMS}

We now consider several techniques for Delaunay triangulatio nin two dimensions. These methods were chosen because its optimal performance in rather different situations. The discussion of the Delaunay algorithms is organized as follows:

\begin{enumerate}
    \item Global Edge Swapping (Lawson)
    \item Divide-and-Conquer Algorithm
    \item Space Marching Method
    \item Incremental Insertion Algorithms
    \begin{enumerate}
        \item Bowyer Algorithm
        \item Watson Algorithm
        \item Green and Sibson Algorithm
    \end{enumerate}
\end{enumerate}

\begin{figure}[H]
    \centering
    \includegraphics[width=10cm]{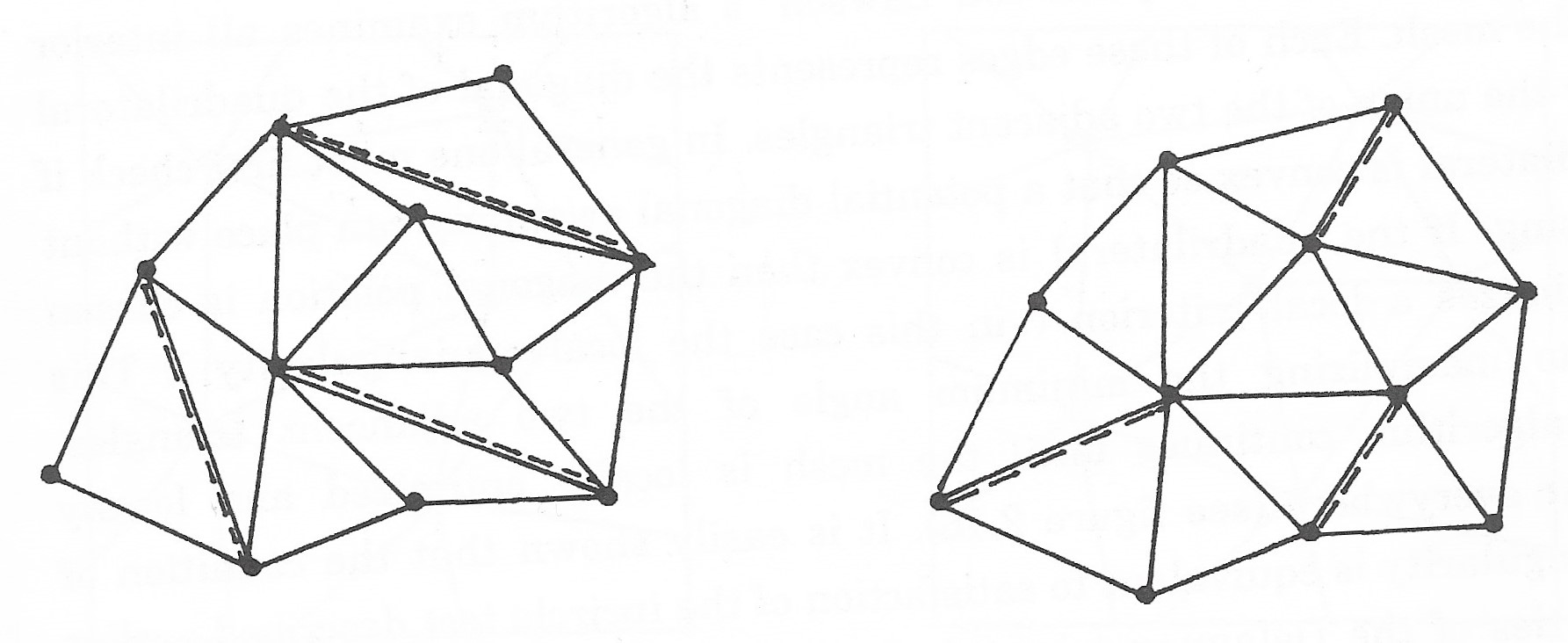}
    \caption{Delaunay triangulation via Edge Swapping.}
    \label{fig:delaunay swapping}
\end{figure}

\paragraph*{Delaunay Triangulation Via Edge Swapping}
This algorithm due to Lawson (1977) \cite{11} assumes that a triangulation exists (not Delaunay) then makes it Delaunay through application of edge swapping such that the equiangularity of a triangulation, A(T), is defied as the ordering of angles A(T) = $[\alpha_1,\alpha_2,\alpha_3,...\alpha_{3n(c)_3}]$ such that $\alpha_i \leq \alpha_j$ if $i < j $. We write A(T*) $<$ A(T) if $\alpha^*_j \leq \alpha_j$ and $\alpha^*_i = \alpha_i$ for $1 \leq i <j$. A triangulation T is globally equiangular if A(T*) $\leq$ A(T) for all triangulations T* of the point set. Lawson's algorithm examines all interior edges of the mesh. Each of these edges represents the diagonal of the quadrilateral formed by the union of the two adjacent triangles. In general on must firs check if the quadrilateral is convex so that a potential diagonal swapping can place without edge crossing. If the quadrilateral is convex then the diagonal position is chosen which optimizes a local criterion (in this case the local equiangularity). This amounts to maximizing the minimum angle of the two adjacent triangles. Lawson's algorithm continues until the mesh is locally optimized and locally equiangular everywhere (see figure \ref{fig:delaunay swapping}). It is easily shown that the condition of local equiangularity is equivalent to satisfaction of the incircle test described earlier (see properties of the Delaunay triangulation). therefore a mesh which is locally equiangular everywhere in a Delaunay triangulation. Note that each new edge swapping (triangulation T*) insures that the global equiangularity increases A(T*)>A(T). Because the triangulation is of finite dimension, this guarantees that the process will terminate in a finite number of steps.

\textbf{Iterative Algorithm:} Triangulation via Lawson's algorithm
\begin{itemize}
    \item swapedge = true
    \item While(swapedge) do
    \begin{itemize}
        \item swapedge= false
        \item Do (all interior edges)
        \begin{itemize}
            \item  If(adjacent triangles form convex quad) then
            \item swap diagonal to form T*
            \begin{itemize}
                \item If( optimization criteria satisfied ) then
                \item T=T* 
                \item swapedge = true 
                \item Endif
            \end{itemize}
            \item  Endif
        \end{itemize}
        \item Enddo
    \end{itemize}
    \item Endwhile
\end{itemize}

When Lawson's algorithm is used for constructing Delaunay triangulations, the test for quadrilateral convexity is not needed. it can be shown that non convex quadrilaterals formed from triangle pairs never violate the circumcircle test. When more general optimization criteria is used (discussed later), the convex check must be performed.

\paragraph*{Divide-and-Conquer Algorithm}
In this algorithm, the nodes are assumed to be prespecified. The idea is to partition the cloud of points T (sorted along a convenient axis) into left (L) and right (R) half planes. Each half plane is then recursively Delaunay triangulated. The two halves mus then be merged together to form a single Delaunay triangulation. Note that we assume that the points have been sorted along the x-axis for purposes of the following discussion (this can be done with O( N log N) complexity, with N number of nodes).

\textbf{Algorithm:} Delaunay triangulation via divide-and-Conquer.
\begin{enumerate}
    \item T partition into two subsets $T_L$ and $T_R$ of nearly equal size.
    \item Delaunay triangulate$T_L$ and $T_R$ into a single Delaunay triangulation.
    \item Merge $T_L$ and $T_R$ into a single Delaunay triangulation.
\end{enumerate}

\begin{figure}[H]
    \centering
    \includegraphics[width=11cm]{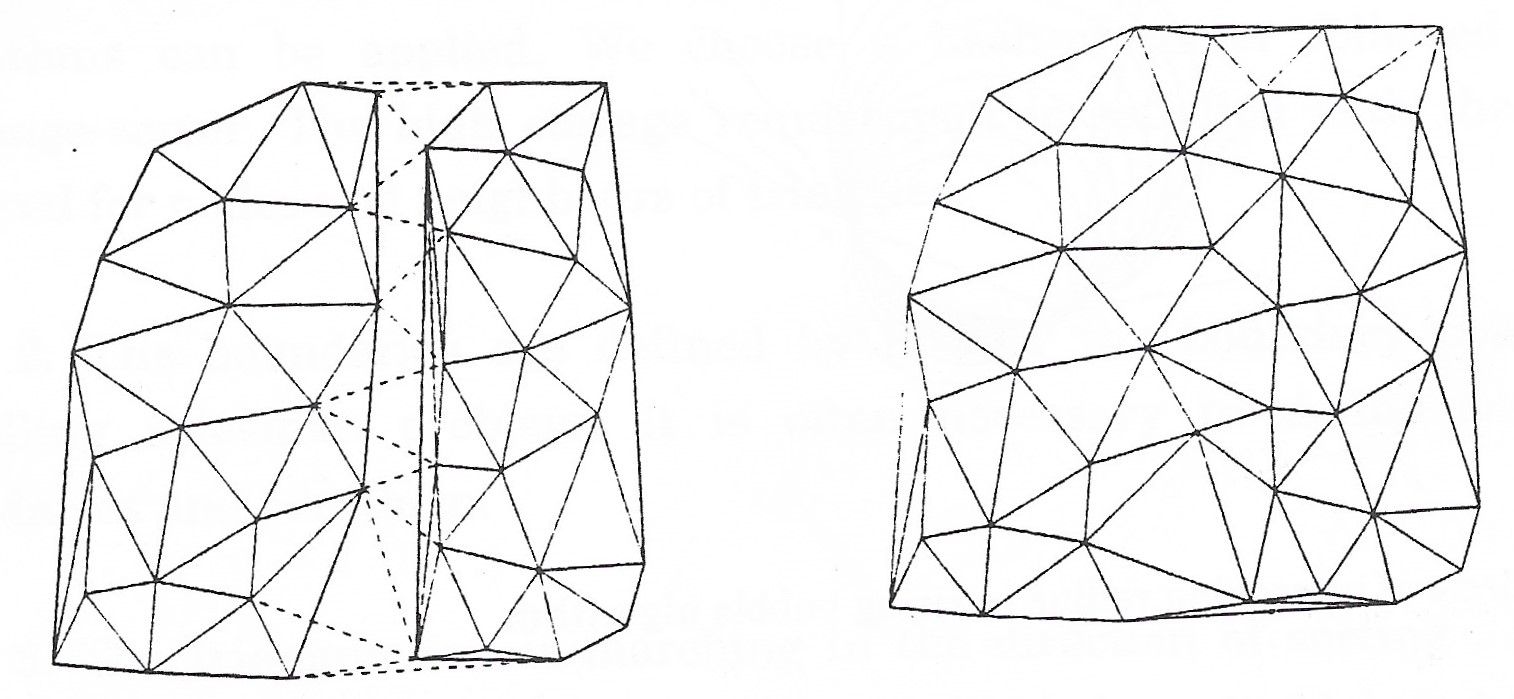}
    \caption{Divide and Conquer.}
    \label{fig:divide and conquer}
\end{figure}

\begin{figure}[H]
    \centering
    \includegraphics[width=6cm]{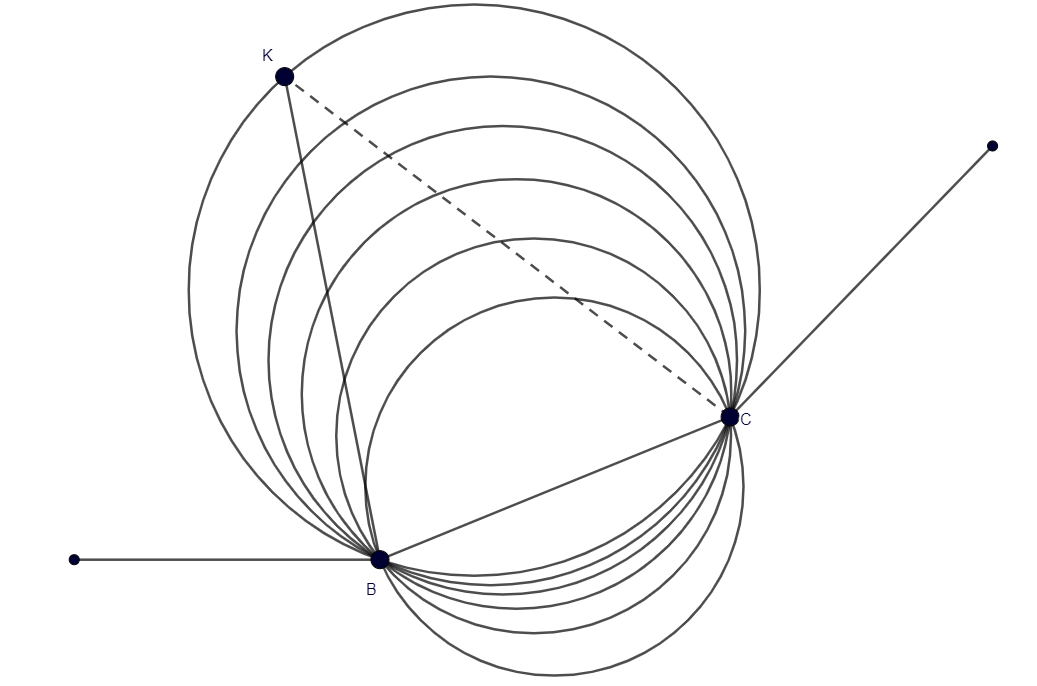}
    \caption{Circle of increasing radius in rising bubble algorithm.}
    \label{fig:circle increasing}
\end{figure}

The only difficult step in the Divide-and-Conquer algorithm is the merging of the left and right triangulations. The process is simplified by noting two properties of the merge:
\begin{itemize}
    \item Only cross edges (L-R or R-L) are created in the merging process. Since vertices are neither added or deleted in the merge process, the need for a new R-R of L-L edge indicates that the original right of left triangulation was defective. (note that the merging process will require the deletion of edges L-L and/or R-R)
    \item Vertices with minimum (maximum) y value in the left and right triangulations always connect as cross edges. this is obvious given that the Delaunay triangulation produces the convex hull of the point cloud.
\end{itemize}
        Given the properties we now outline the "rising bubble" \cite{66} merge algorithm. This algorithm produces cross edges in ascending y-order. The algorithm begins by forming a cross edge by connecting vertices of left and right triangulations with minimum y-value (property 2). This forms the initial cross edge for the rising bubble algorithm. More generally consider the situation in Which we have a cross edge between A and B with endpoints above the half plane formed by a line passing through A-B, see figure \ref{fig:circle increasing}.
    
    This figure depicts a continuously transformed circle of increasing radius passing through the points A and B. Eventually the circle increases in size and encounters a point C from the left or tight triangulation ( in this case, point C is in the left triangulation). A new cross edge (dashed line in figure \ref{fig:circle increasing}) is then formed by connecting this point to a vertex of A-B in the other half triangulation. Given the new cross edge, the process can then be repeated and terminates when the L-L or R-R edges can take place during or after the addition of cross edges. Properly implemented, the merge can be carried out in linear time, O(N). Denoting T(N) as the total running time, step 2 is completed in approximately 2T(N/2). this the total running time is described by the recurrence $T(N) = 2T(N/2) + O(N) = O(N log N)$.
    
    \paragraph*{Space Marching Method}
    The Space Marching Method, elaborated by Vilsmeier/Hänel (1991) \cite{59}, is based primarily on the Basic Delaunay triangulation and offers an enhanced version of the former.

    The disadvantage of the method above is the large computational time due to the extensive search. This time is proportional to $N^2$, where N is the number of points. For a large amount of points this method is therefore not suitable. In order to avoid extensive search the method can be modified by first sorting all points and performing the triangulation in a space marching direction.

    \textbf{Algorithm:} Space Marching Method

    \begin{enumerate}
        \item All points, boundary nodes included, are sorted in a Cartesian direction. It is useful to choose the wider direction. Depending on the problem different sorting algorithms can be applied. We choose a hash-presorter followed by a simple exchange-sorter. The high storage requirement is satisfied with the storage later required for nodes and neighbours of triangles.
        \item The boundaries are defined by linking the boundary points, solving a travelling salesman problem. It is often necessary to define sections on the boundaries and join them..
        
        \item The triangulation is marching in the direction of sorting. Let N1 be the lowest address of a point, which is not yet behind the marching front and N2 be a limiting address, greater than N1. Th front edges, for which a new triangle is sought are those, which have two points with addresses lower or equal to N2. The branch of points to be examined as possible partners for generating a new triangle is limited to those with addresses between N1 and N2. The criterion applied, to find the suitable point is the same as in the basic Delaunay Version. 
        
        \item  Some checks with the new triangles should be performed, especially to avoid crossing of edges of the new triangles with other edges from the front. This may happen on the inner boundaries and in cases where the Delaunay- criterion is not unique (more than three points on the same circle). If problems appear the new triangle is skipped. But to avoid crossing it is sufficient to check all front edges with at least one point, with an address lower or equal to N2.
        
        \item The new edges are inserted in the list of front edges. Two edges, touching each other, are deleted from the front list, while the neighbourhood of the forming triangles is adjusted.
        
        \item  The process continues until the front list is empty, while the addresses N1 and N2 are adjusted successively as required.
    \end{enumerate}
     
\begin{figure}[ht]
      \begin{minipage}[trim=10 0 0 10mm]{0.7\linewidth}
        \includegraphics[width=\linewidth]{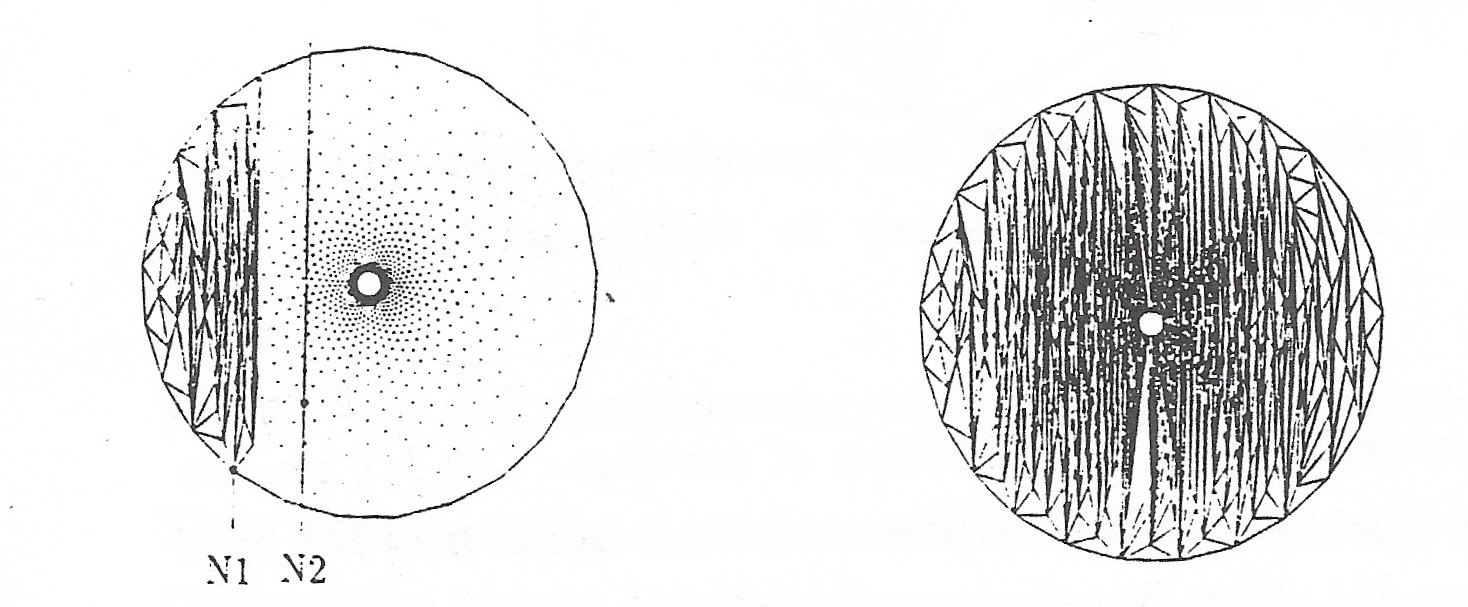}
        \caption{Space marching algorithm}
        \label{fig:space marching algorithm}
      \end{minipage}
      \hfill
      \begin{minipage}[trim=10 0 0 10mm]{0.3\linewidth}
        \includegraphics[width=\linewidth]{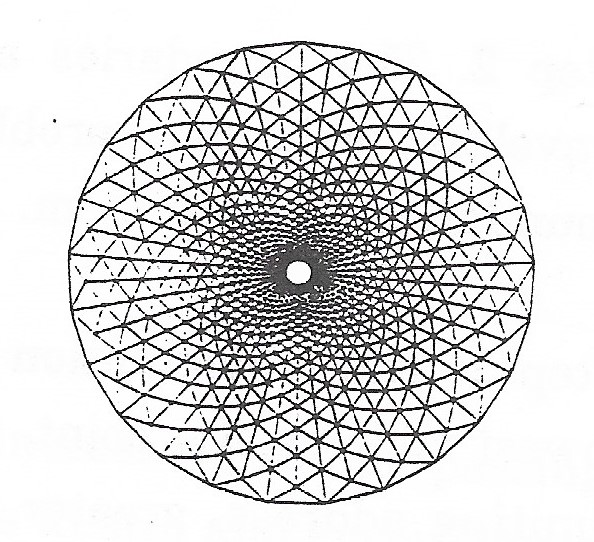}
        \label{fig: marching final grid}
        \caption{Final Grid}
      \end{minipage}%
    \end{figure}

    The output of the method above is a closed triangulation where all nodes and neighbours of the triangles are known, but which does not to have to satisfy Delaunay's criterion due to the reduce choice of points when forming a new triangle (figure \ref{fig:space marching algorithm} ). Therefore remeshing is required (see remeshing techniques).
    
    The major part of computational work, also for the space marching method, is caused  by the triangulation. The searching all the time is proportional to the difference N2-N1. Let us consider the number points on the triangulate d front to be of order. O($\sqrt{N}$) and let us take the same amount of new points, not yet triangulated. The amount N2-N1 will still be of the order O($\sqrt{N}$) and the triangulation time will be of the order O($N\sqrt{N}$). A lower amount of new points reduces the amount N2-N1 and, new point, and to do that only if no new triangle can be built using only the points of the front themselves. The number of front edges with addresses lower or equal to N2 will soon be in the order O(1) as well as the order of the amount N2-N1, so that the order of the triangulation approaches to O(N).
    
    \paragraph*{Incremental Insertion Algorithms}
    For simplicity, assume that the node to be added lies within bounding polygon of the existing triangulation. If we desire a triangulation form a new set of nodes, three initial phantom points can always be added which define a triangle large enough to enclose all points to be inserted. In addition, interior boundaries are enough to enclose all points to be inserted. In addition, interior boundaries are usually temporally ignore for purpose of the Delaunay triangulation. This introduces the task of point location in the triangulation. Some incremental algorithms require knowing any triangle whose circumcircle contains the node. In either case, two extremes arise in this regard. Typical mesh adaptation and refinement algorithms determine the particular cell for node insertion as part of the mesh adaptation algorithm, thereby reducing the burden of point location. In the other extreme, initial triangulations of randomly distributes nodes usually requires advanced searching techniques for point location to achieve asymptotically optimal complexity O(N log N). Search algorithms based on quad-tree and split-tree data structures work extremely well in this case. Alternatively, search techniques based on "walking" algorithms are frequently used because of their simplicity. These methods work extremely well when successively added point are close together. The basic idea is start at the location in the mesh of the previous inserted point and move on edge (or cell) at a time in the general direction of the newly added point. In the worst case, each point insertion requires O(N) walks. This would result in a wort case overall complexity O($N^2$). For randomly distribute points, the average point insertion requires O($N^{0.5}$) walks which give an overall complexity O($N^{1.5}$). For many applications where successive points tend to be close together, the number of walks is roughly constant and these simple algorithms can be very competitive. Using any of these techniques, we can proceed with the insertion algorithms.
    \begin{itemize}
        \item \textbf{ BOWYER'S ALGORITHM}

    The basic idea in Bowyer's algorithm is to insert a new node into an existing Voronoi diagram (for example node Q in figure \ref{fig:bowyer algorithm}). determine its territory, delete any edges completely contained in the territory, the add new edges and reconfigure existing edges in the diagram. The following is Bowyer's algorithm essentially as presented by Bowyer (see Ref. \cite{68} for complete details.).

    \textbf{Algorithm:} Incremental Delaunay triangulation, Bowyer \cite{68}.
    
        \begin{enumerate}
            \item Find any existing vertex in the Voronoi diagram closer to the new point than to its forming points. This vertex will be deleted in the new Voronoi diagram.
            \item Perform tree search to find remaining set of deletable vertices that are closer to the new point that to their forming points.
            \item  Find the set P of forming points corresponding to the erasable vertices.
            \item Delete edges of the Voronoi which can be described by pairs of vertices in the set if both forming points of the edges to be deleted are contained in P.
            \item Calculate the new vertices of the Voronoi, compute their forming points and neighboring vertices, and update the Voronoi data structure.
        \end{enumerate}
        Implementation details and suggested data structures are given in the paper by Bowyer.
    
    
    \begin{figure}[H]
    \centering
  \begin{subfigure}[b]{0.3\textwidth}
    \includegraphics[width=\textwidth]{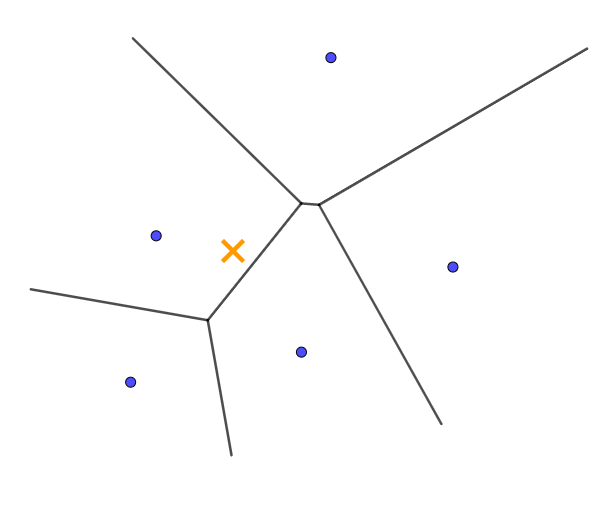}
  \end{subfigure}
  \hspace{1cm}
  \begin{subfigure}[b]{0.3\textwidth}
    \includegraphics[width=\textwidth]{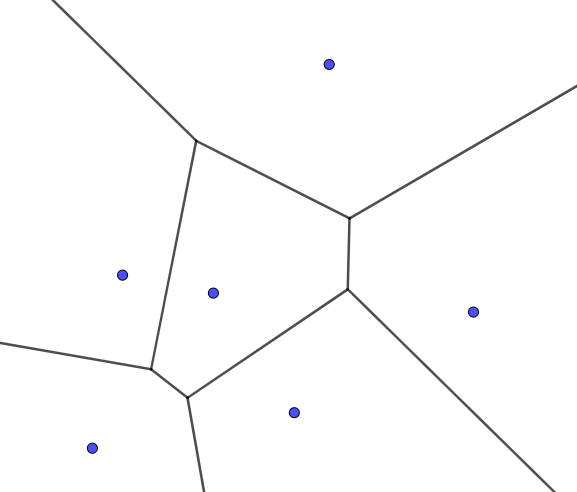}
  \end{subfigure}
  \caption{Bowyer's agorithm.}
  \label{fig:bowyer algorithm}
\end{figure}
    
    \item  \textbf{WATSON'S ALGORITHM}

    Implementation of the Watson \cite{69} algorithm is relatively straightforward. The first step is to insert a new node into an existing Delaunay triangulation and to find any triangle (the root) such that the new node lies interior the that triangles circumcircle. Starting at the root, a tree search is performed to find all triangles with circumcircle containing the new node. this is accomplished by recursively checking triangle neighbours. The resulting set of deletable triangles exposes a polygonal cavity surrounding node Q with all the vertices pf the polygon visible to  node Q. The interior of the cavity is then retriangulate by connecting the vertices of the polygon to node Q (see figure \ref{fig:watson algorithm}). this completes the algorithm. A through account of Watson's algorithm is given by Baker \cite{70} where he considers issues associated with constrained triangulations.

    \textbf{Algorithm:} Incremental Delaunay Insertion, Watson \cite{69}
    \begin{enumerate}
        \item  Inset new node Q into existing Delaunay triangulation.
        \item Find any triangle with circumcircle containing node Q.
        \item  Perform tree search to find remaining set of deletable triangles with circumcircle containing node Q.
         \item Construct list of edges associated with deletable triangles. Delete all edges from the list that appear more that once.
         \item Connect remaining edges to node Q and update Delaunay data structure.
    \end{enumerate}

     \begin{figure}[H]
       \centering
       \begin{subfigure}[b]{0.3\textwidth}
         \includegraphics[width=\textwidth]{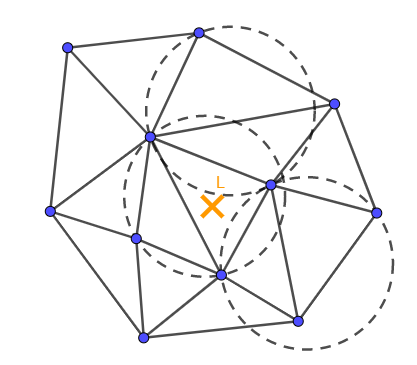}
         \caption{Delaunay triang. with node added in the cross.}
       \end{subfigure}
       \hspace{2cm}
       \begin{subfigure}[b]{0.3\textwidth}
          \includegraphics[width=\textwidth]{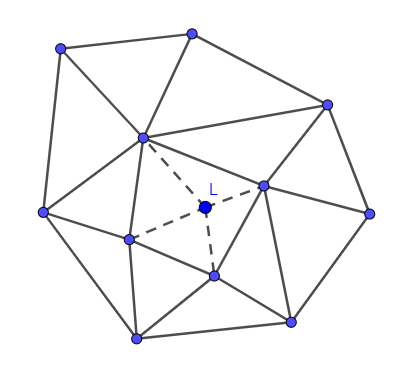}
          \caption{Triangulation after delete the invalid edges and re-connection.}
       \end{subfigure}
       \caption{Watson's algorithm.}
       \label{fig:watson algorithm}
     \end{figure}

    \item   \textbf{GREEN AND SIBSON'S ALGORITHM}

    The algorithm due to Green and Sibson [67] is very similar to the Watson's algorithm. The primary difference is the use of local edge swapping to reconfigure the triangulation. The first step is location, i.e. find the triangle containing point Q. Once this is done, three edges are then created connecting Q to the vertices of this triangle as shown in figure \ref{fig:green algorithm}. If the point falls on an edge, then the edge is deleted and four edges are created connecting to vertices of the newly created quadrilateral. Using the circumcircle criteria it can be shown that the newly created edges (3 or 4) are automatically Delaunay. Unfortunately, some of the original edges are now incorrect. We need to somehow find all "suspect edges" which could possibly fail the circle test. Given that this can be done (described below), each suspect edge is viewed as a diagonal of the quadrilateral formed from the two adjacent triangle. The circumcircle test is applied to either one of the two adjacent triangles of the quadrilateral. If the fourth point of the quadrilateral is interior to the circumcircle, the suspect edge is the swapped as shown in figure \ref{fig:green algorithm} (b), two more edges then become suspect. At given time we can immediately identify all suspect edges. To do this, first consider the subset of all triangles which share Q as a vertex. Once can guarantee at all times that all initial edges incident to Q are Delaunay and any edge made incident to Q by swapping must be Delaunay. Therefore, we need only consider the remaining edges of this subset which form a polygon about Q as suspect and subject to the incircle test. the process terminates when all suspect edges pass the circumcircle test.
    
    \begin{figure}[H]
      \centering
      \begin{subfigure}[b]{0.3\textwidth}
        \includegraphics[width=\textwidth]{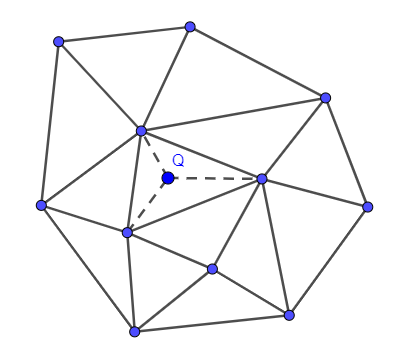}
        \caption{ Insert of vertex.}
      \end{subfigure}
      \hspace{1cm}
      \begin{subfigure}[b]{0.3\textwidth}
        \includegraphics[width=\textwidth]{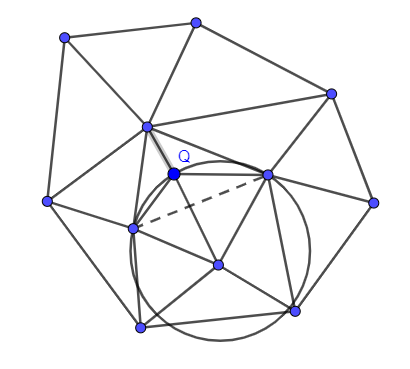}
        \caption{Swapping of suspect edge.}
      \end{subfigure}
      \caption{Green Algorithm}
      \label{fig:green algorithm}
    \end{figure}

    \textbf{Algorithm:} Incremental Delaunay Triangulation, Green and Sibson \cite{67}
    \begin{enumerate}
        \item Locate existing cell enclosing point Q.
        \item Insert site and connect to 3 or 4 surrounding vertices.
        \item Identify new suspect edges.
        \item Perform edge swapping of all suspect edges falling the incircle test.
        \item Identify new suspect edges.
        \item If new suspect edges have been created, go to step 3.
    \end{enumerate}
    
    the Green and Sibson algorithm can be implemented using standard recursive programming techniques. The heart of the algorithm is the recursive procedure which would take the following form for the configuration shown in figure \ref{fig:green algorithm2}.
    
    \begin{itemize}
        \item procedure swap [$v_q,v_1,v_2,v_3,edges$]
        \item if (incircle [$v_q,v_1,v_2,v_3$] = TRUE) then
        \begin{itemize}
            \item call reconfig\_edges [$v_q,v_1,v_2,v_3,edges$]
            \item call swap [$v_q,v_1,v_4,v_2,edges$]
            \item call swap [$v_q,v_1,v_5,v_3,edges$]
        \end{itemize}
        \item endif 
        \item end procedure
    \end{itemize}
            
    This example illustrates an important point. The nature of Delaunay triangulation guarantees that edge swapped incident to Q will be final edges of the Delaunay recursive procedure. In a later section we will consider incremental insertion and edge crossing for generating non-Delaunay triangulations based on other swapping criteria. this algorithm can also be programmed recursively but requires backward propagation in the recursive implementation. For the Delaunay triangulation algorithm, the insertion algorithm would simplify to the following three steps:
    
    Recursive Algorithm: Incremental Delaunay Triangulation, Greeen and Sibson
    \begin{enumerate}
        \item Locate existing cell enclosing point Q.
        \item Insert site and connect to surrounding vertices.
        \item Perform recursive edge swapping on newly formed cells (3 or 4)
    \end{enumerate}

     \begin{figure}[H]
    \centering
    \includegraphics[width=10cm]{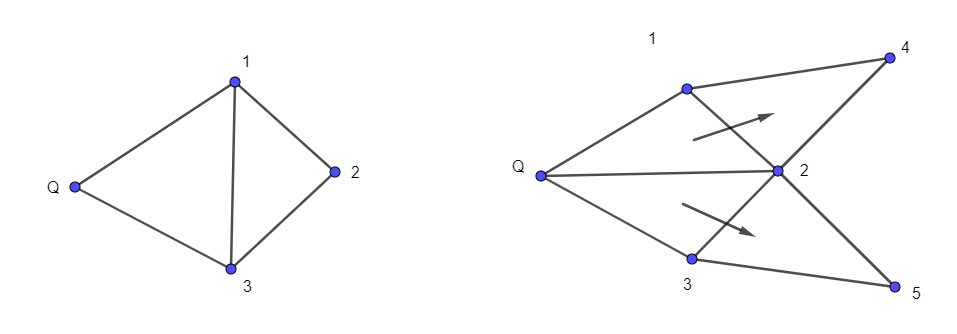}
    \caption{Edge swapping with forward propagation.}
    \label{fig:green algorithm2}
\end{figure}

    \end{itemize}

    \paragraph*{Implementation of Delaunay Algorithm}
    \begin{itemize}
        \item Baker(1990), \cite{5}: A method for generating tetrahedral meshes is described. The algorithm is based on the Delaunay triangulation and can treat objects of essentially arbitrary complexity.
        
        \item Mitty (1991) \cite{72}: Adaptive mesh refinement on unstructured meshes in three-dimensions is applied to obtain a sharp resolution of oblique shock waves. Meshes are generated through the application  Bowye's algorithm to yield a Delaunay tessellation of the space.
        
        \item Mavriplis (1991), \cite{73}: A method for generating and adaptive refining a highly stretched unstructured mesh, suitable for the computation of high-Reynolds-number viscous flows about arbitrary two-dimensional geometries has here been developed. The method is based on the Delaunay triangulation of a predetermined set of points.
        \item Taniguchi (1991), \cite{74}: Delaunay-based grid generation for a three dimensional body with complex boundary geometry.
        \item Meshkat (1991), \cite{75}: Three- Dimensional automatic unstructured grid generation based on Delaunay tetrahedization.
        \item Rebay (1993),\cite{76}: Efficient unstructured mesh generation by means of Delaunay triangulation and Bowyer-Watson algorithm.
    \end{itemize}
    
    \subsubsection{ADVANCING FRONT ALGORITHMS}
    We will now try to study in depth the different advancing front methods (AFM), the second approach to the construction of triangular and tetrahedral grids in computational fluid dynamics(CFD). In this section we will include Huet's "methode de front" as a kind of advancing front algorithm. the discussion of the advancing front algorithms is organised as follows:
    
    \begin{itemize}
        \item Original Advancing Front Method
        \item Advancing Front versions
        \item "Methode de front"
        \item Implementation of the Advancing Front algorithms
    \end{itemize}
    
    \paragraph*{Original Advancing Front Method} \label{chap:afm}
    The best developed and quite likely most widely used 2D and 3D grid generation algorithm is the Advancing Front Method (AFM) by Peraire first described in \cite{24}. As a departure from Delaunay based grid generation it is based on heuristic principles on how to place and connect vertices to create smooth and regular meshes. Moreover, AFM allows to create meshes with cells that are locally stretched in a certain direction,k a feature that many researchers using Delaunay based concepts struggled with \cite{25}. Subsequently the AFM has been successfully extended to 3D tetrahedralisations \cite{15},\cite{17}.
    
    We will first of all present this 'original' version and later of comment the different enhancements
    
\begin{figure}[H]
    \centering
    \includegraphics[width=8cm]{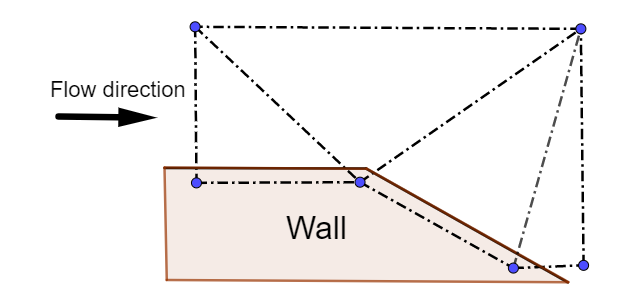}
    \caption{Coarse background grid used t cover the computational domain (shaded) fr a problem of supersonic expansion around a corner.}
    \label{fig:background 1}
\end{figure}

    \subparagraph{The Background Grid}
    For the method to be described here, the process of generating a mesh over a two-dimensional region of arbitrary shape is started by constructing a by hand a coarse background grid of 3-noded triangular elements which completely covers the solution domain of interest. This illustrated in figure \ref{fig:background 1} which shows a possible background grid consisting of only four elements, for a problem of expansion flow around a corner.
    
\begin{figure}[H]
    \centering
    \includegraphics[width=9cm, height=3cm]{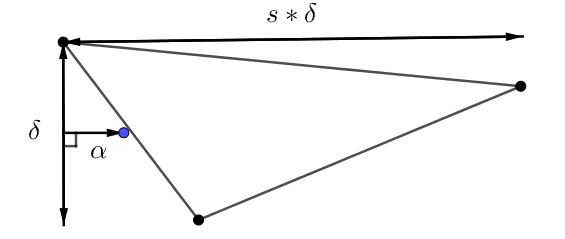}
    \caption{$\delta$, $\alpha$ and $s$ definition in a triangle of the mesh.}
    \label{fig:background}
    \end{figure}

    For the elements to be generated, it is convenient to define a node spacing $\delta$, the value of a stretching parameter $s$ and a direction of stretching $\alpha$. The generated elements will the have typical length $s \delta$ in the direction parallel to $\alpha$ and a typical length $\delta$ in the direction normal to $\alpha$(see figure \ref{fig:background}). The background grid is used to provide a piece wise linear spatial distribution for these parameters over the grid to be generated. Thus, at each node on the background grid, the nodal values of $\delta$, $s$ and $\alpha$ must be specified. During the generation process the local values of these quantities will be obtained by linear interpolation, over the triangles of the background grid, between the specified nodal values. For the initial mesh, the location of one-dimensional features is not known in general and so the value $s = 1$ (i.e. no stretching) is normally specified. The node spacing $\delta$ can also be defined to be uniform but a variation of $\delta$ can be achieved (by suitable construction of the background grid) if it is apparent that increased mesh resolution is required in certain regions of the flow domain, e.g., in the vicinity of the corner in the problem of figure \ref{fig:background}. Note that if $\delta$ is required to be uniform initially and no stretching has been specified, the the background gird need only consist of a single element which completely covers the solution domain.
    
    \subparagraph{Generation of Boundary Nodes}
    The boundary of the solution domain is represented by the union of closed loops of curved segments and boundary nodes are placed at the points of intersection of these segments. for simply connected regions there is only one closed loop, whereas for multi-connected regions there will be as many internal loops as the number of openings inside the domain. The segments of exterior boundary are defined in a anti-clockwise fashion. This means that, as the boundary curve is traversed, the region to be triangulated always lies to the left. Before beginning the process of generation triangles within the region of interest, the positioning of additional nodes on the boundaries of the region has to be performed. Each boundary segment is considered in turn and nodal points are generated on the boundary segments, with the spacing of the points being determined by interpolated values for $\delta,s$ and $\alpha$.
    
    \subparagraph{The Triangle Generation}
    The mesh generation algorithm utilises the concept of a generation front in a form which is very similar to that proposed by S.H. Lo \cite{26}. At the start of the process the front consists of the sequence of straight line segments which connect consecutive boundary nodes. During the generation process, any straight line segment which is available to form an element side is termed active, whereas any segment which is no longer active is removed from the front. Thus, while the domain boundary will always remain the same, the generation front will change continuously and ha st o be updated whenever a new element is formed.
    
    The following steps are involved in the process of generating a new triangle in the mesh:
    \begin{enumerate}
    \item An active segment of the front is chosen to generate a new triangle. This decision can be make choosing the smaller edge (this improves the mesh quality if the spacing delta varies along the domain) or the first active segment (always follows the loop of creation).
    \item When a segment is chosen is time to abstract the generation of the new triangle from the domain. This is done through the calculation of $\delta_M$, $\alpha_M$ and $s_M$ in the middle point (M) of the initial segment ($\overline{AB}$), which will be the base of the new triangle. This is done, as it has been said before, by interpolating over the background grid. Also a local rotation of coordinates is made so that $\alpha_M$ goes along the $x_1$-axis. In this new coordinate system, a triangle as regular as possible will be generated.
    \item to avoid excessive distortion in the generation of the new triangle, the new delta $\delta_1$ is calculated according to 
    \begin{equation}
        \delta_1 =   \begin{cases}
      0.55 AB &  \delta_m < 0.55AB\\
      \delta_M &  0.55AB<\delta_m < 2AB \\
      2AB &  2AB< \delta_m
    \end{cases}
    \end{equation}
    With $\delta_1$ will be created the third vertex of the triangle, the point C.

\item To consider possible incompatibilities with the front, all the points, belonging to the front, inside the circle with centre at C and radius nAB, being $n=5$ according to Peraire but n can have any other value. All the points inside are enumerated increasingly in a list according to their distance to C, being $N_1$ the closest one.

\item Set C at the head of the list unless:
\begin{equation}
    AN_1 < 1.5 \delta_1  \text{  and  }   BN_1 < 1.5 \delta_1
\end{equation}
\item  In this last case, the new point $N_j$ is taken to be the new vertex taking into account that no other node is contained in the triangle (just C) and that the line $MN_j$ does not intersect any other active segment. The new element is formed, the coordinates are transformed back to the original space and the front is updated as indicated previously.
\end{enumerate}

The triangle generation process ceases when the number of active sides in the front is reduced to zero.

\subparagraph{Searching Algorithm}

Each time the values of $\delta_M, s_M$ and $\alpha_M$ are required during the stage (b) above, they have to be obtained by locating point M within an element of the background grid. An efficient search algorithm has been implemented which requires, for each element "e" of the background grid, the knowledge of the three surrounding elements which have sides in common with element "e". Given the coordinate M and a starting element of the background grid, the three area coordinates \cite{27} of M are determined.

    \begin{figure}[H]
    \centering
    \includegraphics[width=9cm]{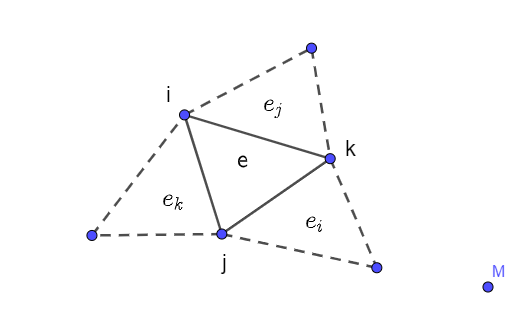}
    \caption{The searching algorithm to locate the point $M$ on the background grid. The area coordinates $L_i$, $L_j$ and $L_k$ of element $e$ are evaluated ant $M$. Here $L_i(M) < L_j(M) < L_k(M)$ and the next element to be checked is element $e_i$.}
    \label{fig:searching}
    \end{figure}

If each area coordinate lies between zero and one the n the element contains the point M. If not, the node for which the area coordinate is a minimum (see figure \ref{fig:searching} is found and this indicates the next element to be checked. In this manner, the necessity of searching over all the elements in the background grid is avoided.

         \begin{figure}[H]
    \centering
    \includegraphics[width=9cm]{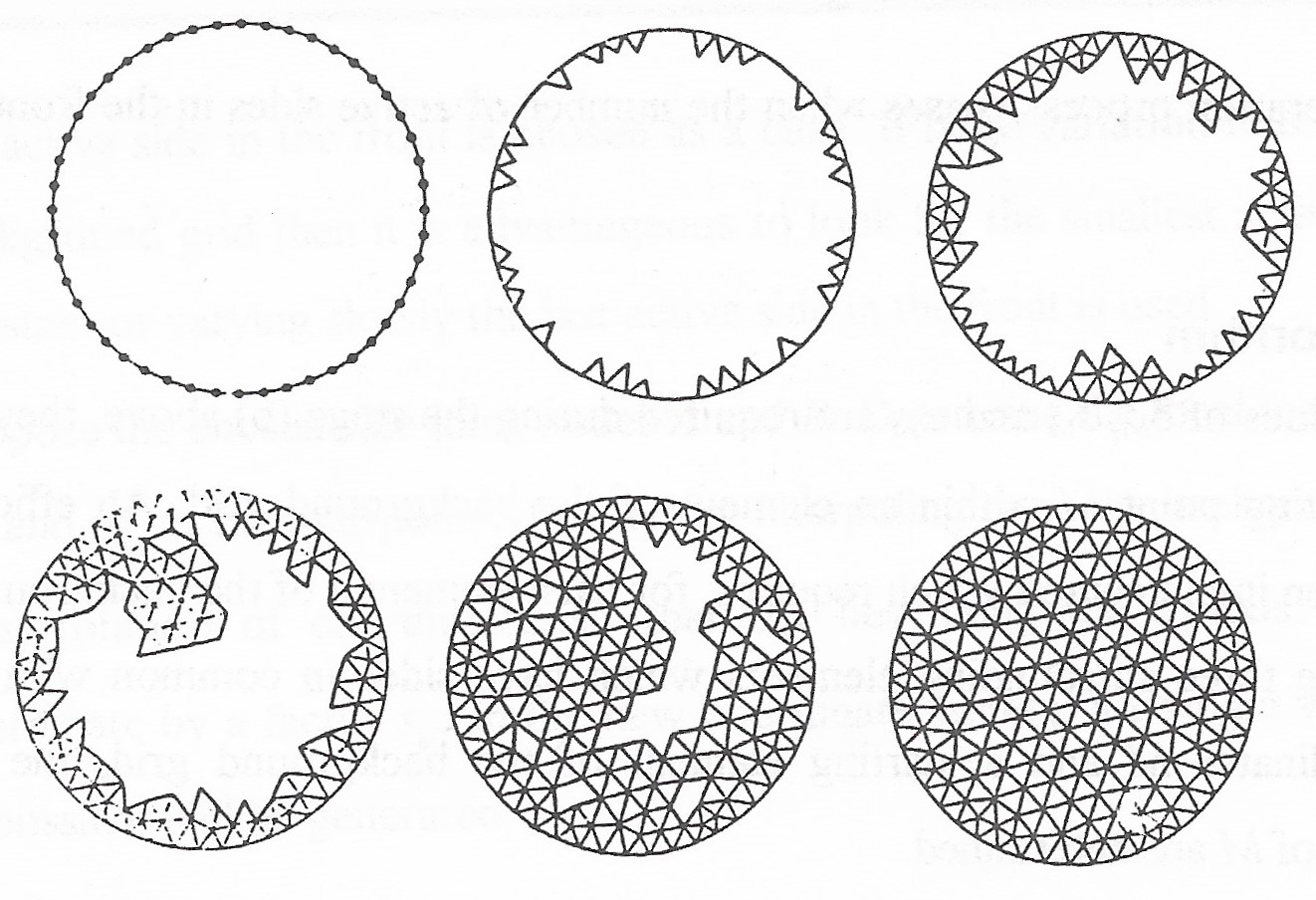}
    \caption{Advancing front technique showing different stages during the triangulation.}
    \label{fig:front meth}
    \end{figure}

\paragraph*{Advancing Front Methods}
The algorithmic procedures to be described for the grid generation are based upon the method originally proposed above \cite{24}. Peraire (1992) \cite{9} presents here a modified triangle generation concept to aid the grind generation procedure, a transformation T, which is a function of $\alpha_i$ and $\delta_i$, is defined. this transformation is represented by a symmetric N x N matrix (where N (=2 or 3), is the number of dimensions) and maps the physical space onto a space in which elements, in the neighbourhood of the point being considered, will be approximately equilateral with unit average size. This new space will be referred to as the normalised space. for a general grid this transformation will be a function of position. The transformation T is the result of superimposing N scaling operations with factors $!/\delta_i$, in each $\alpha_i$ direction. Thus
\begin{equation}
    T(\alpha_i,\delta_i)= \sum^N_{i=1}\frac{1}{\delta_i} \alpha_i \bigotimes \alpha_i
\end{equation}
where $\bigotimes$ denotes the tensor product of two vectors.

\begin{figure}[H]
    \centering
    \includegraphics[width=11cm]{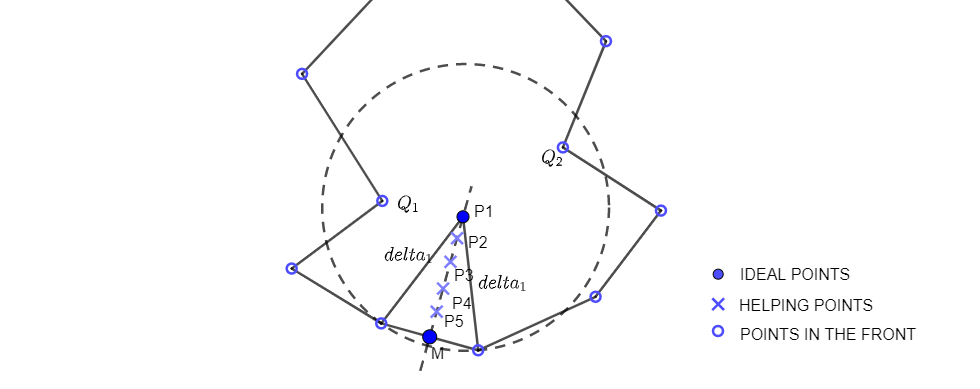}
    \caption{The generation of a new triangle.}
    \label{fig:generation triangle}
    \end{figure}

In the process of generation a new triangle the following steps are involved (see figure \ref{fig:generation triangle}):
\begin{itemize}
    \item Select a side AB of the front to be used  as a base for the triangle to be generated. Here, the criterion is to choose the shortest side. this is especially advantageous when generating irregular grids.
    
    \item Interpolate from the background grid the transformation t at the centre of the side \underline{M} and apply it to the nodes in the front which are relevant to the triangulation. In this implementation Peraire defines the relevant points to be all those which lie inside the circle of centre \underline{M} and radius three times the length of the side being considered. Let \underline{Â}, \underline{\^B}, \underline{\^M} denote the positions in the normalized space of the points, \underline{A}, \underline{B} and \underline{M} respectively.
    
    \item Determine, in the normalised space, the ideal position of $\underline{\hat{P}_1}$ for the vertex of the triangular element. The point $\underline{\hat{P}_1}$is located on the line perpendicular to the side that passes through the point \underline{\^M} and at the distance $\delta_1$ from the points \underline{Â} and \underline{\^B}. the direction in which $\underline{\hat{P}_1}$ is generated is determined by the orientation of the side. The value $\delta_1$ is chosen according to the inequalities of the basic procedure (see above), where \underline{Â} and \underline{\^B} correspond here to A and B in this inequalities. only in situations where the side AB happens to have characteristics very different from those specified by the background grid will the value of $\delta_1$ be different from unity. However, the above inequalities must be taken into account to ensure geometrical comparability. this expression is purely empirical and different inequalities could be devised to serve the same purpose.
    
    \item Select other possible candidates for the vertex and order them in a list. Two types f points considered:
    \begin{itemize}
        \item all the nodes $\underline{\hat{Q}_1}$, $\underline{\hat{Q}_2}$,... in the current generation front which are, in the normalised space, interior to a circle with centre $\underline{\hat{P}_1}$ and radius r = $\delta_1$, and
        \item the set of points  $\underline{\hat{P}_1}$,..., $\underline{\hat{P}_5}$ generated along the height  $\underline{\hat{P}_1}$\underline{\^M}.
        
    \end{itemize}
    
    For each point $\underline{\hat{Q}_i}$, construct the circle with centre $\underline{\hat{Q}_i}$, on the line defined by points $\underline{\hat{P}_1}$ and \underline{\^M} and which passes through the points $\underline{\hat{Q}_i}$,  \underline{Â} and \underline{\^B}. The positions of the centres $\underline{\hat{Q}_i}$, of these circles on the line  $\underline{\hat{P}_1}$\underline{\^M} defines an ordering of the $\underline{\hat{Q}_i}$ points. A list is created which contains all the $\underline{\hat{Q}_i}$ points with the further point from $\underline{\hat{P}_1}$ appearing at the head of the list. The points $\underline{\hat{P}_1}$,..., $\underline{\hat{P}_5}$ are added at the end of the list. 
    
    \item Select the best connecting point. This is the first point in the ordered list which gives a consistent triangle.. consistency is guaranteed by ensuring that none of the newly created sides intersects with any of the existing sides in the front.
    \item Finally, if a new node is created, its coordinates in the physical space are obtained by using the inverse tranformation $T^{-1}$.
    \item Sore the new triangle and update the front by adding/removing the relevant sides.
    
    \end{itemize}
    
    Löner (1992) \cite{28} proposes in his algorithm version of the advancing front method some very useful enhancements in three dimensions that also can also be implemented in two dimensions.
    
    \emph{Checking the intersection of faces}
    
    The most important ingredient of the advancing from generator is a reliable and fast algorithm for checking whether two faces intersect each other. Experience from practical applications indicates that even slight changes in this portion of the generator greatly influence the final mesh. As with the so many other problems in computational geometry, checking whether two faces intersect each other seems trivial for the eye, but is complicated to code. The checking algorithm is based on the following observation: two triangular faces do not intersect if no side of either face intersects the other face. The idea is then to build all possible side-face combinations between any two faces and check them in turn. If no intersection is found, then the faces do not cross. For each two faces, six side-face combinations are possible. Considering that on average about 40 close faces need to be checked, this way of checking the crossing of faces is very CPU-intensive. When it was first implemented, this portion of the grid generation code took more than 80\% of the CPU time required. In order to reduce the work load, a three-layered approach was subsequently adopted:
    \begin{itemize}
        \item \emph{Min/Max-search:} The idea here is to disregard all face-face combinations where the distance between faces exceeds some prescribed minimum distance. this can be accomplished by checking the maximum and minimum value for the coordinates of each face.
        \item \emph{Local element coordinates:} The purpose of checking for face-crossings is to determine whether the newly formed tetrahedron breaks already giver faces. The idea is to extend the previous Min/Max-criterion with the shape functions of the new tetrahedron.
        \item In-depth analysis of side-face combinations: All the faces remaining after the filtering process of steps a) and b) are analysed using side-face combinations as explained above.
    \end{itemize}
    Each of these three filters requires about an order of magnitude more CPU-time than the preceding one. When implemented in this way, the face-crossing check requires only 25\% of the total grid generation time.

    \emph{Data structures to minimise search overheads}
    
    The operations that could potentially reduce the efficiency of the algorithm to O($N^{1.5}$) or even O($N^2$) are:
    \begin{enumerate}[label=\alph*)]
        \item finding the next face to be deleted
        \item Finding the closest given points to a new points
        \item Finding the faces adjacent to a given point
        \item Finding for any given location the values of generation parameters from the background grid. This is an interpolation problem on unstructured grids.
    \end{enumerate}
    
    The verb 'find' appears in all of these operations. The main task is to design the best data structures for performing the search operations a)-d) as efficiently as possible. These data structures are typically binary trees of more complex trees. They were developed in the 1960s for Computer Science applications. Many variations are possible \cite{29}. As with flow solvers, there does not seem to be a clearly defined optimal data structure that all current grid generators use. For each of the data structures currently employed, one can find pathological cases where the performance of the tree-search degrades considerably. The data structures that have been used are:
    \begin{itemize}
        \item Heam-list (\cite{29}, \cite{30}, \cite{31}, \cite{32}), to find the next face to be deleted from the front.
        \item Quad-tree (2-D) and Octrees(3-D) (\cite{29}, \cite{30}, \cite{31}), to locate points that are close to any given location.
        \item N-trees, to determine which faces are adjacent to a points.
    \end{itemize}
    
    Combining these data structures, one can also derive an optimal interpolation algorithm for unstructured grids \cite{34}.

    \emph{Additional techniques to increase speed}
    
    There are some additional techniques that can be used to improve the performance of the advancing front grid generator. The most important of these are:
    \begin{enumerate}[label=\alph*]
        \item  \emph{Filtering:} Typically, the number of close points and faces is so far too conservative, i.e. large. As an example, consider the search for close points. there may be up to eight points inside an octant, but of these only on may be close to the face to be taken out. the idea is to filter out these 'distant' faces and point in order to avoid extra work afterwards. While the search operations are difficult to vectorize, these filtering operations lend themselves to vectorization in a straightforward way, leading to considerable overall reduction in CPU requirements.
        \item \emph{Automatic reduction of unused points:} As the front advances into the domain and more tetrahedrons are generated, the number of tree-levels increases. Thsi automatically implies an increase in CPU-time, as more steps are required to reach the lower levels of the trees. In order to reduce this CPU-increase as much as possible, all trees are automatically restructured. All point which are completely surrounded by tetrahedrons are eliminated from the trees. This procedure has proven to be extremely effective. It reduces the asymptotic complexity of the grid generator to less than O(N log N). In fact, in most practical cases one observes a linear O(N) asymptotic complexity, as CPU is traded between subroutine call overheads and less close faces on average for large problems.
        \item \emph{Global h-refinement:} While the basic advancing front algorithm is a scalar algorithm, h-refinement can be completely vectorized. Therefore, the grid generation process can be made considerably faster by fist generating a coarser, by stretched mesh and then  refining globally this first mesh with classic h-refinements (\cite{35}, \cite{36}). Typical speed-ups achieved by using this approach are 1:6 to 1:7.
    \end{enumerate}
    
    \paragraph*{Methode de front}
    Huet's (1990)\cite{88} "methode de front" is a especial sort of advancing front algorithm. In this approach the nodes are also inserted step by step, but the node and the triangle creation procedure follows a different criteria. This algorithm has been also implemented in three dimension.
    
\begin{figure}[H]
    \centering
    \includegraphics[width=6cm]{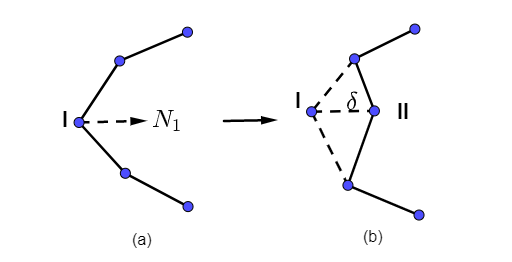}
    \caption{Methode de 'front'.}
    \label{fig:methode de front}
    \end{figure}

    For any node x of the oriented curve front C a normal direction Nx, which divides the angle built by the neighbouring edges of the node x, and a scalar value CVx, which measures the concavity, i.e. shape complexity, could be defined (see figure \ref{fig:methode de front} (a)). Starting from a node I of the front a new node II is created in the direction of Nx and with the node spacing dist. This new node II is connected to I and to the both nodes surrounding node I non the front( see figure \ref{fig:methode de front} (b)).
    
\begin{figure}[H]
    \centering
    \includegraphics[width=10cm]{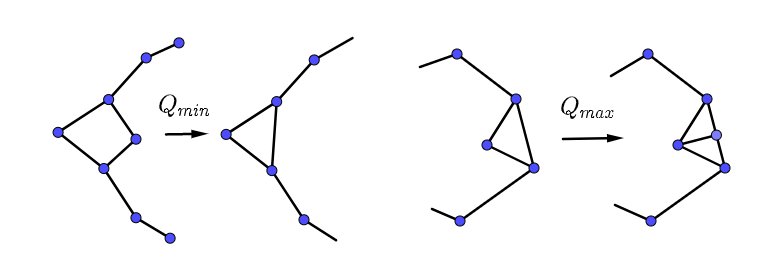}
    \caption{Maximum and minimum size for created elements.}
    \label{fig:maxmin size}
    \end{figure}

    The node II replaces I on the active front and re relative data are updated. First a control of the node creation is realized. A node spacing dist is chosen, that guarantees the creation of triangles as regular as possible. Two different coefficients Qmin and Qmax are introduced, which determine the minimum and maximum size for the elements of the triangulation that will be accepted ( see figure \ref{fig:maxmin size})
    
\paragraph*{Implementation Of Advancing Front Methods}
\begin{itemize}
    \item Löhner (1990), \cite{101}: Grid generation and adaptive refinement techniques suitable for the simulation of strongly unsteady flows past geometrically complex bodies in 3-D are described.
    \item Formaggia (1991), \cite{102}: Implementation of the Advancing front method in different three dimensional configurations.
    \item Cabello (1991), \cite{103}: A variation method for the optimization of  the directional stretched elements created by AFM.
    \item Siden 81990=, \cite{104}: Numerical simulation of the two-dimensional viscous compressible flow in blade cascades using a solution-adaptive unstructured mesh (AFM).
    \item Hwang (1990), \cite{105}: Locally implicit total-variation-diminishing schemes on unstructured triangular meshes (AFM).
    \item Chang (1993), \cite{106}: High- resolution temperature distribution of the transient heat conduction inside an arbitrary domain containing any number of anisotropic materials (AFM).
\end{itemize}

\subsubsection{STEINER ALGORITHMS}
\textbf{Definition:} A Steiner triangulation is any triangulation that adds additional sites to an existing triangulation to improve some measure of grid quality.

\begin{figure}[H]
    \centering
    \includegraphics[width=7cm]{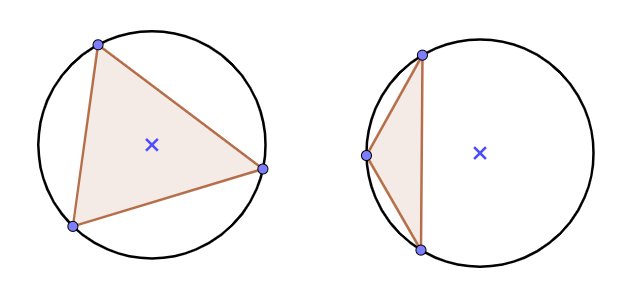}
    \caption{Inserting node at circumcentre of acute and obtuse triangles.}
    \label{fig:inserting node}
    \end{figure}

Technically speaking, the method of advancing front grid generation discussed in these notes would be a special type of Steiner triangulation. The insertion algorithms described earlier also provide a simple mechanism for generating Steiner triangulations. Holmes \cite{89} demonstrated the feasibility of inserting nodes at circumcenters of Delaunay triangles into an existing 2-D triangulation to improve measures of the grid quality. This has the desired effect of placing the new node in a position that guarantees that no other node in the triangulation can lie closer that the radius of the circumcircle, see figure \ref{fig:inserting node}. In a loose sense, the new node is placed as far away from other nearby nodes as conservatively possible.

Warren \cite{90} and Anderson \cite{91} further demonstrated the utility of this type of Steiner triangulation in the generation and adaptive refinement of 2-D meshes. the algorithm developed by Wiltberger \cite{92} also permit Steiner triangulations based on either Min Max (see later) or MaxMin (Delaunay) insertion. Only in the latter case is the insertion at triangle circumcenters truly justifiable.

\begin{figure}[H]
    \centering
    \includegraphics[width=7cm]{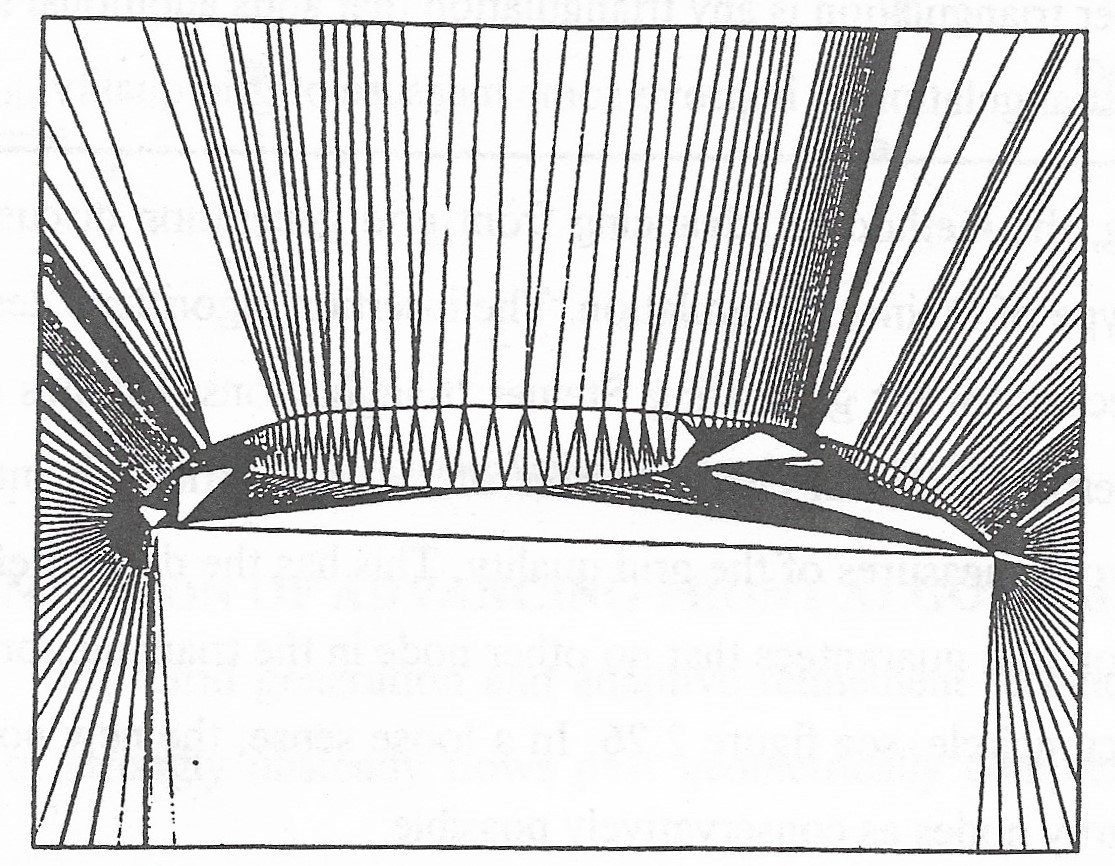}
    \caption{Initial triangulation of the boundary points.}
    \label{fig:initial triangu}
    \end{figure}

The 2-D Steiner point grid generation algorithm described in (\cite{89}, \cite{90}, \cite{91}) consists of the following steps. The first step is the Delaunay triangulation of the boundary data. Usually three or four points are placed in the far field with convex hull enclosing all the boundary points. starting with a triangulation of these points, nodes corresponding to boundary curves are incrementally inserted using Watson's algorithm in (\cite{89}, \cite{90}, \cite{91}) and Green and Sibson's algorithm in \cite{67} as shown in figure \ref{fig:initial triangu}. The initial triangulation does not guarantee that all boundary edges are members of the triangulation. This can be remedied in a variety of ways. One technique adds additional points ot the triangulation so as to guarantee that the resulting Delaunay triangulation contains all the desired boundary edges, see reference \cite{70}. Another approach performs local edge swapping so as to produce a constrained Delaunay triangulation which guarantees that all boundary edges are actual edges of the triangulation.

\begin{figure}[H]
    \centering
    \includegraphics[width=7cm]{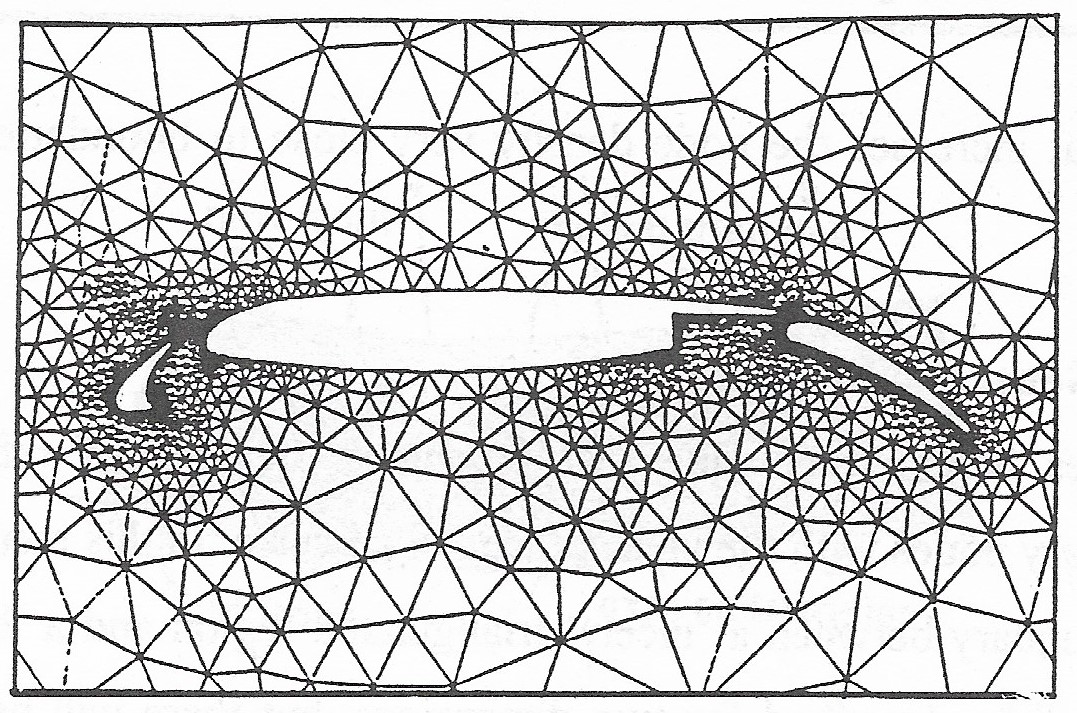}
    \caption{Steiner triangulation with nodes inserted at the circumcentres to reduce maximum cell aspect ratio.}
    \label{fig:steiner triang}
    \end{figure}

In either event, the boundary edges are marked so that they cannot be removed as the triangulation is refined. The algorithms described in (\cite{89}, \cite{90}, \cite{91}) interrogate triangles in an arbitrary order (this makes the triangulation non unique). The user must specify some measures of quality for triangle refinement (aspect ratio, area, containment circle radius, for example) and a threshold value for the measure . If a triangle fails to meet the threshold value, the triangulation is refined by placing a new node at the circumcenter of the failed triangle via Watson's algorithm. Some care must be taken to insure that measures are chosen which are guaranteed to be reduced when the refinement takes place. Using thresholds in this way does not give the user direct control over the actual number of triangle generated in the process of Steiner refinement. Wiltberger takes a different approach by maintaining a dynamic heap data structure of the quality measure. (Heap structures are very efficient way of keeping a sorted list of entries whit insertion and query time O(logN) of N entries. ) The triangle with the largest value of the specified measure will be located at the top of the heap at all times during the triangulation this make implementation of a Steiner triangulation which minimizes the maximum value of the measure very efficient (and unique). In this implementation, the user can either specify the number of triangles to be generated or a threshold value of the measure. Note that multiple measures can be refined lexicographically. Figure \ref{fig:steiner triang} shows a Steiner triangulation using the Wiltberger algorithm with MaxMin insertion and refinement based on maximum aspect ratio.

\paragraph*{Holmes Algorithm} \label{chap:steiner}

\begin{figure}[H]
    \centering
    \includegraphics[width=7cm]{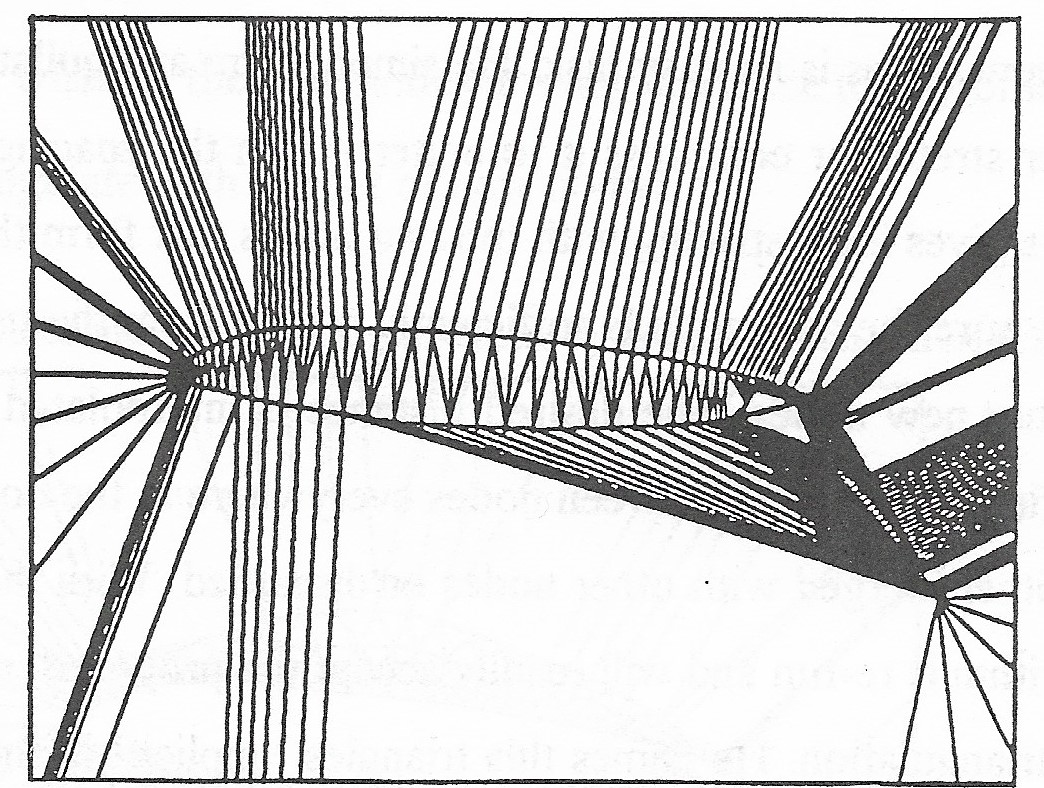}
    \caption{Triangulation of boundary nodes. The nodes on the boundary of a three element airfoil are connected to nodes on the outer boundary (not shown).}
    \label{fig:triangulation of boundary nodes}
    \end{figure}

In itself Delaunay triangulation does to provide an interior point cloud. A particular approach to overcome this deficiency has been outlined by Holmes \cite{6}. A Delaunay triangulation of the boundary nodes is taken as an initial grid. Figure \ref{fig:triangulation of boundary nodes} gives such a triangulation of a three element aerofoil configuration. The initial triangulation consists of larges, very skewed triangles that are found to exceed certain thresholds of maximum area of maximum skewness. Holmes proposes to measure skewness as the ratio of the radius of the circumscribed circle over twice the radius of the inscribed circle.

Once such a bad triangle is detected it will be refined by insertion of a new node at the circumcenter of that triangle. Refinement is performed on the largest triangle in the grid until all triangles are smaller than  a first threshold value. The refinement continues on the skewed triangles starting with the one having the largest circumcircle. The final grid is obtained after all skewed triangles are smaller that a second area threshold. Refining on skewness yields and implicit mechanism that increases the node density very close to boundaries with finer node spacing.

\paragraph*{Frontal Node Generation/ Müller}

Müller \cite{100} introduces here a technique that combines the idea of refining a Delaunay triangulation of boundary nodes with the ideas of frontal advance. In this Delaunay triangulation of boundary nodes with the ideas of frontal advance. In this method, the front will take the form of a boundary between a "nicely" triangulated region and a "badly" triangulate region. The method will be described for two dimensions, but there will be few obstacles to implementing it in three.

If one looks at a Delaunay triangulation of a set of boundary nodes (see figure \ref{fig:triangulation of boundary nodes}), one can observe that almost every boundary face is either the short face of a triangle with one very acute angle or else one of two short faces in a triangle with one very obtuse angle. In accordance with our idea of a front that divides nice triangles from bad triangles, we take the boundary to define the initial position of such a front. To begin with, we have no nice triangles, but we will introduce a layer of well positioned nodes that will allow the front to advance.

The construction of the new nodes is an easy task, We simply form an equilateral triangle with the frontal face and either stretch or compress it to better match the spacing requirements of the background mesh, this gives ideal spacing with the two nodes that form the face. A further check is required to make sure the new node is sufficiently distant from the remaining nodes in the grind and from the other new nodes. The desired distance is interpolate din a background mesh that uniquely specifies the distance between nodes everywhere in the domain. Nodes that exhibit bad spacing are either merged with other nodes or discarded. With these new nodes in place, the Delaunay algorithm is rerun and will readily accept the proposed, nice triangles as t presents skewness in its triangulation. We will name this triangles "implicit" in the following. Also nice triangles between the new node will be formed as the y are guaranteed to be spaced nicely as well, the "implicit" triangles. In the rest of the domain Delaunay still has to construct acute cells, though with slightly improved shape.

    \begin{figure}[H]
    \centering
    \includegraphics[width=9cm]{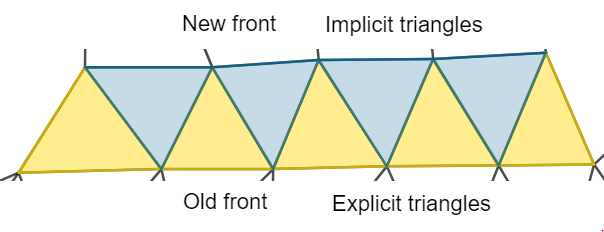}
    \caption{Explicit triangles (striped) and implicit triangles (squared) that are formed along the old front and build the new front.}
    \label{fig:new front built}
    \end{figure}

Again the short faces of these acute cells denote a frontier between the region with nice cells and the region still waiting to be refined, see figure \ref{fig:new front built}, and the process can be repeated until all bad triangles have vanished. Hence, the algorithm can be cast into the following steps:

\begin{enumerate}
    \item Detect all bad triangles in the grid and find their short faces.
    \item Find a set of nodes to form nice triangles with the short faces.
    \item Check whether the new nodes are not too close to any other node already introduced into the structure.
    \item Check whether the new nodes are not too close to any other new node.
    \item Re-triangulate with the set of new nodes.
\end{enumerate}

The steps will be repeated until no more improvement by node insertion can be achieved.

    \begin{figure}[H]
    \centering
    \includegraphics[width=8cm]{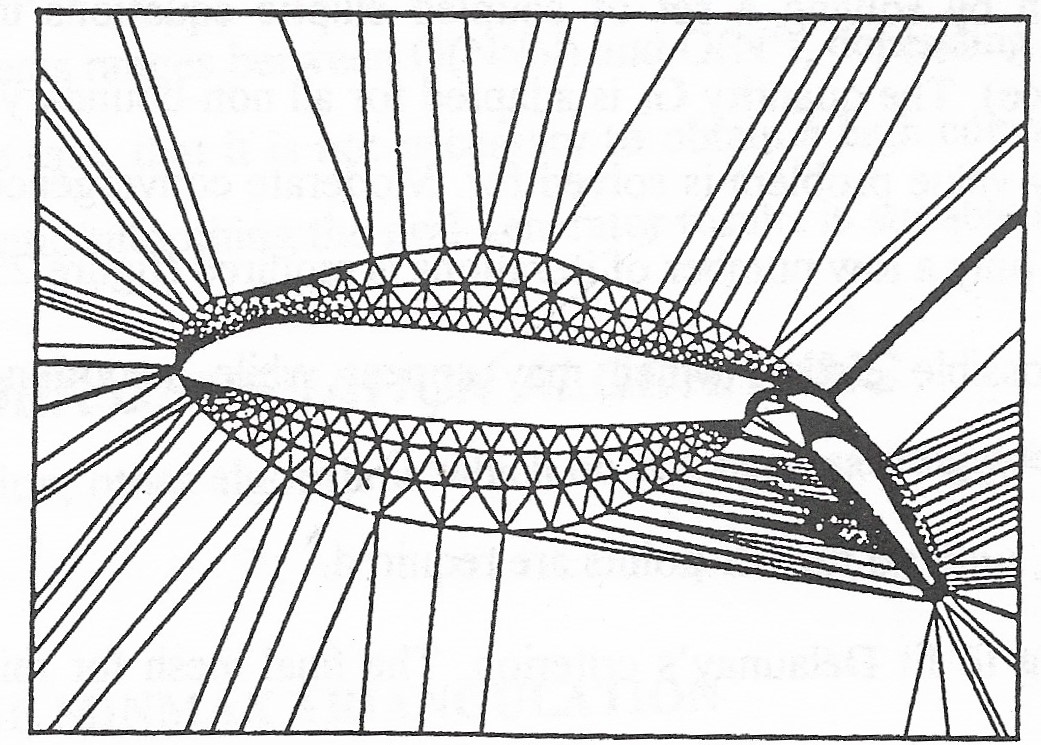}
    \caption{Grid around three element airfoil after three rows of nodes have been inserted.}
    \label{fig:grid around}
    \end{figure}

\begin{figure}[H]
    \centering
    \includegraphics[width=9cm]{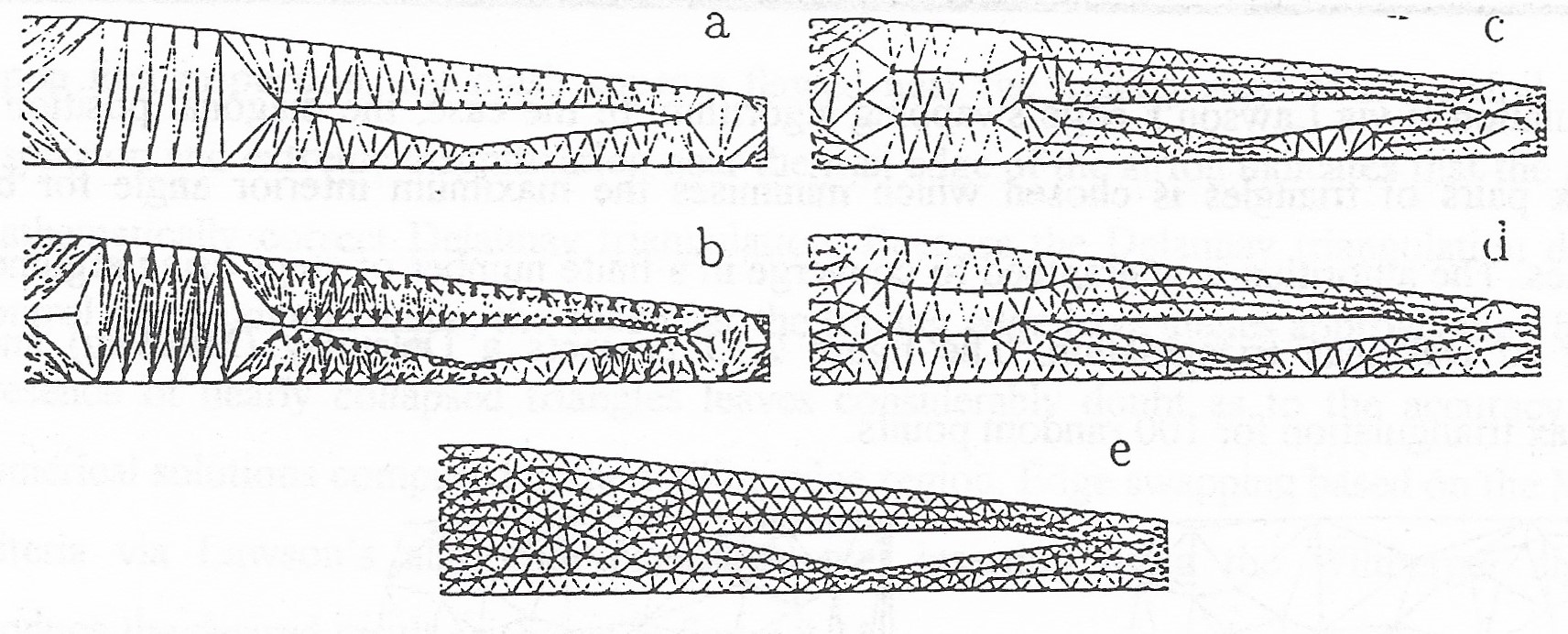}
    \caption{Development of a mesh for a scram get inlet.}
    \label{fig:scram get}
    \end{figure}

\paragraph*{Vilsmeier/Hänel Algorithm}
This algorithm is an extension of the basic Holmes algorithm. The generation starts on a coarse mesh, containing only boundary points. the final mesh is obtained recursively by refining, remeshing and smoothing steps \cite{59}. In detail it is:
\begin{enumerate}
    \item The given points on the boundaries are triangulated using the Delaunay algorithm (see figure \ref{fig:scram get} a).
    \item Store the length of the longer of both boundary edges for each boundary point and multiply it with a statistic factor greater than 1. This quantity will be called the limiting edge length Gk. It will never be changed for the boundary points.
    
    \item 
    Insert a new point in the center of all triangles with at least one edge longer than the minimum value of Gk for the two points forming the edge. Predict Gk for the new points as the average o the values Gk form the three points forming hte triangle which is refined.
    \item Connect every new point by forming three new triangles and by deleting the old one which has been reined. Adapt all neighbouring relation (figure \ref{fig:scram get} b).
    
    \item Apply remeshing (edge swappping). The aim of the remeshing applied in this step is not to satisfy the Delaunays's criterion but to obtain a triangle with six triangles surrounding one point. while no extreme angles should appear. It is obvious, that the desire aim is not achievable for all the points ( figure \ref{fig:scram get} c).
    
    \item Smooth the mesh by solving a set of couple elliptic equations in a recursive algorithm. The quantity Gk is adapted fo all non-boundary points, what means that a boundary value problem is solve for. Moderate convergence is sufficient in this step. Therefore only a low number of iterations is required (figure \ref{fig:scram get} d).
    \item Correction of possible error, which may appear while smoothing in step f), especially in concave polygons.
    \item Continue at C; until no further points are requires.
    \item Perform remeshing to fit Delaunay's criterion. The final mesh for this example is shown in figure \ref{fig:scram get} e.

\end{enumerate}

\emph{Computational time}

The computational time depends not only on the final grid number of points, but also on the geometry. If corrections (step 7) have to be performed the computational time rises dramatically.

The time spent to triangulate the boundary points (step 1) is negligible. The time for introducing new points (step 3) is of the order O(N) for the whole triangulation, as well as the connection of the newly introduced point (step 4). The CPU time for the remeshing tool has shown to be difficult to predict. The number of loops is usually very small, but affects a large number of triangle in each loop. We will say here that the order of the computational time for remeshing is still O(1) for each triangle and therefore O(N) for the whole triangulation.

The most expensive tool is the elliptic smoothing (step 6). The behaviour solving boundary value problems ranges between O($N\sqrt{N}$) and O($N^2$), depending on the acceleration techniques. It should be said, that it is not necessary to obtain a high convergence while smoothing. Even without elliptic smoothing the grid generator results in suitable meshes for CFD applications.

\subsubsection{OTHER GENERATION ALGORITHMS.}
In this section, other algorithms which do not necessarily produce Delaunay triangulations are explored.

\begin{figure}[H]
    \centering
    \includegraphics[width=11cm]{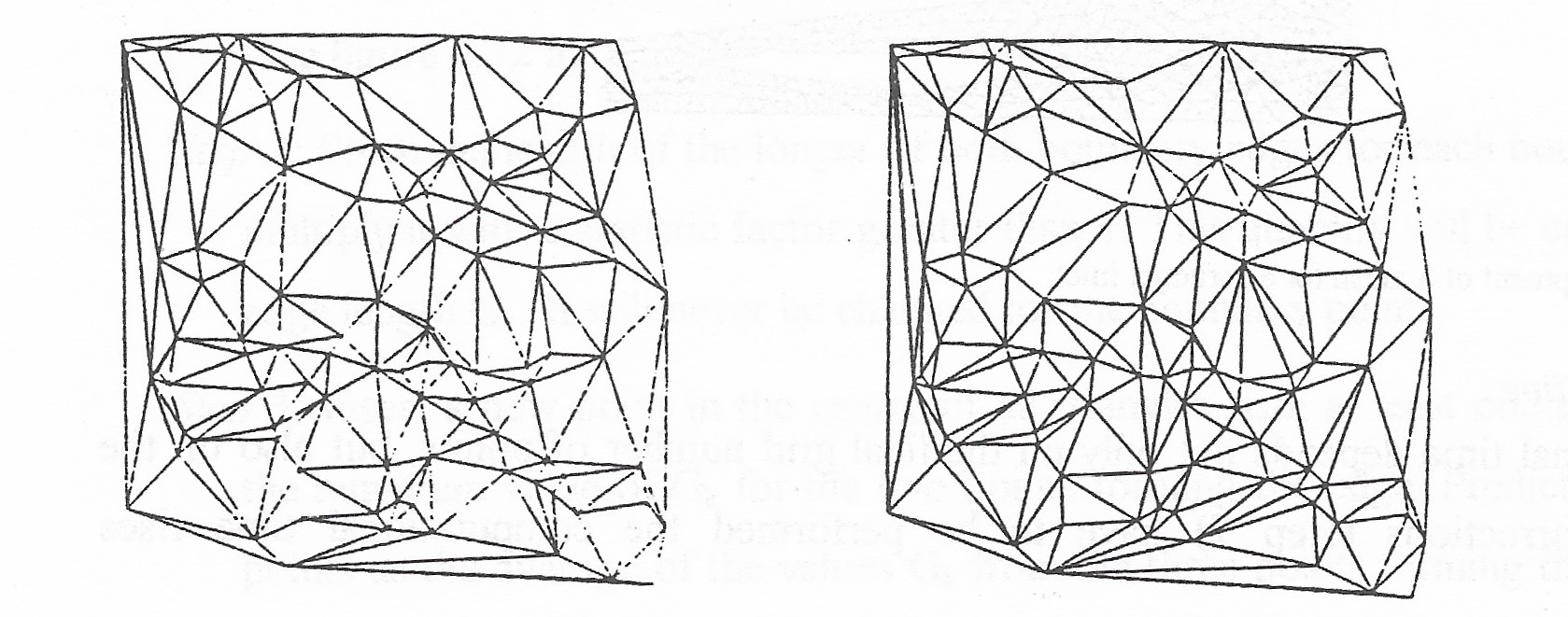}
    \caption{Delaunay triangulation and MinMax triangulation.}
    \label{fig:examples_}
    \end{figure}

\paragraph*{The MinMax Triangulation}
As Babuska and Aziz \cite{94} point out, from the point of view of finite elements the MaxMin (Delaunay
 triangulation is not essential. What is essential is that no angle can be close to 180\degree. In other words, triangulations which minimize the maximum angle are more desirable. These triangulations are referred to as MinMax triangulations. One way to generate a 2-D MinMax triangulation is via Lawson's edge swapping algorithm. In this case, the diagonal position for convex pairs of triangles is chosen which minimises the maximum interior angle for both triangles. The algorithm is guaranteed to converge in a finite number of steps using arguments similar to Delaunay triangulation. The figure \ref{fig:examples_} presents a Delaunay (MaxMin) and an MinMax triangulation for 100 random points.
 
\begin{figure}[H]
    \centering
    \includegraphics[width=9cm]{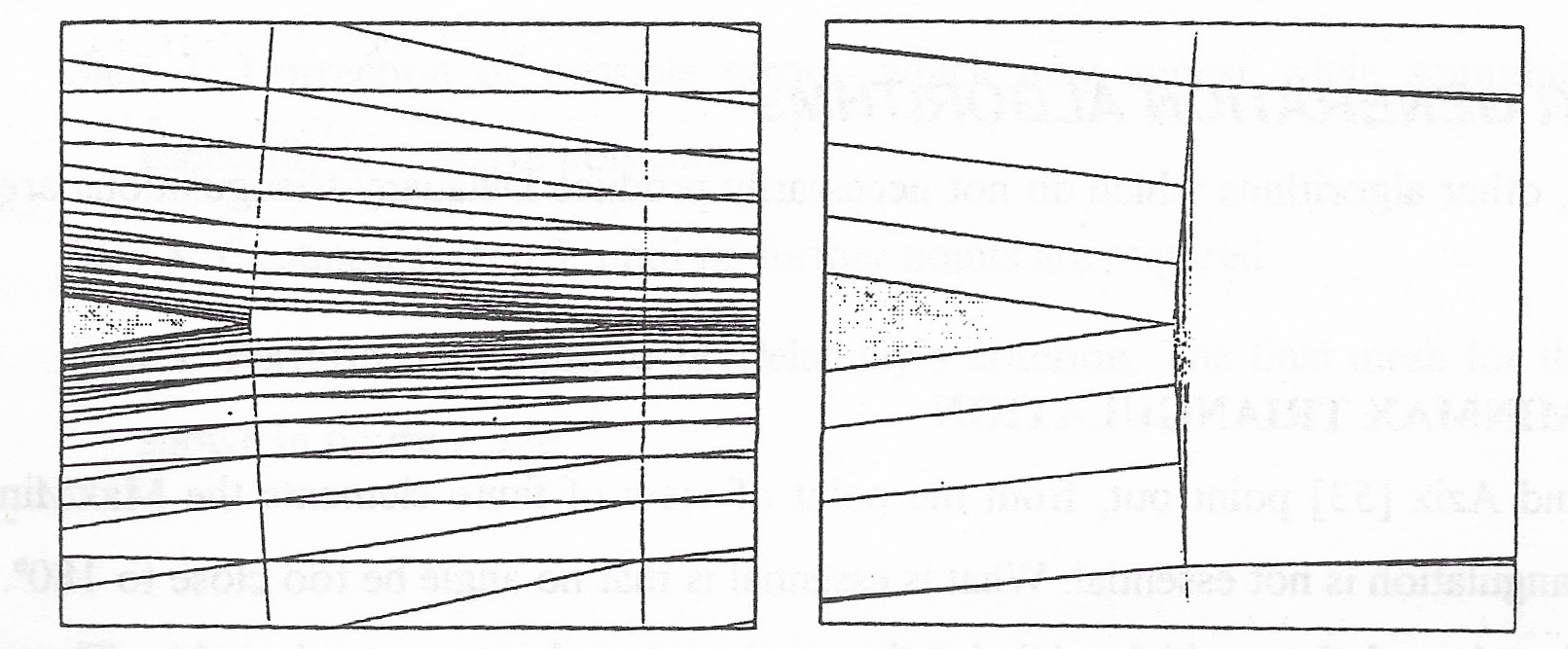}
    \caption{Delaunay triangulation near trailing edge of airfoil.}
    \label{fig:del}
    \end{figure}

 Note that application of local MinMax optimization via Lawson`s algorithm may only result in a mesh which is locally optimal and not necessarily at a global minimum. Attaining a globally optimal MinMax triangulation is a much more difficult task. The bet algorithm to present date (Edelsbrunner, Tan and Waupotitsch  \cite{95}) has a high complexity of O($N^2log N$). Wiltberger \cite{92} has implemented a version of the Green and Sibson algorithm \cite{67} which has been modified to produce locally optimal MinMax triangulations using incremental insertion on local edge swapping. The algorithm is implemented using recursive programming with complete forward and backward propagation. This is a necessary step to produce locally optimized meshes. The MinMax triangulation ha proven to be very useful in CFD. Figure \ref{fig:del} shows the Delaunay triangulation near the trailing edge region of an airfoil with extreme point clustering.
 
\begin{figure}[H]
    \centering
    \includegraphics[width=6cm]{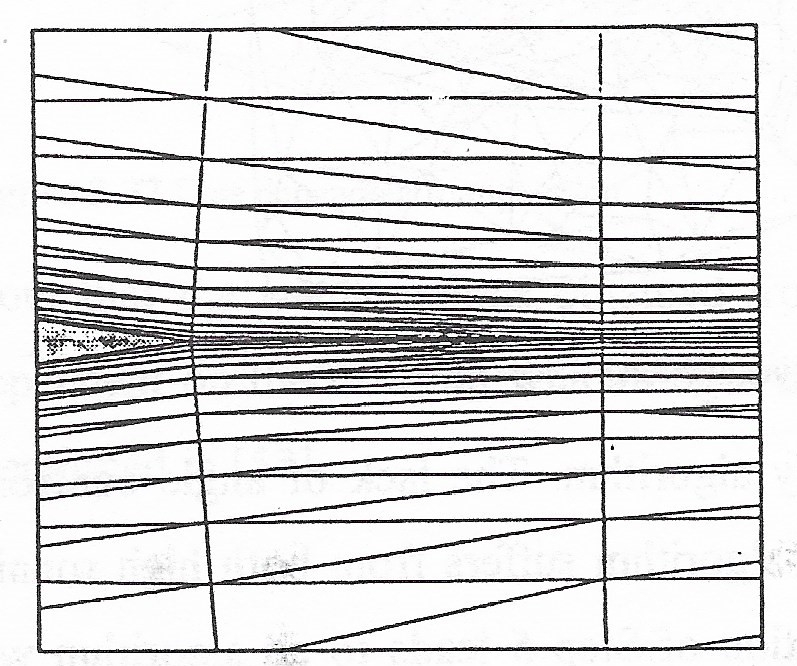}
    \caption{MinMax triangulation near trailing edge of airfoil.}
    \label{fig:minmax_}
    \end{figure}

Upon first inspection, the mesh appears flawed near the trailing edge of the airfoil. Further inspection and extreme magnification near the trail edge of the airfoil indicates that the grid is a mathematically correct Delaunay triangulation does not control the maximum angles, the cells near the trailing edge have angles approaching 180\degree. The presence of nearly collapsed triangles leaves considerably doubt as to the accuracy of any numerical solutions computed in the trailing edges region. Edge swapping based on the MinMax criteria via Lawson's algorithm produce the desired result as shown in figure \ref{fig:minmax_}.

\paragraph*{Greedy Triangulations}

A greedy method is one that never undoes what it did earlier. The greedy triangulation continually adds edges compatible with the current triangulation (edge crossing no allowed) until the triangulation is complete, i.e. Euler's formula is satisfied. One objective of a triangulation might be to choose a set of edges with shortest total length. The best that the greedy algorithm can do is adopt a local criterion whereby only the shortest edge available at that moment is considered for addition to the current triangulation. (this does not lead to a triangulation with shortest total length). Note that greedy triangulation easily accommodates constrained triangulations containing interior boundaries an a non convex outer boundary. In this case the boundary edges are simply listed first in the ordering of candidate edges. The entire algorithm is outlined below.

\textbf{Algorithm:} Greedy Triangulation
\begin{itemize}
    \item  Initialize triangulation T as empty.
    \item Compute $n_2$ candidate edges.
    \item Order pool of condidate edges from ordered pool.
    \item Remove current edge $e_S$ from ordered pool
    \item If( $e_S$ does not intersect edges of T) add $e_S$ to T.
    \item If(Euler's formula not satisfied) go to Step 4.
\end{itemize}

    \begin{figure}[H]
    \centering
    \includegraphics[width=12cm]{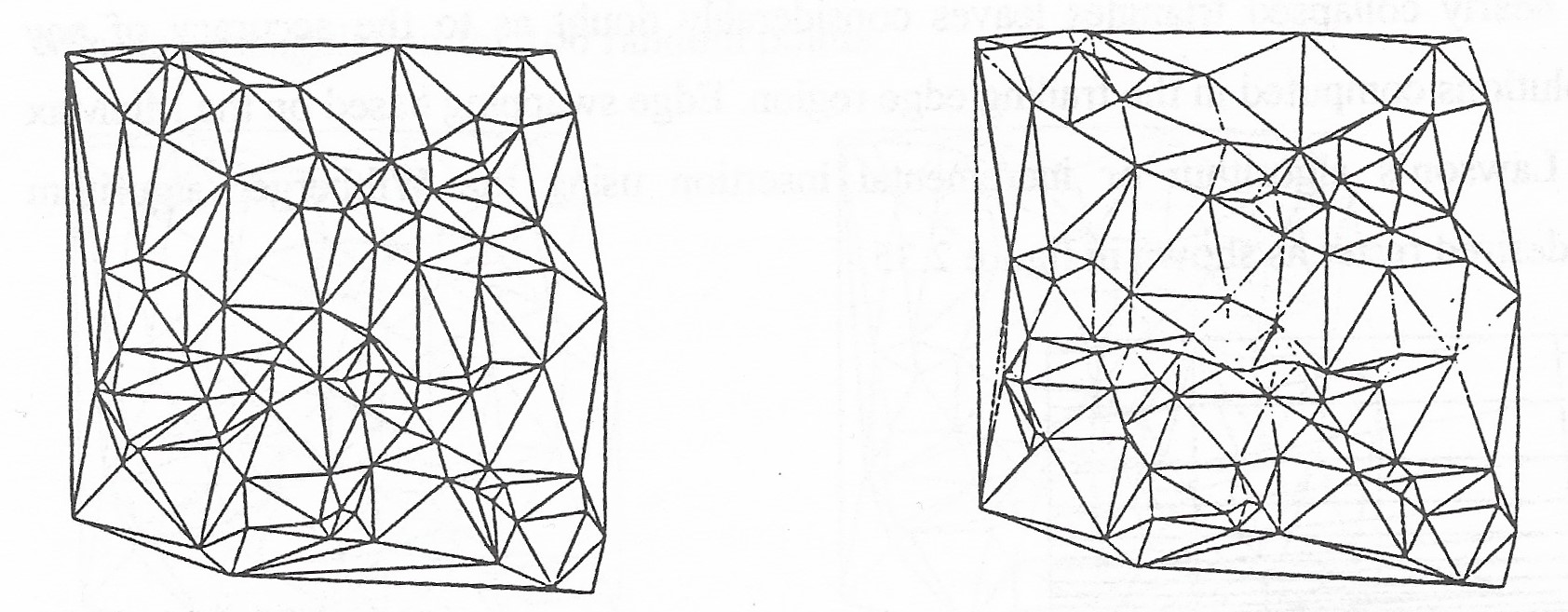}
    \caption{Delaunay triangulation and Greedy triangulation.}
    \label{fig:greedy}
    \end{figure}

Figure \ref{fig:greedy} contrasts the Delaunay and the Greedy algorithm. The lack of angle control is easily seen in the greedy triangulation. The greedy algorithm suffers from both high running time as well as storage. In fact a naive implementation of Step 5 leas to an algorithm with O($N^3$) complexity. Efficient implementation techniques are given in Gilbert \cite{96} with the result that the complexity can b reduced to O($N^2 $log N) with O($N^2$) storage.

\paragraph*{Data Dependent Triangulation}

Unlike mesh adaptation, a data dependent triangulation assumes that the number and position of vertices is fixed and unchanging. Of all possible triangulation of these vertices, the goal is to find the best triangulation under data dependent constraints. In Nira, Levin and Rippa \cite{97}, they consider several data dependent constraints together with piece-wise linear interpolation. In order to determine if a new mesh is "better" than a previous one, a local cost function is defined for each interior edge. Two choices which prove to be particular effective are the JND (Jump in Normal Derivatives) and the ABN (Angle Between Normals).

\begin{figure}[H]
    \centering
    \includegraphics[width=5cm]{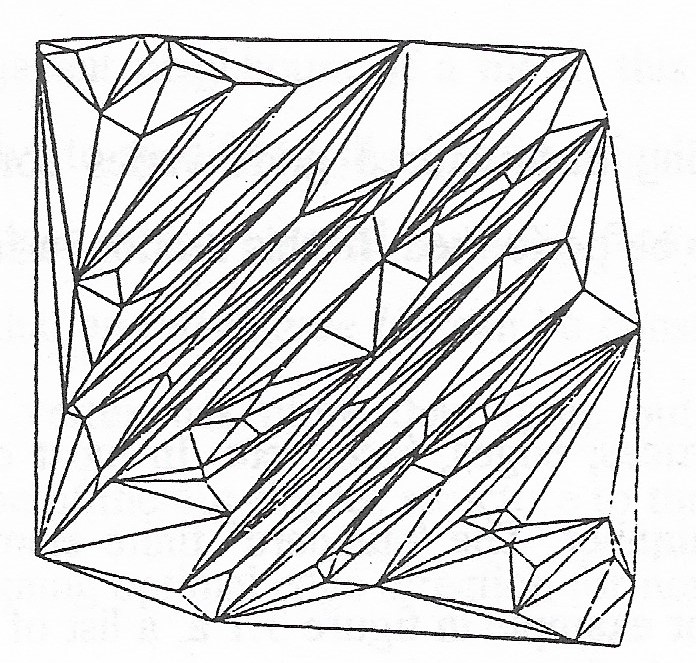}
    \caption{Data dependent triangulation.}
    \label{fig:data dependent tri}
    \end{figure}

Recall that a Delaunay triangulation would result if the cons function is chosen which maximizes the minimum angle between adjacent triangles (Lawson's algorithm): Although it would be desirable to obtain a global optimum for all cost function, this could be very costly in many cases. An alternate strategy is to abandon the pursuit of a globally optimal triangulation in favor of a locally optimal triangulation. Once again Lawson's algorithm is used. Note that in using Lawson's algorithm, when require that the global measure decrease at each edge swap. This is not as simple as before since each edge cost function. Nevertheless, this domain of influence is very small and easily found. In figure \ref{fig:data dependent tri} the data dependent triangulation is plotted.

Note that the triangulations obtained from this method are not globally optimal and highly dependent on the order in which edges are accessed. Several possible ordering strategies are mentioned in \cite{98}.

\subsubsection{DATA STRUCTURES}
\begin{figure}[H]
    \centering
    \includegraphics[width=12cm]{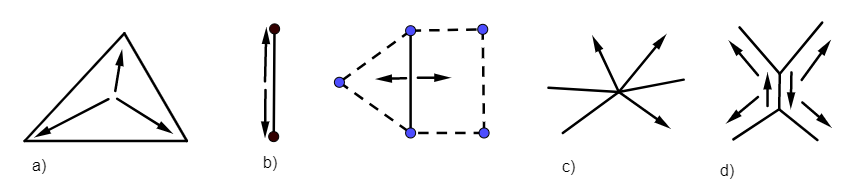}
    \caption{Data structure: a) FE data structure b) Edge structure c) Out-degree structure d) Quad-edge structure.}
    \label{fig:data structure}
    \end{figure}

The choice of data structures used in representing unstructured grids varies considerably depending on the type of algorithmic operations to be performed. In this section, a few of the most common data structures will be discussed. The mesh is assumed to have a numbering of vertices, edges, faces, etc. In most cases, the physical coordinates are simply listed by vertex number. The "standard" finite element (FE) data structure lists connectivity of each element. For example in figure \ref{fig:data structure}, a list of the three vertices of each triangle would be given.

The Fe structure extends naturally to three dimensions. The FE structure is used extensive in finite element solvers for solids and fluids.

For planar meshes, another typical structure is the edge structure (figure \ref{fig:data structure}) b which lists connectivity of vertices and adjacent faces by storing a quadruple for each edge consisting of the origin and destination of each edge as well as the two faces (cells) that share that edge. This structure allows easy traversal through a mesh which can be useful in certain grid generation algorithms. (This traversal is not easily done using FE structure.) The extension to three dimensions is face wise (vertices of a face are given as well as the two neighboring volumes( and requires distinction between different face types.

A third data structure provide connectivity via vertex lists as shown in figure \ref{fig:data structure} c. The brute force approach is list to all adjacent neighbours for each vertex(usually as a linked list). Many sparse matrix solver packages specify non zeros of a matrix using row of column storage schemes which list all non zero entries of a given row or column.For discretizations involving only adjacent neighbours of a mesh, this would be identical to specifying a vertex list. An alternative to specifying all adjacent neighbours is to direct edges of the mesh. In this case only those edges which pointed outward from a vertex are listed. In the next section, it will be shown that an out-degree list can e constructed for planar meshes by directing a graph such that no vertex has more than three outgoing edges. This is asymptotically optimal. The extension of the out-degree structure to three dimensions is not straightforward and algorithms for obtaining optimal edge direction for non planar graphs are still under development.

The last structure considered here is the quad edge structure proposed by Guibas and Stolfi \cite{66}, see figure ===. Each edge is stored as pair of directed edges. Each of the directed edges stores its origin and pointers to the next and previous directed edge of the region to its left. The quad tree structure is extremely useful in grid generation where distinctions between topological and geometrical aspects are crucial. The structure has been extended to three dimensional arrangements by Dobkin and Laslo \cite{108} and Brisson \cite{109}.

From the previous sections it is apparent that a successful implementation of the generation algorithm will require the use of data structures which enable certain sorting and searching operations to be performed efficiently. For instance the generation front of the AFM will require a data structure which allows for the efficient insertion/deletion of the sides and which also allows for the efficient identification of the sides which intersects with a prescribed region in space.

The problem of determining the members of a set of N nodes which lies inside a prescribed sub-region of an n dimensional space is known as geometrical searching. Several algorithms have been proposed (\cite{110}, \cite{111},\cite{112}, \cite{113}) which solve the above of equivalent problems with a computational expense proportional to log(N). The problem complexity increases considerably when instead of considering points, one dials with finite size objects such line segments, triangles or tetrahedral in three dimensions. A common problems encountered here, name geometric intersection, consists of finding the objects which overlap a certain sub-region of the space being considered. Algorithms for solving this problem in two dimensions exist \cite{81} and have been applied in determining the intersection between geometrical objects in the plane. To our knowledge, the only algorithm capable of solving this problem in three dimensions is based on the use o alternate digital tree (ADT) \cite{107}.

\subsubsection{GRID ADAPTATION}
Numerical methods used for the analysis of problems involving compressible flow must be capable of providing adequate definition of the narrow regions of high gradients (e.g. shocks) which frequently occur and which are normally found to be embedded in large areas in which the flow variables vary slowly. As the location of these high gradients regions is not known to the analyst a priory, it is apparent that adaptive mesh methods, with a posterior error estimators, will have an important role to play in the development of efficient solution techniques for such problems. to date, successful grid adaptation techniques for such problems. To enrichment of mesh movement. This techniques have generally been implemented in conjunction with explicit time integration procedures and have been designed for analysis of steady state problems.

\paragraph*{Mesh Movement}
Mesh movement can be accomplished b advancing the solution towards steady state and, at certain times, replacing the mesh sides by springs of a certain stiffness. The nodes are then moved until the spring system is in equilibrium. the spring strengths are normally based on the local solution gradient
.


In two dimensions such techniques have been used with finite volume methods on quadrilaterals \cite{45} and finite elements methods on triangles \cite{9} . Since new nodes are not added and nodal connectivities are not changed, a drawback of this method is that the accuracy of the final computation is limited by the structure and resolution of the initial grid.

\paragraph*{Mesh Enrichment}

\begin{figure}[H]
    \centering
    \includegraphics[width=6cm]{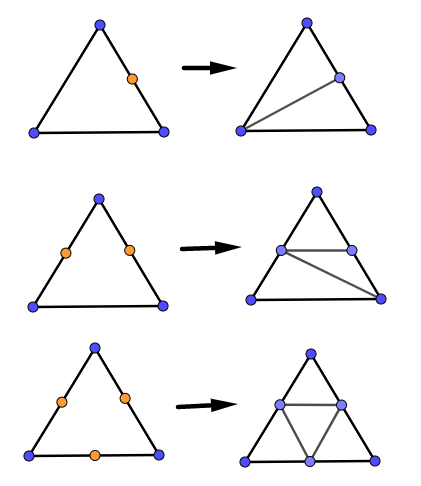}
    \caption{The grid enrichment process showing three possible refinement cases for a triangular element}
    \label{fig:enrichment}
    \end{figure}

Mesh enrichment algorithms have been used in finite element methods with triangles and tetrahedral (\cite{71}, \cite{77}) and finite volume methods with quadrilaterals (\cite{93}, \cite{99}). The basic approach is to advance the solution toward steady state on an initial (coarse= grid and the to use an error or refinement indicator to mark the computational cells which should be refined. the marked cells are automatically subdivided, the computation proceeds and the process is repeated until the analyst is satisfied with the solution quality (see figure \ref{fig:enrichment}).

Although this method has proved to be effective in practice, it also suffers from certain drawbacks, e.g., in the solution of two-dimensional problems the areas in the vicinity of one-dimensional flow features, such as shocks and boundary layers , are not refined in a efficient manner and the number of elements employed increases rapidly with each refinement (see figure ==).

    \begin{figure}[H]
    \centering
    \includegraphics[width=12cm]{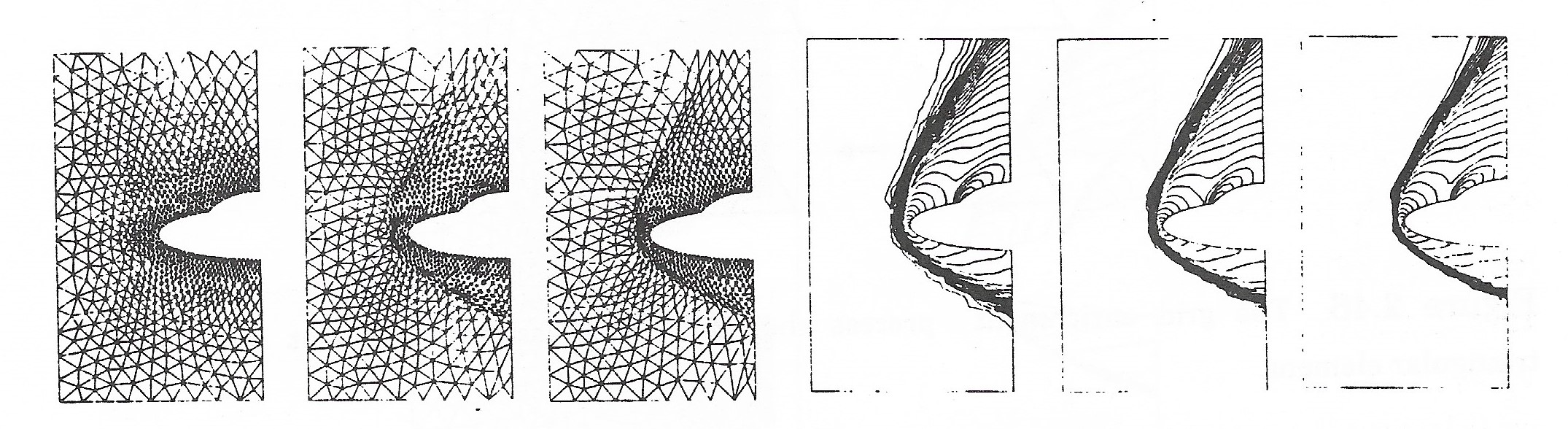}
    \caption{Supersonic flow past a double ellipse at a free stream Mach number of 8.15 and at an angle of attack of thirty degrees showing a sequence of grids and solution obtained following the use of adaptive grid movement.}
    \label{fig:supersonic}
    \end{figure}

\begin{figure}[H]
    \centering
    \includegraphics[width=6cm]{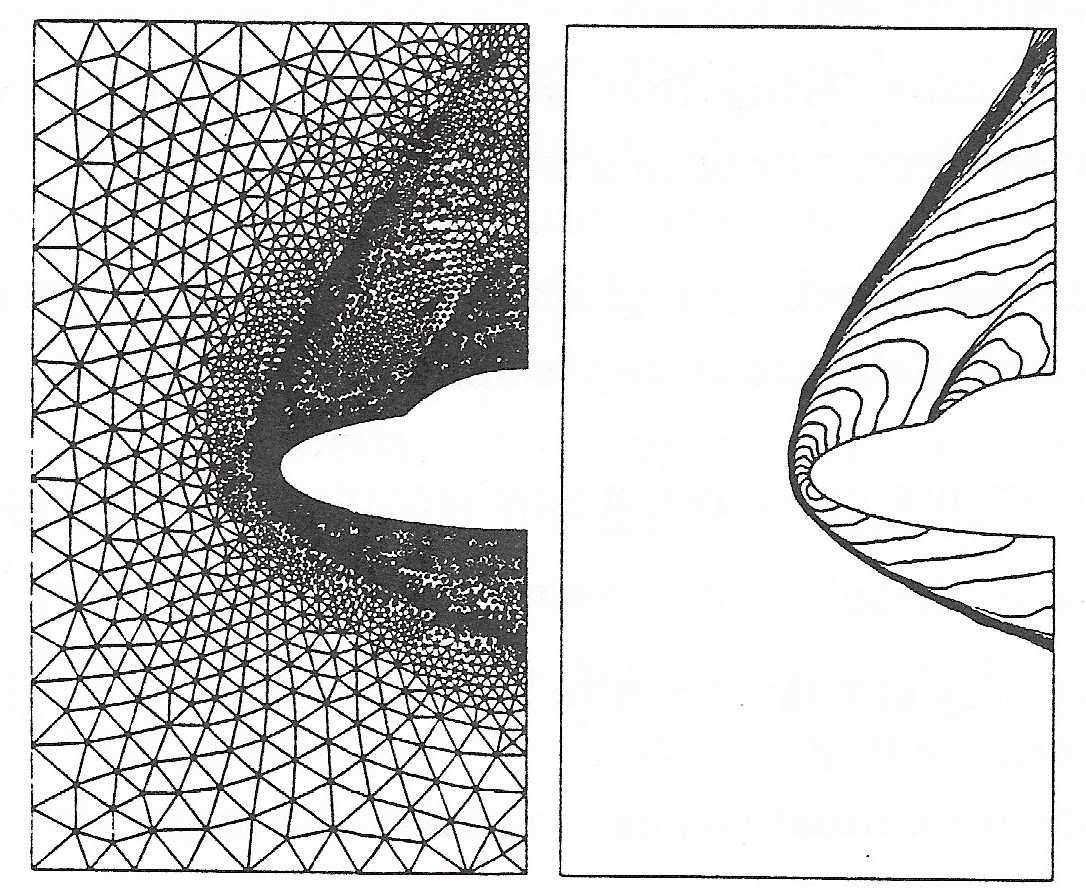}
    \caption{Supersonic flow past a double ellipse at a free stream Mach number of 8.15 and at an angle of attack of thirty degrees showing a sequence of grids and solution obtained following the application of adaptive grid movement to a previously enriched grid.}
    \label{fig:supersonic flow 2}
    \end{figure}

As possible remedy to this problem is to combine grid enrichment and grid movement procedures. This is demonstrated in figure \ref{fig:supersonic flow 2} which shows the application of the movement procedure to the final enriched grid of figure \ref{fig:supersonic}.

%% file: Meshgen2_0.tex
\section{MeshGen 2.0}
Meshgen 2.0. is a Windows app created to create and plot a mesh in the domain designed by the user. It has been programmed in Qt Creator, using C++. This Objected Oriented Language allows to optimize the use of the memory while running the app, however it makes harder to explain the structure of the program. In order to do this, the first thing explained will be how the structure behind the user interface behaves.
\section{Meshgen Structure}
The App project is structured in Classes. This classes are the following ones:
\begin{itemize}
    \item \textbf{MainWindow class}. This is the principal class, where all the interface buttons and interactions are defined. This class calls all the others and all the operations are done here.
    \item\textbf{TextEditReader class}. This class reads the Boundary segments written in the Text Edit on the right of the interface and can pass this information to other classes
    \item \textbf{LienzoCreator class}.This is where all Drawings are done in the MiddleWidget of the interface.
    \item \textbf{Dialog class}. Is the ui(user interface) class that creates the Dialog that pops up when the user wants to change the Boundary and Mesh width settings.
    \item \textbf{Trimpack class} is the class that is used as "bridge" between fortran code and C++ code.
    \item \textbf{TRIMPACK.} The rest of files (.f or .lib) the Triangular unstructured Mesh generation Package given by Juan Manuel Tizón.
\end{itemize}

\begin{figure}[H]
    \centering
    \includegraphics[width=12cm]{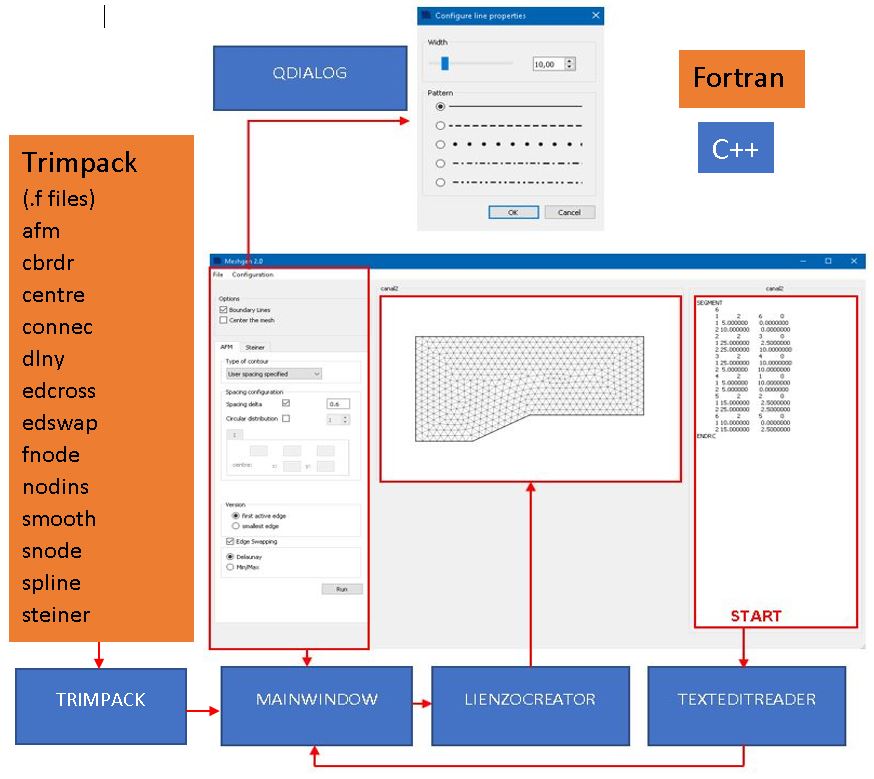}
    \caption{Information Flux}
    \label{fig:InfoFlux}
\end{figure}

The information flux works like in the Figure \ref{fig:InfoFlux}. When the user opens a .mg file (where the domain shape is defined), it appears on the right white window (Text Edit) of the interface. The program will read whatever is inside the Text Edit, so if the user wants to change something, it can be changed. But when closing the app a window will pop up asking if the user wants to save the changes made in the file or not.As it has been explained before, the information flux starts in the Text Edit on the right and the class in charge of reading this information and send it to MainWindow Class is TextEditReader Class.

The information about the spacing, the versions of the unstructured functions or the smoothing/edge crossing is read by the own MainWindow Class. 

Once the domain and the meshing properties are defined the user can run the program, this is when TRIMPACK (fortran) functions are called, and the drawing is made. The class that paints the Middle Widget is LienzoCreator Class. The "configuration Menu" at the top of the Interface, allows the user to change color, size and style of the mesh and boundary lines.

\section{User Interface}
In this section the use of the program will be explained. What makes each button and what impact has each one to the creation of the unstructured mesh.
\subsection{File Menu}
\begin{figure}[H]
    \centering
    \includegraphics{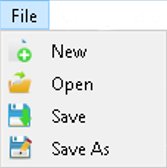}
    \caption{File Menu}
    \label{fig:FileMenu}
\end{figure}
This menu located at the top left corner of the Interface is responsible for all the management of the .mg file with the domain's boundaries. It can be created a new file, opened an existing one, save the changes the user has made in the Text Edit on the right of the interface or save it under another name. 
In addition when the user changes the file in the Text edit Widget, if tries to close the App it will pop up a window asking for saving the changes or removing them. 

\subsection{input file: .mg format}
The first thing the user has to open to use the app is the domain shape file. This file has to be written in a certain way in order that the program can read the information from it.

The .mg file is structured with different "record sections".  This method of writing a text file allows the program to read multiple information from the file. The sections starts with a key name chosen by the programmer and ends with "ENDRC"
In this case, due to the nature of the app it is only needed one "record section", the one where the boundary lines are defined: the SEGMENT section. The structure of the file will be explained with the simplest example possible: a square.

The first thing to know is that the exterior lines of the domain have to be oriented from left to right, and the inner lines from right to left (always defining \textbf{in order} the edges). This orientation helps the mesh generator to know in what side of the segment has to create the mesh.

\begin{figure}[H]
\begin{minipage}[trim=10 0 0 10mm]{0.45\linewidth}
\includegraphics[width=\linewidth]{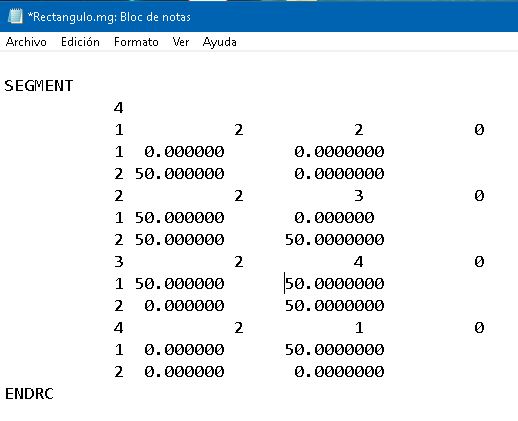}
\caption{rectangle.mg file}
\end{minipage}
\hfill
\begin{minipage}[trim=10 0 0 10mm]{0.45\linewidth}
\includegraphics[width=\linewidth]{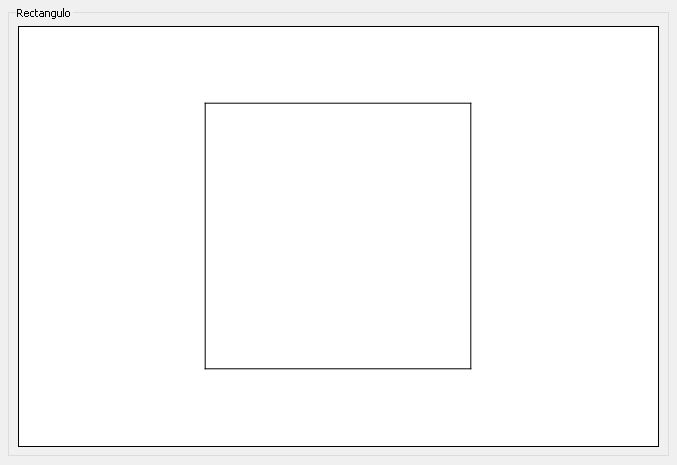}
\caption{square }
\end{minipage}%
\end{figure}
It is really important to know that the separation between numbers has to be spaces, \textbf{NEVER} tabs because if not the program will fail reading the file. 

-"SEGMENT" as explained before, starts the record of the domain boundary section

- "4" the number of edges that has the domain.

- "1" number of the segment. "2" number of points that defines the segment (this allows the user to define curved edges with multiple points inside a segment). "2" the segment to which connects the current segment. "0" always has to be 0.

- "1" number of the node. "0.000" x position of the point. "0.00000" y position of the point.
- "2" number of the node. "50.000" x position of the point. "0.00000" y position of the point.

This, draws the segment 1 from the point (0,0) to (50,0). The segments don't have to be written in order while the segment to which are connected is the correct one. This last three lines format repeats another 3 times, one for each segment.

- "ENDRC" this line ends the record of the segment definition.

\subsection{Options buttons}
\begin{figure}[H]
    \centering
    \includegraphics{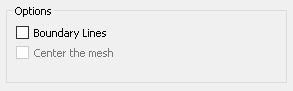}
    \caption{Options section}
\end{figure}
Once opened the .mg file, if the user clicks on the first check button the Boundary lines of the domain are drawn. While the second button centers the mesh in the middle of the Middle Widget.
\subsection{AFM page}
It consists of four sections:
\subsubsection{Type of contour}
\begin{figure}[H]
    \centering
    \includegraphics{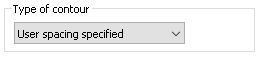}
    \caption{AFM Type of contour Section}
\end{figure}
The first one is for specify the type of contour used. There are three different types of contours but only one is avaliable in the application.
\begin{itemize}
    \item \textbf{User spacing specified}. Is the one used in this application. At the beginning of the mesh creation process, reads the spacing defined by the user and divides the boundary segments with this spacing.
    \item \textbf{User defined boundary}. It is enabled in this application. The boundaries are not divided by the program, them have to be divided by the user. In this way the user can specify the exactly point in which a mesh point is needed.
    \item \textbf{Free Nodes}. Is enabled in this application. The user defines a point cloud and the own program sets the external boundaries and creates the mesh using the points defined by the user.
\end{itemize}
\subsubsection{Spacing Configuration}
\begin{figure}[H]
    \centering
    \includegraphics[width=5cm]{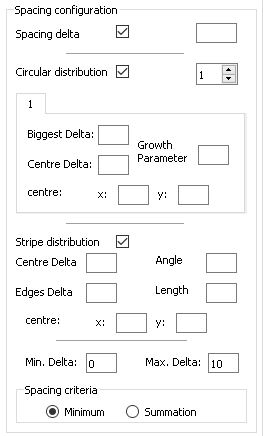}
    \caption{AFM Spacing Section}
\end{figure}
Spacing section is where the user defines the mesh size and distribution, there are several ways of defining the mesh desired.
This section is explained in more detail in the section Spacing functions (\ref{Spacing Functions}).
\subsubsection{AFM version}
\begin{figure}[H]
    \centering
    \includegraphics{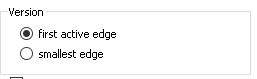}
    \caption{AFM Version Section}
\end{figure}
In the third section the user defines the version of AFM used, there are two versions:
\begin{itemize}
    \item \textbf{First Active Edge}. This version takes the firs active edge in the stack to triangulate with it. This version is recommended with constant spacing factor in the domain.
    \item \textbf{Smallest Edge}. This version takes the smallest edge in the stack to triangulate with it. This version is recommended with changing spacing factor in the domain. In sophisticated meshes this is a better choice.
\end{itemize}
The difference between this two versions can be easily observed in the next figures .
\begin{figure}[H]
\begin{minipage}[trim=10 0 0 10mm]{0.45\linewidth}
\centering
\includegraphics[width=5cm]{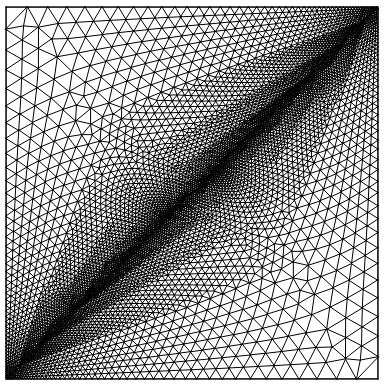}
\caption{first active edge version}
\label{figure:faeexample}
\end{minipage}
\hfill
\begin{minipage}[trim=10 0 0 10mm]{0.45\linewidth}
\centering
\includegraphics[width=5cm]{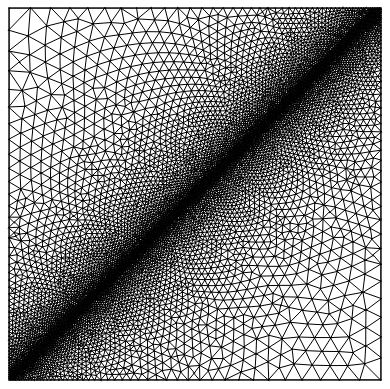}
\caption{smallest edge example}
\label{figure:seexample}
\end{minipage}%
\end{figure}
\subsubsection{Edge Swapping}
\begin{figure}[H]
    \centering
    \includegraphics{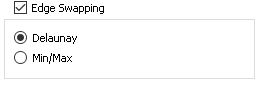}
    \caption{AFM Edge Swapping Section}
\end{figure}

In the last section of the AFM page is the EdgeSwapping Section. Here the user choose if wants to make edge swapping and what type, there are two options:
\begin{itemize}
    \item \textbf{Delaunay Criteria}. This version maximizes the minimum angle in the two possible triangulation configurations
    \item \textbf{Min/Max Criteria}. This version minimizes the maximum angle in the two possible triangulation configurations
\end{itemize}

\subsection{Steiner page}
\begin{figure}[H]
    \centering
    \includegraphics[width=5cm]{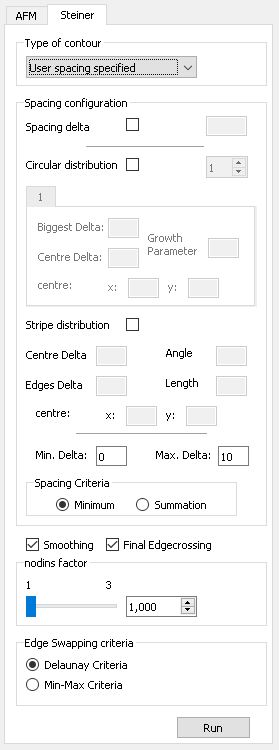}
    \caption{Steiner Page}
\end{figure}
It consists of five sections:
\subsubsection{Type of contour}
\begin{figure}[H]
    \centering
    \includegraphics{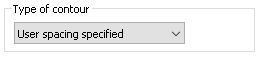}
    \caption{Steiner Type of contour Section}
\end{figure}
The first one is for specify the type of contour used. There are three different types of contours but only one is available in the application. Is the same of AFM method.
\begin{itemize}
    \item \textbf{User spacing specified}. Is the one used in this application. At the beginning of the mesh creation process, reads the spacing defined by the user and divides the boundary segments with this spacing.
    \item \textbf{User defined boundary}. The boundaries are not divided by the program, them have to be divided by the user. In this way the user can specify the exactly point in which a mesh point is needed.
    \item \textbf{Free Nodes}. Is enabled in this application. The user defines a point cloud and the own program sets the external boundaries and creates the mesh using the points defined by the user.
\end{itemize}
\subsubsection{Spacing Configuration}
\begin{figure}[H]
    \centering
    \includegraphics[width=5cm]{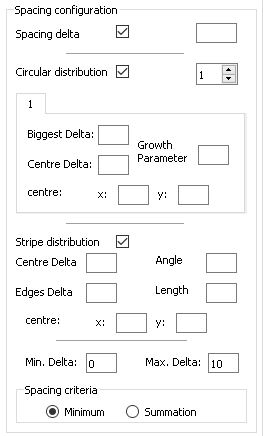}
    \caption{Steiner Spacing Section}
\end{figure}
Spacing section is where the user defines the mesh size and distribution, there are several ways of defining the mesh desired.
This section is explained in more detail in the section Spacing functions (\ref{Spacing Functions}).

\subsubsection{Smoothing \& Final Edge crossing Buttons}
\begin{figure}[H]
    \centering
    \includegraphics{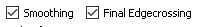}
    \caption{Steiner Smoothing and Edge crossing Section}
\end{figure}
The third section is composed by two buttons:
\begin{itemize}
    \item \textbf{Smoothing}. Grid smoothing alters the position of the interior nodes without changing the topology of the grid.
    \item \textbf{Final Edge crossing}. Checks a possible crossing between edges and fixes it.
\end{itemize}
\subsubsection{Steiner Factor}
\begin{figure}[H]
    \centering
    \includegraphics{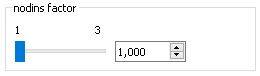}
    \caption{Steiner Factor Section}
\end{figure}
This factor controls the size of the triangles of the mesh. If factor value is 1 all triangles will be smaller that a delta triangle (the one defined by the spacing section in each point of the mesh) and if factor value is 3 all triangles are greater the delta triangle.
\subsubsection{Edge Swapping Criteria}
\begin{figure}[H]
    \centering
    \includegraphics{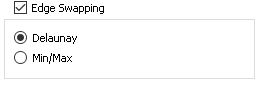}
    \caption{Steiner Edge Swapping Section}
\end{figure}
In the last section of the Steiner page is the EdgeSwapping Section. Here the user choose if wants to make edge swapping and what type, there are two options:
\begin{itemize}
    \item \textbf{Delaunay Criteria}. This version maximizes the minimum angle in the two possible triangulation configurations
    \item \textbf{Min/Max Criteria}. This version minimizes the maximum angle in the two possible triangulation configurations
\end{itemize}

\section{Spacing Functions} \label{Spacing Functions} \label{chapc:spc}
In the unstructured meshes it can be defined the size of each triangle in each point of the domain with simple functions. In this program are defined three different spacing distributions:
\begin{itemize}
    \item Uniform Spacing Delta. 
    \item Circular distribution. 
    \item Stripe distribution.
\end{itemize}
\subsection*{Uniform Spacing Delta}
This distribution generates an uniform mesh in all the domain. The user sets the value in the spacing delta section.
\begin{figure}[H]
    \centering
    \includegraphics{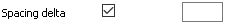}
    \caption{Spacing Delta Section}
\end{figure}
There is no top limitation in the size of the triangles, however there is a lower limitation at 0.001. This limitation exists because the drawer precision doesn't reach more decimals. But the program can create this kind of meshes. The usual value of this delta for a good-quality mesh is one magnitude order less than the sides of the domain.
\begin{figure}[H]
\begin{minipage}[trim=10 0 0 10mm]{0.45\linewidth}
\centering
\includegraphics[width=7cm]{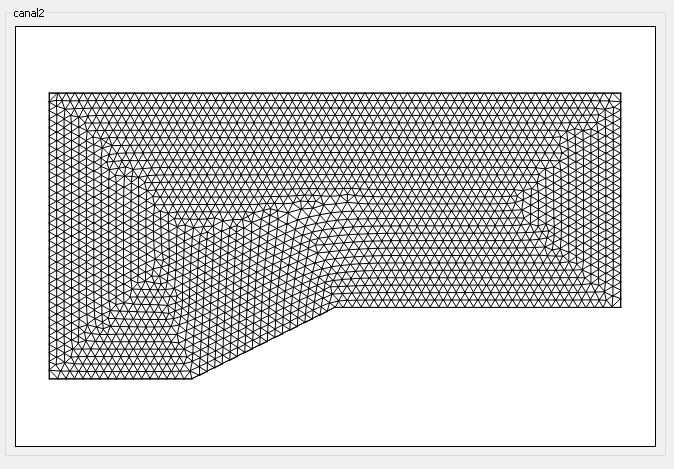}
\caption{AFM uniform spacing canal2.mg delta value: 0.3}
\end{minipage}
\hfill
\begin{minipage}[trim=10 0 0 10mm]{0.45\linewidth}
\centering
\includegraphics[width=7cm]{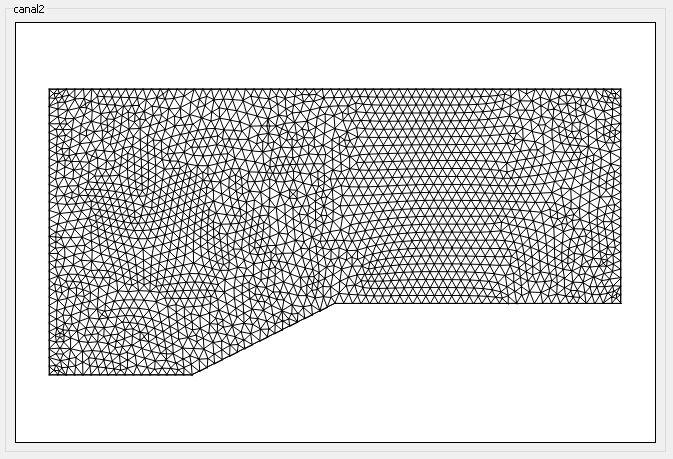}
\caption{Steiner uniform spacing canal2.mg delta value: 0.3}
\end{minipage}%
\end{figure}

\subsection*{Circular Distribution}
This distribution generates a growing size mesh from a centre chosen by the user.
\begin{figure}[H]
    \centering
    \includegraphics{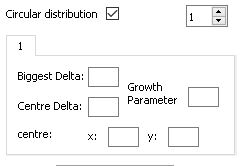}
    \caption{Circular Distribution Section}
\end{figure}
The inputs used in this distribution are:
\begin{itemize}
    \item $\delta_A$: "Biggest Delta". It is the maximum value that reaches the mesh.
    \item $\delta_B$: "Centre Delta". It is the value in the centre of circular distribution.
    \item $\beta$: "Growth Parameter". It defines the growing parameter of the method. When $\beta$ grows, the compacted mesh is smaller.
    \item $X_s$: "centre: x". It sets the x coordinate of the distribution's centre.
    \item $Y_s$: "centre: y". It sets the y coordinate of the distribution's centre.
\end{itemize}

The equation that controls this distribution is:
\begin{equation}
    \delta = \delta_A + (\delta_B - \delta_A)*e^{-\beta*r^2}
\end{equation}
Being $r=\sqrt{(X-X_s)^2 + (Y-Y_s)^2}$, the distance between the point in which the mesh is being drew and the centre of the distribution. With this function delta is constrain between $\delta_A$ and $\delta_B$.

Analyzing the limits of this function:
\begin{equation}
    \lim_{r\to 0} \delta = \delta_B
\end{equation}
\begin{equation}
    \lim_{-\beta*r^2\to \infty} \delta = \delta_A
\end{equation}

The $\beta$ variable, as it has been said before regulates the speed at which the mesh grows. This can be easily seen representing the same function but changing the value of $\beta$.
\begin{figure}[H]
\begin{minipage}[trim=10 0 0 10mm]{0.45\linewidth}
\textbf{r (abscissa axis) - delta (ordinate axis) $\beta = 1$}
\centering
\includegraphics[width=7cm]{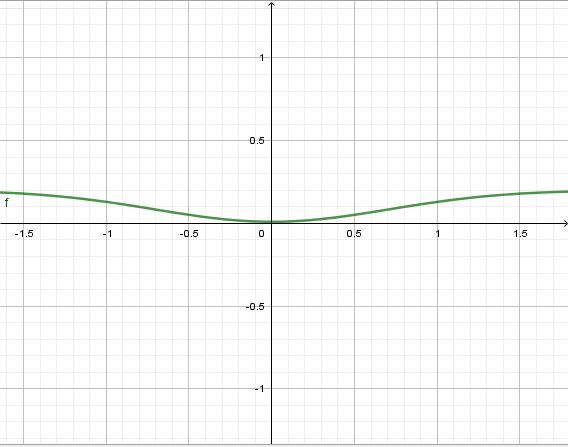}
\caption{circular distribution function with  $\delta_A = 0.2, \delta_B=0.01$}
\end{minipage}
\hfill
\begin{minipage}[trim=10 0 0 10mm]{0.45\linewidth}
\textbf{r (abscissa axis) - delta (ordinate axis) $\beta = 0.1$}
\centering
\includegraphics[width=7cm]{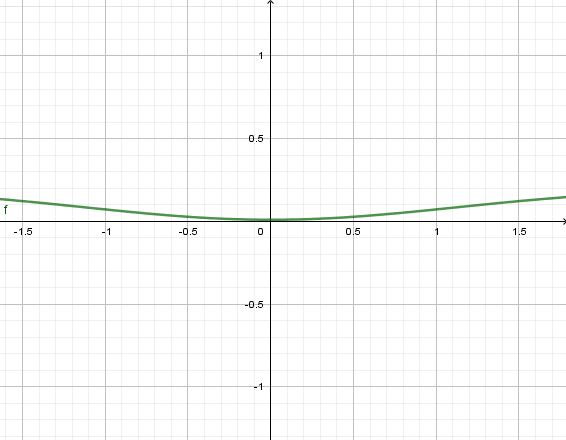}
\caption{circular distribution function with  $\delta_A = 0.2, \delta_B= 0.01 $ }
\end{minipage}%
\end{figure}

This two examples results in the following meshes:
\begin{figure}[H]
\begin{minipage}[trim=10 0 0 10mm]{0.45\linewidth}
\textbf{circular distribution with $\beta = 1$}
\centering
\includegraphics[width=7cm]{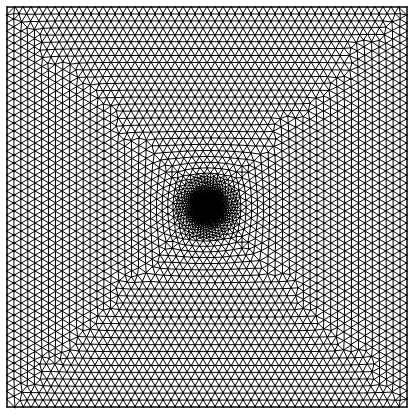}
\caption{$\delta_A = 0.2, \delta_B=0.01$}
\end{minipage}
\hfill
\begin{minipage}[trim=10 0 0 10mm]{0.45\linewidth}
\textbf{circular distribution with $\beta = 0.1$}
\centering
\includegraphics[width=7cm]{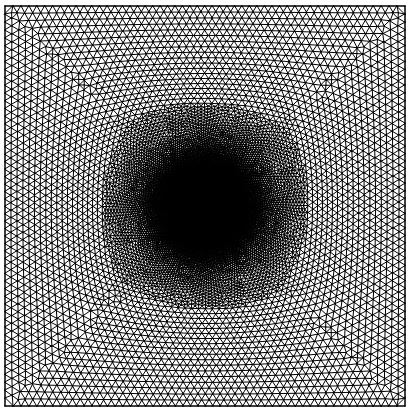}
\caption{$\delta_A = 0.2, \delta_B= 0.01 $ }
\end{minipage}%
\end{figure}
\subsection*{Stripe distribution}
\begin{figure}[H]
    \centering
    \includegraphics[width=5.5cm]{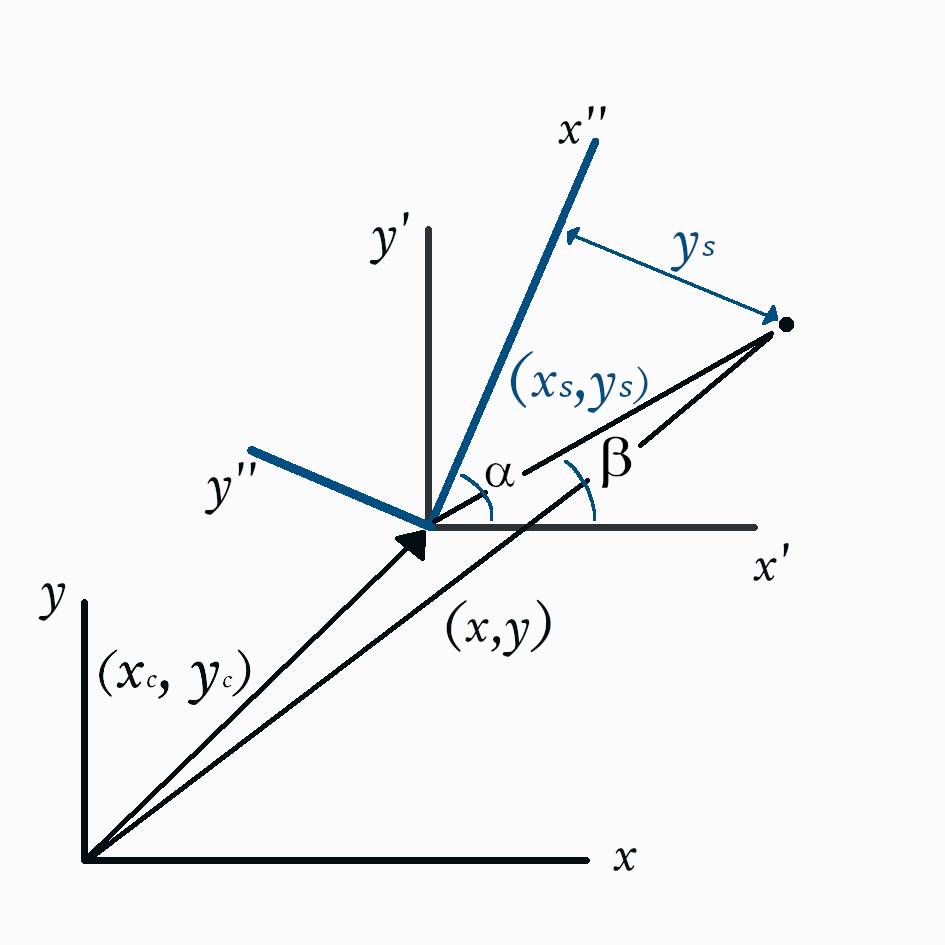}
    \caption{Axis used in Stripe distribution}
    \label{fig:stripe axis}
\end{figure}

This distribution generates a growing mesh from a line with an specified inclination by the user.
\begin{figure}[H]
    \centering
    \includegraphics{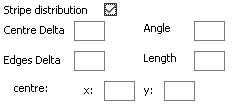}
    \caption{Stripe Distribution Section}
\end{figure}
The inputs used in this distribution are:
\begin{itemize}
    \item $\delta_A$: "Centre Delta". It is the value in the centre of circular distribution.
    \item $\delta_B$: "Edges Delta". It is the maximum value that reaches the mesh.
    \item $\alpha$: "Angle". It defines the angle of the compact stripe
    \item $L$: "Length". Sets the region size in which the mesh is more compact.
    \item $X_C$: "centre: x". It sets the x coordinate of the distribution's centre.
    \item $Y_C$: "centre: y". It sets the y coordinate of the distribution's centre.
\end{itemize}
This function is more complex than the previous ones. The idea is that the spacing delta grows linearly while gets far from a line defined by the user (x''). To make this possible it is needed a change of axis as the figure \ref{fig:stripe axis}.

The equation that controls this distribution is:
\begin{equation}
    \delta = \delta_A + \delta_B * \frac{y_s}{L}
\end{equation}
Being $y_s$:
\begin{equation}
    y_s =/\sqrt{(x-x_c)^2+(y-y_c)^2}*sin(\beta-\alpha)/
\end{equation}

This spacing delta size distribution is linear so it is not top delimited, while the minimum is $\delta_A$.

Two examples of this type of distributions:
\begin{figure}[H]
\begin{minipage}[trim=10 0 0 10mm]{0.45\linewidth}
\textbf{Canal2.mg}
\centering
\includegraphics[width=7cm]{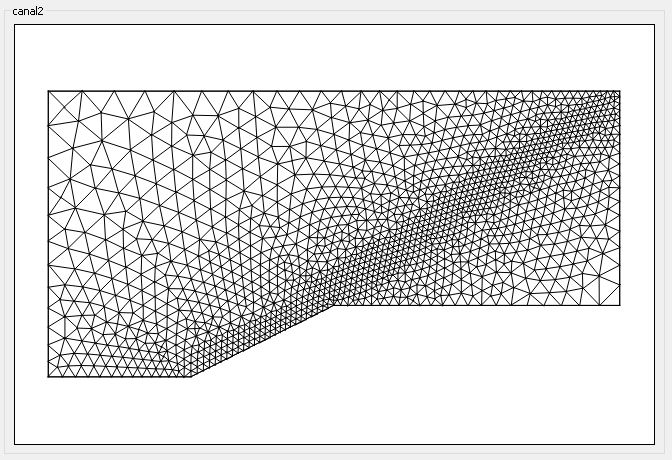}
\end{minipage}
\hfill
\begin{minipage}[trim=10 0 0 10mm]{0.45\linewidth}
\textbf{NACA 2412.mg}
\centering
\includegraphics[width=7cm]{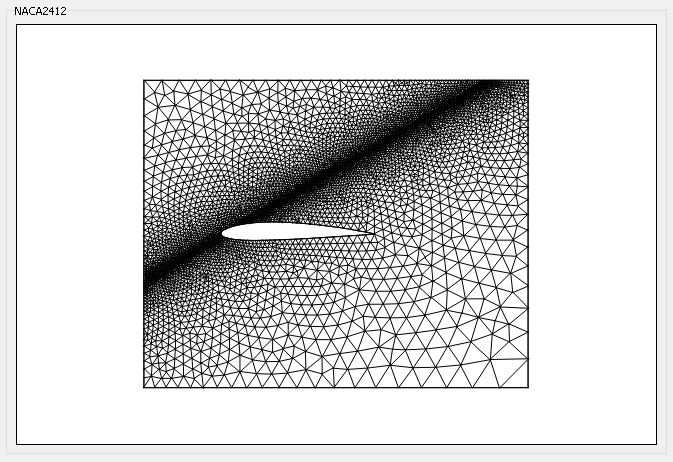}
\end{minipage}%
\end{figure}

Another good distribution that can be implemented to this program is a cone shape, that could be useful at the time of study shock waves produced inside a combustion chamber.

%% file: Trimpack.tex
\section{TRIMPACK}

The TRIM-Pack (\textbf{TRI}angular unstructured \textbf{M}esh generation\textbf{ PACK}age) is a library of elements (FORTRAN subroutines) for generating and refining methods of triangular unstructured grids. It also includes elements of verification and evaluation for this existing grids.

Special attention was paid designing the subroutines and the data structure. This allows a posterior extension, for example with new generation of adaptation and refinement subroutines, achieving a high flexibility of usage.

In this section, a detailed description and evaluation of the selected mesh generation algorithms will be given which form with some necessary tools the TRIngular unstructured Mesh generation PACKage.

\subsection{Contents of Trimpack}
\subsubsection{Generation algorithms}

\paragraph{\textbf{dlny}}

This subroutine realises a constricted Delaunay triangulation for a set of nodes and its corresponding boundaries, however a triangulation without boundaries is possible, too.

If boundaries are given, the dlny-subroutine is supported by an edge-crossing algorithm, which checks the veracity and feasibility of the routine decisions. This is necessary in account of the possible complexity that may have the boundaries.

For a deeper description go to \ref{par:dlny}

\paragraph{\textbf{afm}}

This subprogram performs the triangulation of a domain between given boundaries by means of the Advancing Front Method (AFM).

\paragraph{\textbf{steiner}}

Steiner generates an usntructured triangular mesh using the Steiner algorithm. After a Delaunay triangulation, this suboutine inserts nodes, swaps edges and smoothes the mesh to improve the mesh quality.

\subsubsection{Grid-refinement algorithms}

\paragraph{\textbf{nodins}}

Nodins adds new nodes in the triangles of the already generated grid to attain a refinement of certain parts or the whole grid. At present, the nodes will be inserted in the centre of gravity of each triangle, but subsequent expansion is easily possible.

\paragraph{\textbf{edswap}}

This subprogram performs a diagonal swapping, following the Delaunay criteria or the Min-max criteria, which changes the connectivities among the given nodes in the grid without altering their position. Edswap assumes that a triangulation exists (not Dealunay) then makes it Delaunay through the application of edge swapping.

\paragraph{\textbf{smooth}}

The subroutine smooth alters the position of the interior (free) nodes without changing the topology of the mesh to improve the mesh quality. In considers every connection between two nodes (if they are no boundary nodes) as a spring of equal elasticity and base-length. The equilibrium position is the result of smooth.

\subsubsection{Tools}

\paragraph{\textbf{cbrdr}}

This subprogram creates boundaries, using splines (spline-subroutine) for the boundary node distribution, which will be supplied by the user through a boundary definition and a node spacing subroutine. If th edomain to be triangulated has no boundaries, cbrdr will create an initial edge to start the triangulation from. Therefor, two subroutines (fnode and snode) provide the two nodes, which define this first edge. In this case -where no boundaries are given- the user has to define some free nodes. Cbrdr can be used for the boundary creation for th edlny-triangulation.

\paragraph{\textbf{connec}}

The connec subroutine generates optionally the following eight connectivity matrix of a given triangulation that may be used for additional calculations: 
\begin{itemize}
    \item Triangles as a function of triangles.
    \item Nodes as a function of nodes.
    \item Triangles as a function of nodes.
    \item Nodes as a function of triangles.
    \item Edges as a function of nodes.
    \item Nodes as a function of edges.
    \item Triangles as a function of edges.
    \item Edges as a function of triangles.
\end{itemize}

\paragraph{\textbf{edcros}}

Edcros provides an indicator, which will warn of a crossing between two edges.

\paragraph{\textbf{spline}}

provides the subroutine cbrdr by means of cubic splines with he finally boundary node distribution.

\subsubsection{Other subroutines}

\paragraph{\textbf{stmsh} (NOT used in this program):}

This subroutine delivers some statistical information about an existing grid:
\begin{itemize}
    \item Number of connectivities nodes-nodes.
    \item Number of connectivities triangles-nodes.
    \item Number of areas of the triangles in percentage of a spacing-triangle area.
    \item Number of lengths of the edges in percentage of a spacing-edge length.
    \item Number of angles of the triangles in percentage of 60º.
    \item Furthermore, stmsh checks the general formula: Nu
\end{itemize}

\paragraph{\textbf{centre}}

Provides the coordinates of the centre of a circle, defined with three nodes.

\paragraph{\textbf{fnode}}

Finds the node with minimum y-coordinate in a given node cloud and return its number -to the subroutine brdr

\paragraph{\textbf{snode}}

Finds a second initial node which will create with the first node the initial edge of the triangulation. This second node will be the one which forms the minimum angle to the horizontal. In This wait it will be quite sure that the first edge is in a right position for the following triangulation.

\subsection{Description}
\subsubsection{Data Structure} \label{func:iw}
Along this library, all the mesh information is saved in an array called $iw$, this array has the following structure:
\begin{figure}[H]
    \centering
    \includegraphics[width=14cm]{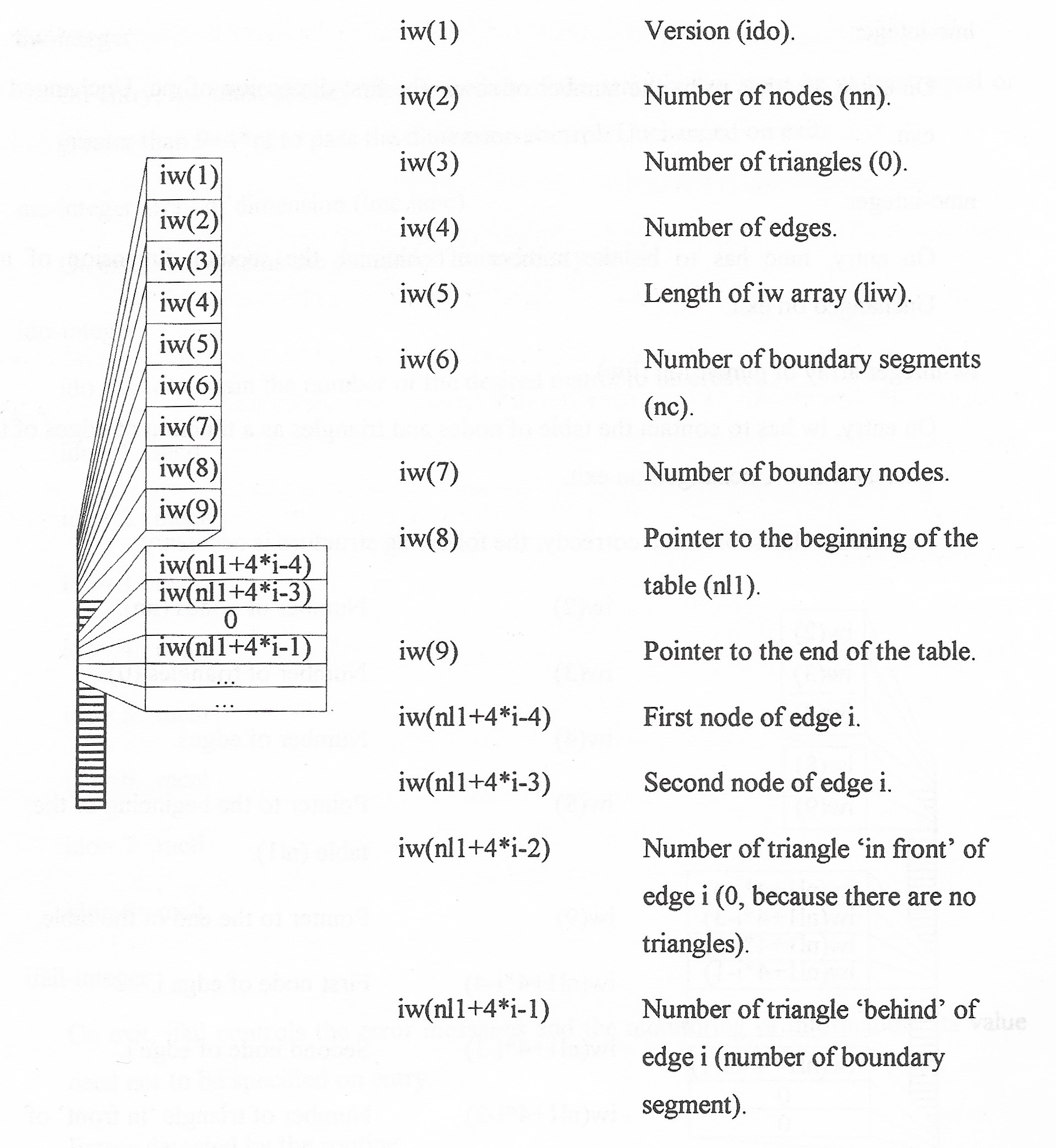}
    \caption{$iw structure$}
    \label{fig:iw}
\end{figure}
Also three subroutines are passed by arguments defined as externals in the main code. This structure allows the user to create or customize the functioning of these subroutines
\paragraph*{drw} \label{func:drw}
The drw subroutine provides the user with the possibility to clear the screen or draw the mesh while is being calculated. The arguments of this subroutine are the following ones.
\begin{center}
    $call$ drw(idrw,x1,x2,y1,y2)
\end{center}

This subroutine draws a line in between the nodes 1 and 2. The int $idrw$, depending of its value, can clean the screen, draw the edges. 
\paragraph*{spc} \label{func:spc}
This subroutine provides the spacing factor which controls the insertion of nodes. The user must specify here the value of delta for the centre of each triangle of the mesh. This value can either be constant of can vaty in the manner the user wants. Its specification is:
\begin{center}
    $call$ spc(x,y,delta)
\end{center}
Where x and y determines the point in the domain and with a formula inside this subroutine the delta in return can be any.
\paragraph*{bnd} \label{func:bnd}
bnd provides the subroutine $brdr$ (\ref{func:brdr}) with the exterior/interior boundary of the triangulation domain. The user must specify in this subroutine the number of describing nodes of the segment with their xb- and yb- coordinates, as well as the next connecting boundary segment (nj) if there is any connection.

It is very important that the boundary segments (and therefore each of the boundary edges) must have a directional order, in the way that the triangulating domain is located \textbf{always on the left} of the boundary edges if the segment is traversed from its first to its second node. Its specification is:
\begin{center}
    $call$ bnd(in,nb,xb,yb,nj)
\end{center}

\begin{center}
\begin{tabular}{ |c|c|c| } 
 \hline
 variable & input & output \\
 \hline
  in - integer & \multicolumn{2}{|c|}{number of boundary segments}  \\ 
 \hline
 nb - integer & - & total number of describing boundary nodes  \\ 
 \hline
 nj - integer & empty & number of the next connecting boundary segment \\
 \hline
 xb - real array (nbmax) &  \multicolumn{2}{|c|}{x-coordinates of all boundary nodes}   \\
 \hline
 yb - real array (nbmax) &  \multicolumn{2}{|c|}{y-coordinates of allboundary nodes }   \\
 \hline
\end{tabular}
\end{center}

\subsubsection{Generation algorithm}
\paragraph*{DLNY:}\label{par:dlny}
The algorithm which is used here is a basic Delaunay triangulation algorithm, whose properties and main ideas are described in chapter \ref{chap:delaunay}.

To launch dlny subroutine, the user has to provide some former information with the program call about the domain that he wants to be triangulated and in gree nodes (if existent):

\begin{center}
    $call$ dlny(nn,nl,nt,x,y,drw,iw,liw,ido,ifail)
\end{center}

\subparagraph{Arguments values}

\begin{center}
\begin{tabular}{ |c|c|c| } 
 \hline
 variable & input & output \\
 \hline
  nn - integer & \multicolumn{2}{|c|}{number of nodes in the domain (boundary $+$ existing free) }  \\ 
 \hline
 nl - integer & number of boundary edges & number of all mesh edges  \\ 
 \hline
 nt - integer & empty & number of mesh triangles  \\
 \hline
 x - real array (nn) &  \multicolumn{2}{|c|}{x-coordinates of all nodes in the domain (boundary and free) }   \\
 \hline
 y - real array (nn) &  \multicolumn{2}{|c|}{y-coordinates of all nodes in the domain (boundary and free) }   \\
 \hline
 drw - subroutine  & \multicolumn{2}{|c|}{drw (idrw,x1,x2,y1,y2) } \\
 \hline
 iw - int array (liw)  & boundary structure & total mesh structure \\
 \hline
 liw - int & \multicolumn{2}{|c|}{drw (dimension of array iw, at least $ liw \geq 5 * \text{max. number of edges (nl)} + 9$} \\
 \hline
 ido - integer & \multicolumn{2}{|c|}{ number of version }  \\
  \hline
 ifail - integer & 0 & error messages \\
 \hline
\end{tabular}
\end{center}

\subparagraph{Ido values}

\begin{center}
\centering
    \begin{tabular}{ |c|c| }
    \hline
        value & specification \\
        \hline
        0 & without boundaries \\
        1 & with boundaries \\
        2 & with boundaries constructed with splines \\
        \hline
    \end{tabular}
\end{center}

For $ido = 1,2$ an edge crossing check will be performed, which is necessary in account of the possible complexity that may have the boundaries.

\subparagraph{Ifail values}

\begin{center}
\centering
    \begin{tabular}{ |c|c| }
    \hline
        value & specification \\
        \hline
        1 & The user must check the final mesh, looking for mistakes \\
        0 & Normal exit \\
        -1 & The input data are not correct, probably liw too small \\
        \hline
    \end{tabular}
\end{center}
\subparagraph{Code description}

At the beginning of the main loop, $nsk$ gives us the number of boundary edges adn the integer work array $iw$ contains the number of nodes and triangles as a function of edges ($iw(nl1...nl2-1$) and an index of each edge (0:active, 1:inactive) (iw(nl2)). The loop will continue until $nsk$ is 0, e.e. the active front is empty and all indices are 1. The first edge of the stack (of active edges) is chosen (nodes $n1$ and $n2$) adn the first candidate node ($inodo$) is proved to be the right one to perform a Delaunay triangle.

Therefore the straight line function of the edge (straight line through both nodes of the edge) is evaluated and it is checked if the candidate node ($inodo$) lies on the correct side of the edge to be a possible Delaunay node. In the case the candidate node ($inodo$) is not located on the correct place, a different(the next) candidate node ($inodo$) is chosen.

Once a correct placed node ($inodo$) has been proved, the provisional triangle ($n1$, $n2$ and $inodo$), i.e. the circumcentre and the radius of the circle through thtse three nodes are calculated. Now the subroutine checks if any other node ($jnodo$) is placed inside the described circle. If the routine finds a node inside the circle, the provisional triangle is disparaged and the node in the circle is elected as next candidate node ($inodo$).

If there are no other nodes inside the circle of the provisional triangle, the routine examines with the subroutine $edcros$ (edgecrossing), if the two new edges (edge $n1-nodo$ and edge $inodo-n2$, remember: direction) cross any other existent edge of the trianglation. If there is an edgecrossing, the provisional triangle is not updated. Otherwise it is a Delaunay triangle and the routine finally checks, if the new edges already exist before updating the work array ($iw$) with the new data (adding a new edge to iw with its two belonging noes $n1$ and $n2$ and its belonging triangle) and taking the next active edge of the stack as base for the search for a third node to build a Delaunay triangle with.

At the end of this routine, the integer work array ($iw$) is reorganized and contains now the information described in \ref{func:iw}

\paragraph*{AFM:}\label{par:afm}
The algorithm which is used here for the Advancing Front Method is based on the method proposed by Peraire et. al. firs described in [24] (see section \ref{chap:afm}).

To start the afm subroutine, the user has to take care for the correct delivery of the necessary information about the domain.

\begin{center}
    $call$ afm(nn,nl,nt,x,y,lnn,drw,iw,liw,ido,ifail)
\end{center}

\subparagraph{Arguments values}

\begin{center}
\begin{tabular}{ |c|c|c| } 
 \hline
 variable & input & output \\
 \hline
  nn - integer & total number of boundary nodes & total number of mesh nodes  \\ 
 \hline
 nl - integer & number of boundary edges & number of all mesh edges  \\ 
 \hline
 nt - integer & empty & number of mesh triangles  \\
 \hline
 x - real array (nn) &  x-coordinates of boundary points & all mesh points x-coords     \\
 \hline
 y - real array (nn) &  u-coordinates of boundary points & all mesh points y-coords   \\
 \hline
 lnn - integer & \multicolumn{2}{|c|}{x and y dimension (estimated number of nodes) }  \\
 \hline
 drw - subroutine & \multicolumn{2}{|c|}{drw (idrw,x1,x2,y1,y2) } \\
 \hline
 iw - int array (liw) & boundary structure & total mesh structure \\
 \hline
 liw - int & \multicolumn{2}{|c|}{drw (dimension of array iw, at least $ liw \geq 5 * \text{max. number of edges (nl)} + 9$} \\
 \hline
 ido - integer & \multicolumn{2}{|c|}{ number of version }  \\
  \hline
 ifail - integer & 0 & error messages \\
 \hline
\end{tabular}
\end{center}

\subparagraph{Ido values}

\begin{center}
\centering
    \begin{tabular}{ |c|c| }
    \hline
        value & specification \\
        \hline
        1 & the first active edge in the stack is taken to triangulate with (recom.with constant spacing factor) \\
        2 & the smallest edge of the stack is taken to triangulate with (recom. with a changing spacing factor) \\
        \hline
    \end{tabular}
\end{center}

\subparagraph{Ifail values}

\begin{center}
\centering
    \begin{tabular}{ |c|c| }
    \hline
        value & specification \\
        \hline
        0 & Normal exit \\
        -1 & The input data are not correct, probably liw too small \\
        -7 & The "edge-cross stack" is too small \\
        -8 & The length of the list of near nodes is too small \\
        \hline
    \end{tabular}
\end{center}
\subparagraph{Code description}

After some previous calculations, such as initialization of variables and the calculation of the maximum possible number of edges to fit in iw (which is used later for the error check), the afm subroutine begins with the main loop.

\begin{figure}[H]
    \centering
    \includegraphics[width=10cm]{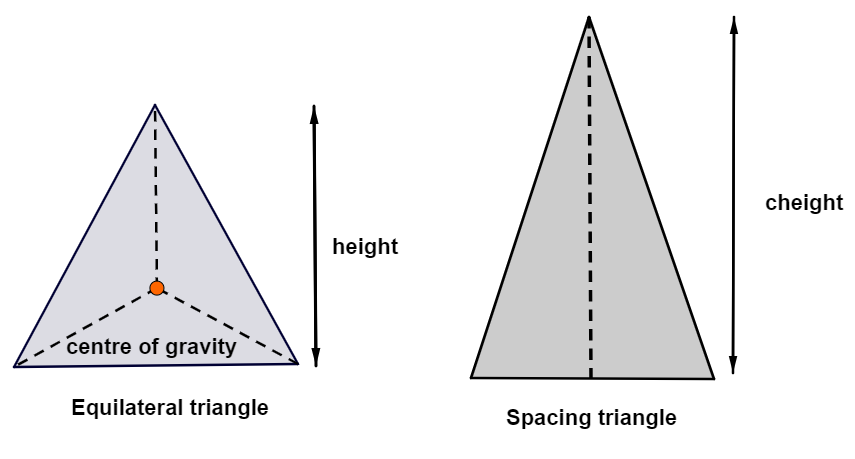}
    \caption{Construction of a new triangle}
    \label{fig:afmtrim}
\end{figure}

First of all, the \textbf{smallest} or the x and y dimension (estimated number of nodes \textbf{first active edge} of the stack iw will be searched to take as bases for the first triangle. This searching routine depends of the value of $ido$ (see above). For this edge, the vertex (called $P1$) for the "perfect" triangle (the triangle with the base $n1-n2$ and a vertex with the  distance $delta1$ from $n1$ and $n2$) is iterated: For this iteration, the height (here called $cheight$) of a triangle, constructed with the $spacing factor$ of the center of gravity of the belonging equilateral triangle is compared with the height of this equilateral triangle (see Figure \ref{fig:afmtrim}).

If these two heights differ more than 1\%, the centre of this \emph{spacing triangle} is taken to obtain the new \emph{spacing factor} and therefore a new \emph{spacing triangle}.

The \emph{spacing factor} is delivered by a spacing subroutine and replaces here the background grid suggested by Peraire. This makes the delivery of the "background grid information" simple and efficient. In this spacing subroutine, the user has to define for each value x and y a factor, which has to be a somehow "average distance" for this node. However, a real "background grid" can be implemented easily to increase the computational time of the searching algorithms. 

If the correct height is iterated and the value of $\delta_M$ ($deltam$) which is the length of the legs of the spacing triangle is calculated, the value of $\delta_1$ ($delta1$) is obtained out of the inequalities of Peraire et. al. (see chapter \ref{chap:afm}). The location of $P1$ is now the vertex position for the triangle with the basis $n1-n2$ and a length of legs of $\delta_1$.

After the calculation of $P1$, the routine determines all the active nodes which lie within the circle with centre at $P1$ and radius $\delta_1$. If an active node has been found (first node of an active edge) and lies on the left side of the basis edge (left, if the edge is traversed from $n1$ to $n2$), the node will be a possible connecting node if it has the right distance from $n1$ and $n2$ (at most $1.5 * \delta_1$). All these nodes, which accomplish these suppositions are ordered in a list in accordance to their distance to the basis edge, while $P1$ is added at its end.

\begin{figure}[H]
    \centering
    \includegraphics[width=8cm]{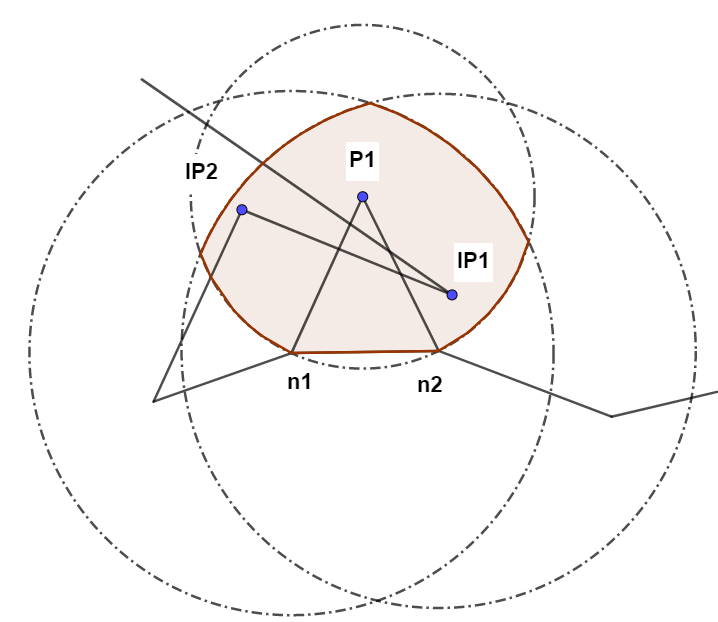}
    \caption{Area fr pints to connect (hatched)}
    \label{fig:afmtrim2}
\end{figure}

\begin{figure}[H]
    \centering
    \includegraphics[width=8cm]{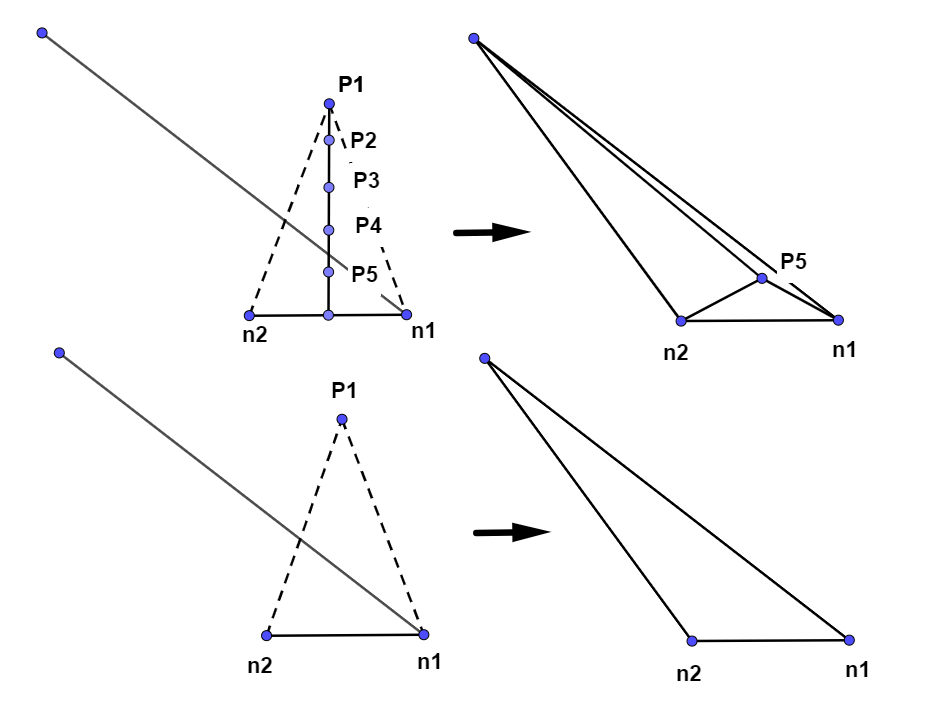}
    \caption{Triangulation example of (a) Peraire, (b)used algorithm}
    \label{fig:afmtrim3}
\end{figure}

\begin{figure}[H]
    \centering
    \includegraphics[width=8cm]{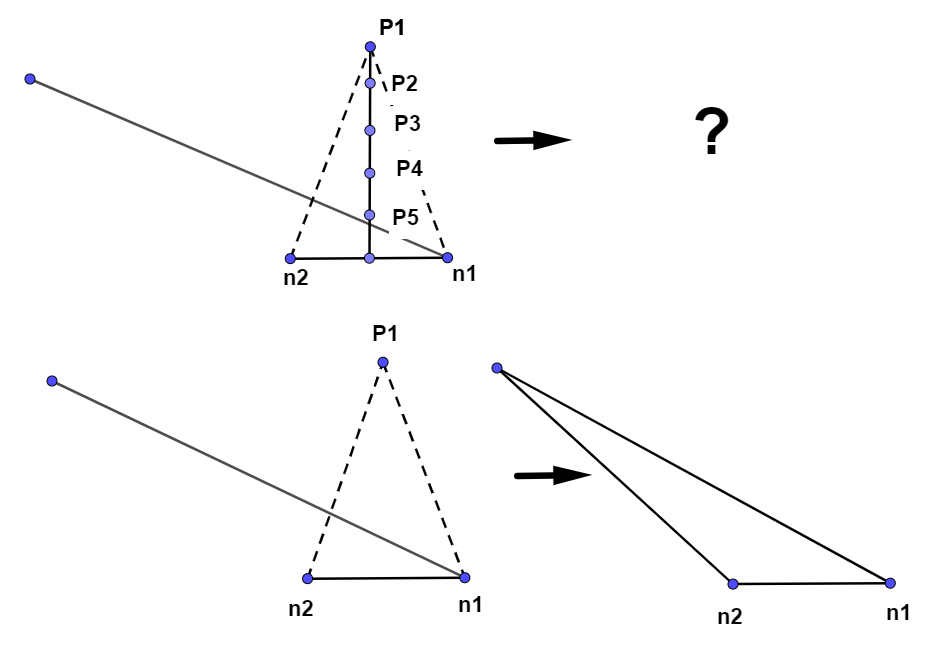}
    \caption{Problem of edgecrossing with the last point in the list of Peraire's algorithm (a) and its solution in the used algorithm (b)}
    \label{fig:afmtrim4}
\end{figure}

In the main loop, the afm routine examines, if the edge with the first node in the list does not intersect any existing side in the front. if there is an edge-crossing, the routine takes the next node in the list and repeats the edge crossing control. if even the last node in the list provokes an edge-crossing, no triangulation with this base will be done now. This is the main difference to the algorithm proposed by Peraire: in his algorithm, long and skew triangles could occur if a node of the list is located very near to an existing edge (see Figure \ref{fig:afmtrim3}), and also the question of edge-crossing with all nodes in the list has not been solved sufficiently (see Figure \ref{fig:afmtrim4}).

Afm takes the next longer edge of the front as basis edge and repeats teh formedrly described algorithm. To do this, a new array $cros$ is necessary to retain this edge for the searching of the next longer edge. If the dimension of this array ($dim$) is too less (occurs if the boundary's geometry is very complex), the routine will return with an ifail = -7 to the main program.

If finally a triangle can be constructed, the basis edge becomes inactive (the value of its belonging space in the part of $iw$ becomes 1) and gets the number of the next triangle "in front". With this new triangle, teh routine controls if its new edges already exist (if a new node was generated) and updates the stack. Here, the possibility is given to draw the new edges like the user wants (see \ref{func:drw}).

Afm repeats this cycle until the number of active edges ($isk$) is 0 and no further triangulation can be done. At the end of this routine, the integer work array ($iw$) is reorganized and contains now the information described in \ref{func:iw}.

\paragraph*{STEINER:}\label{par:steiner}
The algorithm which is used here is similar to the basic Holmes algorithm mentioned in section \ref{chap:steiner}.

To start the steiner subroutine, the user has to take care for the correct delivery of the necessary information about the domain.

\begin{center}
    $call$ steiner(nn,nl,nt,x,y,lnn,spc,drw,iw,liw,factor,ido,ifail)
\end{center}

\subparagraph{Arguments values}

\begin{center}
\begin{tabular}{ |c|c|c| } 
 \hline
 variable & input & output \\
 \hline
  nn - integer & free nodes + boundary nodes & free nodes+ boundary nodes + inserted nodes \\ 
 \hline
 nl - integer & number of boundary edges & number of all mesh edges  \\ 
 \hline
 nt - integer & empty & number of mesh triangles  \\
 \hline
 x - real array (nn) &  \multicolumn{2}{|c|}{x-coordinates of all nodes in the domain (boundary and free) }   \\
 \hline
 y - real array (nn) &  \multicolumn{2}{|c|}{y-coordinates of all nodes in the domain (boundary and free) }   \\
 \hline
 lnn - integer & \multicolumn{2}{|c|}{ x and y dimension (estimated number of nodes) } \\
 \hline
  spc - subroutine \ref{func:spc} & \multicolumn{2}{|c|}{spc (x,y,delta)} \\
 \hline
 drw - subroutine \ref{func:drw} & \multicolumn{2}{|c|}{drw (idrw,x1,x2,y1,y2) } \\
 \hline
 iw - int array (liw) \ref{func:iw} & boundary structure & total mesh structure \\
 \hline
 liw - int & \multicolumn{2}{|c|}{drw (dimension of array iw, at least $ liw \geq 5 * \text{max. number of edges (nl)} + 9$} \\
 \hline
 factor - real & \multicolumn{2}{|c|}{controls the size of the triangles} \\
 \hline
 ido - integer & \multicolumn{2}{|c|}{number of version and specific final smoothing mechanism \ref{chap:ido}}  \\
  \hline
 ifail - integer & 0 & error messages \\
 \hline
\end{tabular}
\end{center}
\subparagraph{Factor values}
This factor can be any number in between 1 and 3. It is the size factor for the subroutine $nodins$ call inside steiner method.
\begin{center}
\centering
    \begin{tabular}{ |c|c| }
    \hline
        value & specification \\
        \hline
        1 & triangles smaller than specified spc (triangle with sidelength delta)\\
        3 & all triangles greater than delta triangle \\
        \hline
    \end{tabular}
\end{center}

\subparagraph{Ido values} \label{chap:ido}

\begin{center}
\centering
    \begin{tabular}{ |c|c|c| }
    \hline
        position & value & specification \\
        \hline
         \multirow{2}{4em}{ido(1)} & 1 & with boundaries \\
         & 2 & with boundaries, constructed with splines \\
        \hline
        \multirow{3}{4em}{ido(1)} & 1 & insets noe in the graviti centre of a triangle \\
         & 2 & not existent yet\\
         & 3 & not existent yet \\
         \hline
        \multirow{2}{4em}{ido(3)} & 0 & no smoothing \\
         & 1 & with smoothing \\
        \hline
        \multirow{2}{4em}{ido(4)} & 0 & no final edge-crossing \\
         & 1 & with final edge-crossing \\
        \hline
    \end{tabular}
\end{center}

\subparagraph{Ifail values}

\begin{center}
\centering
    \begin{tabular}{ |c|c| }
    \hline
        value & specification \\
        \hline
        1 & The user must check the final mesh, looking for mistakes \\
        0 & Normal exit \\
        -1 & The input data are not correct, probably liw too small \\
        -2 & The expected number of nodes (lnn) too small \\
        -3 & Too many nodes in boundary segment \\
        -4 & The delivered value of $nn$, $nt$ or $nl$ is not correct \\
        -5 & The transferred dimension of the connectivity matrix (columns, $nmc$) is too small \\
        -6 & The transferred dimension of the connectivity matrix (rows, $lmc$)is too small \\
        -7 & Edge-crossing in the final mesh \\
        \hline
    \end{tabular}
\end{center}

\subparagraph{Code description}

This subroutine consists of the calls to \emph{dlny}, \emph{nodins}, \emph{edswap} and \emph{smooth}, always followed by an ifail error check.

Steiner will return to the main program whether there is an ifail less than zero or no node insertion occurred, and therefore no further mesh refinement is necessary. Just before returning to the main program, a further edge-crossing control validates the mesh. if here is a positive ifail, the routine prints it on the screen indeed, but will also go on.

\subsubsection{Grid refinement algorithms}

\paragraph{Nodins:}

The algorithm which is used here realises a \emph{posterior node insertion} (mentioned in chapter \ref{chap:remesh}) and can be used for a posterior grid refinement (remeshing). This means, that a former triangulation has to be existent but it makes no difference in which way it was established.

Besides of he grid information, the user has to transmit some further information to launch this subroutine:

\begin{center}
    $call$ nodins(x,y,nn,lnn,nt,nl,iw,liw,spc,drw,factor,ido,ifail)
\end{center}

\subparagraph{Arguments values}
\begin{center}
\begin{tabular}{ |c|c|c| } 
 \hline
 variable & input & output \\
 \hline
 x - real array (lnn) &  \multicolumn{2}{|c|}{x-coordinates of all nodes in the domain (boundary and free) }   \\
 \hline
 y - real array (lnn) &  \multicolumn{2}{|c|}{y-coordinates of all nodes in the domain (boundary and free) }   \\
 \hline
  nn - integer & number of nodes in the domain & updated number of nodes in the domain  \\ 
 \hline
 lnn - integer & \multicolumn{2}{|c|}{must specify the estimated final x and y size (nn + all new inserted nodes)} \\
  \hline
 nt - integer & number of mesh triangles & updated number of mesh triangles  \\
 \hline
 nl - integer & number of mesh edges & updated number of mesh edges  \\ 
 \hline
  iw - int array (liw)  & boundary structure & total mesh structure \\
 \hline
 liw - int & \multicolumn{2}{|c|}{drw (dimension of array iw, at least $ liw \geq 9 + 4*\text{nl[on entry]} + 7* \text{nt [on entry]}$} \\
 \hline
 drw - subroutine  & \multicolumn{2}{|c|}{drw (idrw,x1,x2,y1,y2) } \\
 \hline
 spc - subroutine \ref{func:spc} & \multicolumn{2}{|c|}{spc (x,y,delta) } \\
 \hline
 factor - real & \multicolumn{2}{|c|}{controls the size of the triangles} \\
 \hline
 ido - integer & \multicolumn{2}{|c|}{number of version and specific final smoothing mechanism \ref{chap:ido}}  \\
  \hline
 ifail - integer & 0 & error messages \\
 \hline
\end{tabular}
\end{center}

\subparagraph{Ido values}
In this subroutine, the $ido$ value sets the position of the node to be inserted.
\begin{center}
\centering
    \begin{tabular}{ |c|c| }
    \hline
        value & specification \\
        \hline
        1 & in the gravity centre \\
        2 & in the centre of the circumcircle \\
        3 & with the distance of delta (provided by spcg) to the vertices \\
        \hline
    \end{tabular}
\end{center}

\subparagraph{Ifail values}

\begin{center}
\centering
    \begin{tabular}{ |c|c| }
    \hline
        value & specification \\
        \hline
        1 & No space (in $lw$, $x$ or $y$) to insert node \\
        0 & Normal exit \\
        -1 & The delivered value of liw is too small to start the subroutine \\
        -4 & The delivered value of $nn$, $nt$ or $nl$ is not correct \\
        \hline
    \end{tabular}
\end{center}

\begin{figure}[H]
    \centering
    \includegraphics[width=7cm]{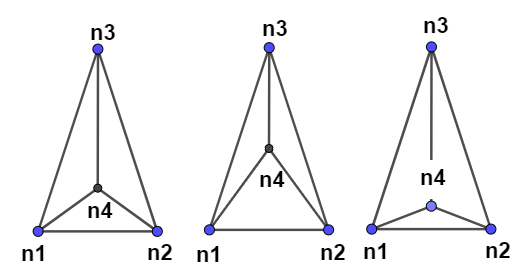}
    \caption{Node insertion possibilities: a) centre of gravity b) circumcircle centre c) spacing distance to n1,n2 or n3}
    \label{fig:afmtrim5}
\end{figure}

\subparagraph{Code description}
To execute this subroutine correctly, the integer work array ($iw$) has to have a certain structure and a minimum length ($liw$), which is defined with the length of the two connectivity-tables, created in the connec subroutine, the length of a helping part which is important for the following calculations and with th eparts of $iw$ which are mentioned above. If the length of the $x$/$y$ arrays ($lnn$) or $iw$ ($liw$) is too small (to get started of to insert all nodes), the routine returns with an ifail. The subroutine also performs an error check of the delivered data before creating the two connectivity tables $mcnt$ and $mclt$ placing them at the end of the array $iw$. The loop over all old triangles begins, calculating the centre of the current triangle and examining, whether a node insertion is necessary or not. Therefor, the external $spc$ delivers a criteria which will be compared with the area of the current triangle. If a node insertion is necessary, the routine defines the number and the position of each edge belonging to this triangle, adds the calculated node to the tables of coordinates $x$ and $y$ and adds the new three edges to the end of $mcnl$. For the position of this node to be inserted, the user has the possibility to select between three versions (see definition of ido and Figure \ref{fig:afmtrim5}).

The following description belongs to $ido = 1$ because the two other versions are not yet existent:
The new three triangles are inserted in $mctl$ in the way, that one new triangle gets the number and place of the old one, while the both remaining new ones are added at the end of $mctl$ in $iw$. Here, the user has the option of drawing the new edges (see \ref{func:drw}). After a necessary modification of $mctl$ (the neighbour triangle has to be changed), the routine takes the next old triangle. Just before the subroutine returns to the main program, it actualises the $iw$ info part with the new number of nodes, triangles and edges and also sets the pointer to the end of the table.

\paragraph{Edswap:}

The algorithm performs a \emph{diagonal edge swapping} (mentioned in chapter \ref{chap:remesh}) and can be used apart from grid refinement also for grid generation algorithms (Lawson, see chapter \ref{chap:delaunay}). in any case, there has to be an existing grid, which do in some parts not correspond to the Delaunay triangulation criteria.

The call to this subroutine needs besides the grid structure also some further information to start correctly:

\begin{center}
    $call$ edswap(x,y,nn,nt,nl,iw,liw,drw,ifail)
\end{center}

\subparagraph{Arguments values}
\begin{center}
\begin{tabular}{ |c|c|c| } 
 \hline
 variable & input & output \\
 \hline
 x - real array (lnn) &  \multicolumn{2}{|c|}{x-coordinates of all nodes in the domain (boundary and free) }   \\
 \hline
 y - real array (lnn) &  \multicolumn{2}{|c|}{y-coordinates of all nodes in the domain (boundary and free) }   \\
 \hline
  nn - integer & \multicolumn{2}{|c|}{ number of nodes in the domain }   \\ 
 \hline
 nt - integer & \multicolumn{2}{|c|}{ number of mesh triangles }    \\
 \hline
 nl - integer &  \multicolumn{2}{|c|}{ number of mesh edges }  \\ 
 \hline
  iw - int array (liw) \ref{func:iw} & boundary structure & total mesh structure \\
 \hline
 liw - int & \multicolumn{2}{|c|}{drw (dimension of array iw, at least $ liw \geq 9 + 4*\text{nl[on entry]} + 7* \text{nt [on entry]}$} \\
 \hline
 drw - subroutine \ref{func:drw} & \multicolumn{2}{|c|}{drw (idrw,x1,x2,y1,y2) } \\
 \hline
 ifail - integer & 0 & error messages \\
 \hline
\end{tabular}
\end{center}

\subparagraph{Ifail values}

\begin{center}
\centering
    \begin{tabular}{ |c|c| }
    \hline
        value & specification \\
        \hline
        0 & Normal exit \\
        -1 & The delivered value of liw is too small to start the subroutine \\
        -4 & The delivered value of $nn$, $nt$ or $nl$ is not correct \\
        \hline
    \end{tabular}
\end{center}

\subparagraph{Code description}
To execute this subroutine correctly, the integer work array ($iw$) has to have a certain structure and a minimum length ($liw$), defined with the length of the two connectivity-tables, which will be created in the connec subroutine, the length of a helping part which is important for the following calculations and with the parts of $iw$, which has been mentioned before.

The subroutine starts initiating pointers for the $iw$ workarray and extracting information out of $iw$. If the lenght of $iw$ ($liw$) is too small, the routine returns with an $ifail$. The subroutine also performs an error check of the delivered data before creating the two connectivity tables $mcnt$ and $mclt$.

\begin{figure}[H]
    \centering
    \includegraphics[width=10cm]{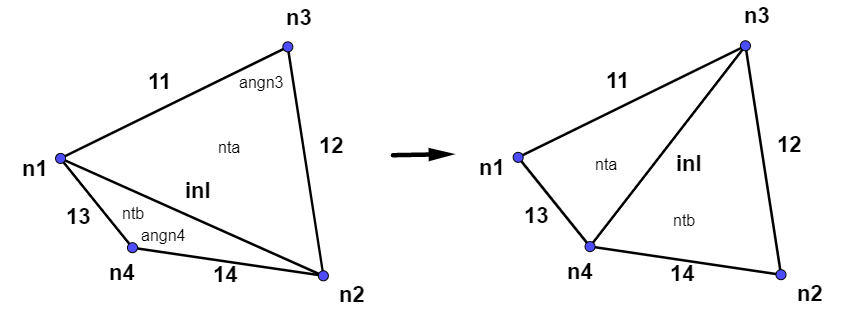}
    \caption{Definition of variables}
    \label{fig:variables..}
\end{figure}

After these two connectivity tables ahve been placed ant the end of $iw$, the algorithm starts examining all quadrilaterals if their existing diagonal is a boundary edge. If not, the number and position of the four nodes of this quadrilateral are defined and the angles "above" and "below" the diagonal are calculated. (For the position of the nodes, triangles, edges adn angles see Figure \ref{fig:variables..}). The edge swapping will be performed, if the sum of this two angles is greater than the sum of the equivalent angles in the new configuration, i.e. the sum of the old configuration is greater than 180\degree. This criterion also exluces nonconvex quadrilaterals which is necessary to avoid edgecrossing.

After the number and position of the edges in this quadrilateral are defined, the routine changes the four connectivity tables contained in $iw$ and repeats the algorithm until the quadrilateral of each edge is checked. At the end of this routine, the possibility is given to drawthe new mesh like the user wants (see \ref{func:drw} subroutine) before returning the main program.

\paragraph{Smooth:}

This subroutine realises a \emph{posterior grid smoothing} according to the \emph{spring model} and can be used for remeshing of grid adaptation (in chapter \ref{chap:remesh}). In any case, there has to be an initial grid, but it makes no difference, in which way it was created.

The call to this subroutine needs besides the grid structure also some further information to start correctly:

\begin{center}
    $call$ smooth(x,y,nn,lmc,iw,liw,drw,ifail)
\end{center}

\subparagraph{Arguments values}
\begin{center}
\begin{tabular}{ |c|c|c| } 
 \hline
 variable & input & output \\
 \hline
 x - real array (lnn) &  \multicolumn{2}{|c|}{x-coordinates of all nodes in the domain (boundary and free) }   \\
 \hline
 y - real array (lnn) &  \multicolumn{2}{|c|}{y-coordinates of all nodes in the domain (boundary and free) }   \\
 \hline
  nn - integer & \multicolumn{2}{|c|}{ number of nodes in the domain }   \\ 
 \hline
 lmc - integer &  \multicolumn{2}{|c|}{ the macimum number of partner nodes which a node in the grid can have }  \\ 
 \hline
  iw - int array (liw) \ref{func:iw} & boundary structure & total mesh structure \\
 \hline
 liw - int & \multicolumn{2}{|c|}{drw (dimension of array iw, at least $ liw \geq 9 + 4*\text{nl[on entry]} + 7* \text{nt [on entry]}$} \\
 \hline
 drw - subroutine \ref{func:drw} & \multicolumn{2}{|c|}{drw (idrw,x1,x2,y1,y2) } \\
 \hline
 ifail - integer & 0 & error messages \\
 \hline
\end{tabular}
\end{center}

\subparagraph{Ifail values}

\begin{center}
\centering
    \begin{tabular}{ |c|c| }
    \hline
        value & specification \\
        \hline
        1 & No free nodes \\
        0 & Normal exit \\
        -1 & The delivered value of liw is too small to start the subroutine \\
        -4 & The delivered value of $nn$, $nt$ or $nl$ is not correct \\
        \hline
    \end{tabular}
\end{center}

\subparagraph{Code description}

To execute this subroutine correctly, the integer work array ($iw$) has to have a certain structure and minimum length ($liw$), defined with the length of the connectivity table, which will be created in the connec subroutine and with the parts of $iw$, which are mentioned above.

If the length of $iw$ ($liw$) is too small, the delivered number of total nodes is not correct, or if there are no free nodes to move, the subroutine returns to the main program with an $ifail$. After this error check and the creation of the $mcnn$ matrix in the $iw$ work array, the routine calculates the average quadratic distance ($avdist$) of the nodes in the domain, which will later serve as the move criteria.

In a loop over all \textbf{free} nodes in the domain, the routine searches the node which has to be moved most, calculates its new coordinates and changes the x/y - coordinate tables.

To calculate the new coordinates of the node, the $spring model$ is used, i.e. each belonging side to the node is considered as spring of \textbf{equal} stiffness. The nodes are iteratively moved until the spring system is in equilibrium (the movement is less than the criteria).

In this algorithm, the "springs
 have equal stiffness, but a stiffness according to the local spacing factor can also be used and its implementation is quite simple. A locally different stiffness would give the user more control over the triangle and mesh structure but it would slow down the execution of the subroutine considerably.
 
\begin{figure}[H]
    \centering
    \includegraphics[width=8cm]{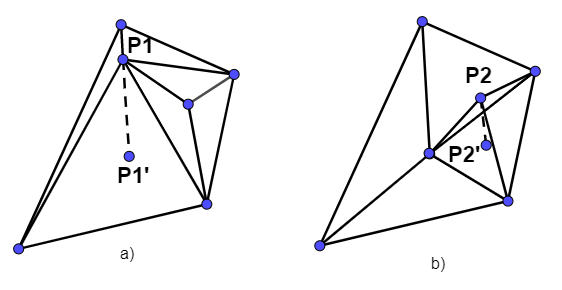}
    \caption{Possible edgecrossing after smoothing: a)Distance P1-P1' is greater than the criteria, b) distance P2-P2' not}
    \label{fig:edgecrossing fig}
\end{figure}

 In fact, an edge crossing provoked by smoothing could occur in rare cases if the criteria is too big (see figure \ref{fig:edgecrossing fig}) and a subsequent edge crossing control should be performed.
 
 Just before returning to the main program, the subroutine offers the possibility to draw the changed mesh like the user wants (see \ref{func:drw} subroutine):

\subsubsection{Tools}

\paragraph{Brdr:}  \label{func:brdr}

The call to the brdr subroutine is the following:

\begin{center}
    $call$ brdr(nn,nl,mb,nc,x,y,lnn,bnd,spc,drw,w,lw,iw,liw,ido,ifail)
\end{center}

\subparagraph{Arguments values}
\begin{center}
\begin{tabular}{ |c|c|c| } 
 \hline
 variable & input & output \\
 \hline
  nn - integer &  number of boundary nodes & total number of boundary nodes (old $+$ new)  \\ 
 \hline
 nl - integer & number of boundary edges & total number of boundary edges (old $+$ new)  \\ 
 \hline
 mb  - integer array (nc) & empty &nn as function of the segment  \\
 \hline
 nc - integer & \multicolumn{2}{|c|}{number of boundary nodes } \\
 \hline
 x - real array (nn) &  x-coordinates of boundary points & all boundary points x-coords     \\
 \hline
 y - real array (nn) &  u-coordinates of boundary points & all boundary points y-coords   \\
 \hline
 lnn - integer & \multicolumn{2}{|c|}{x and y dimension (estimated (nn [on entry] + added nodes)) }  \\
 \hline
 bnd - subroutine \ref{func:bnd} &\multicolumn{2}{|c|}{bnd (in,nb,xb,yb,nj) } \\
 \hline
 spc - subroutine \ref{func:spc} & \multicolumn{2}{|c|}{spc (x,y,delta) } \\
 \hline
 drw - subroutine \ref{func:drw} & \multicolumn{2}{|c|}{drw (idrw,x1,x2,y1,y2) } \\
 \hline
 w - real array (lw) & \multicolumn{2}{|c|}{used as real working space} \\
 \hline
 lw - integer & \multicolumn{2}{|c|}{ w dimension lw $\geq 4 *$ maximum nodes per segment  }\\
 \hline
 iw - int array (liw) \ref{func:iw} & \multicolumn{2}{|c|}{boundary structure  }  \\
 \hline
 liw - int & \multicolumn{2}{|c|}{drw (dimension of array iw, at least $ liw \geq 5 * \text{max. number of edges (nl)} + 9$} \\
 \hline
 ido - integer & \multicolumn{2}{|c|}{ number of version }  \\
  \hline
 ifail - integer & 0 & error messages \\
 \hline
\end{tabular}
\end{center}

\subparagraph{Ido values}
In this subroutine, the $ido$ value contains the number of the version of the desired dlny triangulation.
\begin{center}
\centering
    \begin{tabular}{ |c|c| }
    \hline
        value & specification \\
        \hline
        0 & without boundaries \\
        1 & with boundaries defined by the user \\
        2 & with boundaries, defined by the user interpolated with splines \\
        \hline
    \end{tabular}
\end{center}

\subparagraph{Ifail values}

\begin{center}
\centering
    \begin{tabular}{ |c|c| }
    \hline
        value & specification \\
        \hline
        0 & Normal exit \\
        -1 & The input data is not correct, liw too small \\
        -2 & The expected number of nodes ($lnn$) is too small \\
        -3 & Too many nodes in the boundary segment \\
        \hline
    \end{tabular}
\end{center}

\subparagraph{Code description}

After some previous calculation and initialisation, the value of $ido$ decides, which boundary distribution sill be done.

For $ido$ = 0, the user wants to triangulate an open domain and an initial edge is necessary to begin the triangulation. This first edge consists of the first node, delivered by the subroutine $fnode$ and the second node, delivered by $snode$. This first edge is implanted in the $iw$ array and the variables for the information part in the $iw$
 array updated.
 
 For $ido = 1$ or $ido  = 2$ , the routine receives in a loop over all boundary segments the number and coordinates of the boundary nodes and the number of the next connecting boundary segment. If $ido=2$, the subroutine splines ($spl$) is called which returns with the member and coordinates of all boundary nodes of the actual boundary segment. The routine implants this information in the stack ($iw$ and the x,y-coordinates tables while using a real work array ($w$) to store the coordinates of all boundary nodes ($xb$,$yb$) of the segment. $w$ is a real work array which stores: 
 \begin{itemize}
     \item w(1... nw2): x-coordinates of the nodes which define the segment, delivered by the bnd subroutine.
     \item w(nw2+1...nw3): y-coordinates of the nodes which define the segment, delivered by the bnd subroutine.
     \item w(nw3+1...nw4): x-coordinates of the boundary nodes of the segment.
     \item w(nw4+1... ): y-coordinates of the boundary nodes of the segment.
 \end{itemize}
 
 After this insertion of boundary nodes, their coordinates are transmitted to the external dew to give the user the possibility to evaluate this information (e.g. plotting the boundary on screen or printer).
 
 At the end of this routine, the integer work array $iw$ is reorganised and contains the boundary information.

\paragraph{Connec:}

Many routines of the TRIM-Pack need further information about the grid and its connectivities. Therefor, this subroutine connec has been established as an independent algorithm.

To start this subroutine, a mesh has to be existent and its structure has to be transmitted:

\begin{center}
    $call$ connec(lmc,nmc,iw,liw,mc,ido,ifail)
\end{center}

\subparagraph{Arguments values}
\begin{center}
\begin{tabular}{ |c|c|c| } 
 \hline
 variable & input & output \\
 \hline
  lmc - integer &   \multicolumn{2}{|c|}{number of rows of mc  }   \\ 
 \hline
 nmc - integer & \multicolumn{2}{|c|}{number of columns of mc  } \\
 \hline
 iw - int array (liw)  & \multicolumn{2}{|c|}{mesh structure  }  \\
 \hline
 liw - int & \multicolumn{2}{|c|}{drw (dimension of array iw, at least $ liw \geq 5 * \text{max. number of edges (nl)} + 9$} \\
 \hline
 mc - int array (lmc,nmc) & \multicolumn{2}{|c|}{contains the demanded connectivity matrix} \\
 \hline
 ido - integer & \multicolumn{2}{|c|}{ number of the desired matrix to be created }  \\
  \hline
 ifail - integer & 0 & error messages \\
 \hline
\end{tabular}
\end{center}

\subparagraph{Ido values}
In this subroutine, the $ido$ value contains the desired matrix to be created.
\begin{center}
\centering
    \begin{tabular}{ |c|c| }
    \hline
        value & specification \\
        \hline
        1 & mctt \\
        2 & mcnn \\
        3 & mctn \\
        4 & mcnt \\
        5 & mcln \\
        6 & mcnl \\
        7 & mctl \\
        8 & mclt \\
        \hline
    \end{tabular}
\end{center}

\subparagraph{Ifail values}

\begin{center}
\centering
    \begin{tabular}{ |c|c| }
    \hline
        value & specification \\
        \hline
        0 & Normal exit \\
        -1 & liw too small \\
        -5 & nmc is too small \\
        -6 & lmc too small \\
        \hline
    \end{tabular}
\end{center}

\subparagraph{Code description}

After the initialisation of some variables, the extraction of necessary information out of the information part of $iw$ and an error check of the minimum length of $iw$, the value of $ido$ decides, which connectivity table will be calculated:
\begin{itemize}
    \item mctt (3,nt) [ido = 1]: matrix connectivity nodes-triangles. After a routine check of the dimensions a loop over all edges in the array $iw$ starts and stores the three "neighbour"-triangles in mc(mi/m2,nt1/nt2) where m1/m2 signifies the number of the current "neighbour"-triangles, and nt1/nt2 the current "centre triangle". Here, the helping space in $iw$ is necessary and contains the "neighbour"-triangle counter (n1/m2) which is the position where to place the triangle in $mc$. 
    
    \item mcnn (?,nn) [ido = 2]: matrix connectivity nodes-nodes. After the dimension check of \textbf{nmc}, the routine initialises the first row of mc which will serve as counter of partner-nodes and furthermore as position, where to place each partner-node in $mc$. Then, the loop over all edges in $iw$ starts relating each node (n1/n2) with its partner-node(n2/n1). Each time, a node is added, the routine checks the dimensio of rows of mc (lmc).
 
    \item mctn (?,tn) [ido = 3]: matrix connectivity triangles-nodes. After the dimension check of \textbf{nmc}, the routine initialises the first row of mc which will serve as counter of triangles which share the node and furthermore as position, taking two connected nodes and their two bordering triangles, examining with another loop if one or both triangles already exist in mc(x,n1/n2) and finally inserts then in $mc$ if necessary with changing of the "full-cells"-counter mc(1,n1/n2).
    
    Each time, a node is added, the routine checks the dimension of rows of mc (lmc).
    
    \item mcnt (?,nt) [ido = 4]: matrix connectivity nodes-triangles. After both dimensions are checked the loop over all edges in $iw$ starts, taking two connected nodes and their two bordering triangles, examining with another loop if one or both nodes already exist in mc(x,nt1/nt2), changing the value of the helping space in iw(ihlp+nt1/nt2) which serves as nodes counter and therefore as position-hint where to place the number of the node in mc, and finally, inserting the number of the node(s) in mc.
    
    \item mcln (?,ln) [ido = 5]: matrix connectivity edges-nodes. After the dimension check of $nmc$, the routine initialises th efirst row of $mc$ which will serve as counter of edges which share the node and furthermore as position-marker, where to place each edge in $mc$. Then, the loop over all edges starts, inserting the values in mc(x,n1/n2) while changing the position-marker and checking the dimension of rows ($lmc$).
    
    \item mcnl (?,nl) [ido = 6]: matrix connectivity nodes-edges. After both dimensions are checked the loop over all edges in $iw$ starts, inserting the values of $iw$ in $mc$. While this table already exists in iw, this routine is relatively simple.
    
    \item mctl (?,tl) [ido = 7]: matrix connectivity triangles-edges. After both dimensions are checked the loop over all edges in $iw$ starts, inserting the values of $iw$ in $mc$. While this table already exists in $iw$, this routine is relatively simple.
    
    \item mclt (?,lt) [ido = 8]: matrix connectivity edges-triangles. After both dimensions are checked the loop over all edges in $iw$ starts, relating every triangle-number, which has a number grater than 0, with its edges, while the helping space in iw serves as counter and therefore as position-marker of $mc$. 
\end{itemize}

\paragraph{Edcross:}

This subroutine is necessary part of the dlny and afm subroutine but can also be used for a final mesh structure check (e.g. after smoothing was performed).

The call is the following:

\begin{center}
    $call$ edcros(xa,ya,xb,yb,x1,y1,x2,y2,indi)
\end{center}

\begin{figure}[H]
    \centering
    \includegraphics[width=5cm]{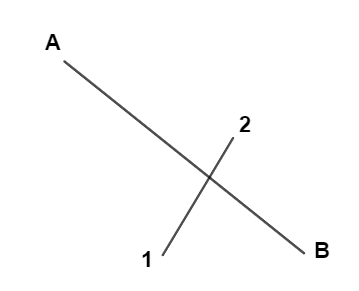}
    \caption{edcros scheme}
    \label{fig:ecros scheme}
\end{figure}

\subparagraph{Arguments values}
\begin{center}
\begin{tabular}{ |c|c|c| } 
 \hline
 variable & input & output \\
 \hline
  xa,ya - real &   \multicolumn{2}{|c|}{coordinates of the node A (edge A-B)  }   \\ 
 \hline
 xb,yb - real &   \multicolumn{2}{|c|}{coordinates of the node B (edge A-B)  }   \\ 
 \hline
  x1,y1 - real &   \multicolumn{2}{|c|}{coordinates of the node 1 (edge 1-2)  }   \\ 
 \hline
  x2,y2 - real &   \multicolumn{2}{|c|}{coordinates of the node 2 (edge 1-2)  }   \\  
 \hline
 indi - integer &  \multicolumn{2}{|c|}{ Indicator of edge crossing }   \\  
 \hline
\end{tabular}
\end{center}

\subparagraph{indi values}
.
\begin{center}
\centering
    \begin{tabular}{ |c|c| }
    \hline
        value & specification \\
        \hline
        0 & there is no edge crossing between the eges A-B and edge 1-2 \\
        1 & there exist an edge crossing \\
        \hline
    \end{tabular}
\end{center}

\subparagraph{Code description}

The subroutine edcros checks a possible crossing between the edge A-B and the edge 1-2. Therefore a coordinate transformation to the A-B system is realised which will simplify the control routines. The edcros-routine is divided in three different cases depending on the situation of the edge 1-2 in relation ot the edge A-B: Vertical, horizontal and any other possibility. If an edge crossing occurs, the subroutine edcros will return to the main program an indicator (indcr) with the value 1. If there is no edge crossing the value of indcr will be 0.

In this subroutine a special parameter zero is defined. Zero is a very small positive number, greater than the round off error of the machine, Zero and -zero define a strip surrounding the real 0 value.l This measure must be taken because of the generated error in the subtractions between real number in the FORTRAN language.

\paragraph{Spline:}

This subroutine is here used only for the boundary node distribution but further use, e.g. in the node-insertion subroutine is also possible.

The call is the following:

\begin{center}
    $call$ spl(n,x,y,m,xx,yy,spc)
\end{center}

\subparagraph{Arguments values}
\begin{center}
\begin{tabular}{ |c|c|c| } 
 \hline
 variable & input & output \\
 \hline
  n - int &   \multicolumn{2}{|c|}{number of describing boundary nodes of the actual boundary segment  }   \\ 
 \hline
 x,y - real array (n) &   \multicolumn{2}{|c|}{coordinates of the describing boundary nodes }   \\ 
 \hline
  m - int &   \multicolumn{2}{|c|}{number of created final boundary nodes  }   \\ 
  \hline
  xx,yy - real array & \multicolumn{2}{|c|}{coordinates of the final created boundary nodes  } \\
 \hline
spc - subroutine \ref{func:spc} & \multicolumn{2}{|c|}{spc (x,y,delta) (must be declared as external) } \\
 \hline
\end{tabular}
\end{center}

\subsubsection{Other Subroutines}

\paragraph{Stmsh:}

This subroutine can be used after the generation of a grid to evaluate its quality or to obtain a criteria whether remeshing is necessary or not.

It is obvious, that this subroutine needs therefore the whole grid information and also the support of the connec subroutine:

\begin{center}
    $call$ stmsh(nn,nt,nl,x,y,liw,iw,kst,ido,ifail)
\end{center}

\subparagraph{Arguments values}
\begin{center}
\begin{tabular}{ |c|c|c| } 
 \hline
 variable & input & output \\
 \hline
  nn - integer & \multicolumn{2}{|c|}{ number of nodes in the domain }   \\ 
 \hline
 nt - integer & \multicolumn{2}{|c|}{ number of mesh triangles }    \\
 \hline
 nl - integer &  \multicolumn{2}{|c|}{ number of mesh edges }  \\ 
 \hline
 x - real array (lnn) &  \multicolumn{2}{|c|}{x-coordinates of all nodes in the domain  }   \\
 \hline
 y - real array (lnn) &  \multicolumn{2}{|c|}{y-coordinates of all nodes in the domain  }   \\
  \hline
 liw - int & \multicolumn{2}{|c|}{drw (dimension of array iw} \\
 \hline
  iw - int array (liw) \ref{func:iw} & boundary structure & total mesh structure \\
 \hline
 kst - int array (21) & \multicolumn{2}{|c|}{contains the statistical information} \\
 \hline
 ido - integer & \multicolumn{2}{|c|}{number of the statistical information} \\
 \hline
 ifail - integer & 0 & error messages\\
 \hline
\end{tabular}
\end{center}

\subparagraph{Ido values}
\begin{center}
\centering
    \begin{tabular}{ |c|c| }
    \hline
        value & specification \\
        \hline
        1 & Number of connectivities nodes-nodes \\
        2 & Number of connectivities triangles-nodes \\
        3 & Number of areas of the triangles in percentage of a spacing-triangle area \\
        4 & Number of lenghts of the edges in percentage of a spacing-edge length \\
        5 & Number of angles of the triangles in percentage of 60\degree \\
        \hline
    \end{tabular}
\end{center}

\subparagraph{Ifail values}
\begin{center}
\centering
    \begin{tabular}{ |c|c| }
    \hline
        value & specification \\
        \hline
        2 & The validation nt + nl = nn + 1 is not correct \\
        0 & Normal exit \\
        -11 & The dimension of kst is too small \\
        \hline
    \end{tabular}
\end{center}

\paragraph{Centre:}

    This subroutine calculates the centre of the circle defined by three points.

\begin{center}
    $call$ centre(xn1,yn1,xn2,yn2,xn3,yn3,xc,yc,r2,mark)
\end{center}

\begin{figure}[H]
    \centering
    \includegraphics[width=5cm]{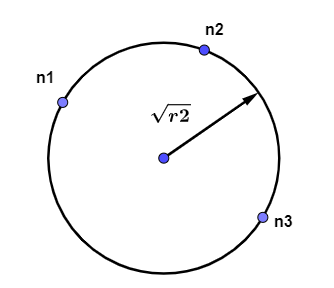}
    \caption{centre shceme}
    \label{fig:centre}
\end{figure}

\subparagraph{Arguments values}
\begin{center}
\begin{tabular}{ |c|c|c| } 
 \hline
 variable & input & output \\
 \hline
  xn1,yn1 & \multicolumn{2}{|c|}{ coordinates of first describing node }   \\ 
 \hline
 xn2,yn2 & \multicolumn{2}{|c|}{ coordinates of second describing node }   \\ 
 \hline
  xn3,yn3 & \multicolumn{2}{|c|}{ coordinates of third describing node }   \\  
 \hline
 xc,yc & \multicolumn{2}{|c|}{ coordinates of the centre }   \\ 
 \hline
  xr2 - real  & \multicolumn{2}{|c|}{ Quadratic radius of the circle }   \\ 
  \hline
  mark & \multicolumn{2}{|c|}{marker of three nodes on a line (mark = 1) }   \\ 
 \hline
\end{tabular}
\end{center}

\paragraph{Fnode:}

    This subroutine searches over all y-coordinates the smallest and returns the belonging node number as n1.

\begin{center}
    $call$ fnode(v,nn,n1)
\end{center}

\subparagraph{Arguments values}
\begin{center}
\begin{tabular}{ |c|c|c| } 
 \hline
 variable & input & output \\
 \hline
  v - real array (nn) & \multicolumn{2}{|c|}{ y-coordinate table }   \\ 
 \hline
 nn - int & \multicolumn{2}{|c|}{ number of all nodes }   \\ 
 \hline
  n1 - int & \multicolumn{2}{|c|}{ number of the first node }   \\  
 \hline
\end{tabular}
\end{center}

\paragraph{Snode:}

    Snode find a second initial node, which will create with the first node the initial edge of the triangulation. This second node will be the one which forms the minimum angle to the horizontal. In this way it will be quite sure that the first edge is in a right position for the following triangulation.

\begin{center}
    $call$ fnode(x,y,nn,n1,n2)
\end{center}

\subparagraph{Arguments values}
\begin{center}
\begin{tabular}{ |c|c|c| } 
 \hline
 variable & input & output \\
 \hline
  x,y - real array (nn) & \multicolumn{2}{|c|}{ x and y coordinates arrays }   \\ 
 \hline
 nn - int & \multicolumn{2}{|c|}{ number of all nodes }   \\ 
 \hline
  n1 - int & \multicolumn{2}{|c|}{ number of the first node }   \\  
 \hline
   n2 - int & \multicolumn{2}{|c|}{ number of the second node }   \\  
 \hline
\end{tabular}
\end{center}
